\documentclass[aps,onecolumn,prd,amsmath,nofootinbib,preprintnumbers,superscriptaddress]{revtex4-1}
\usepackage[utf8]{inputenc}
\usepackage{epsfig}
\usepackage{slashed}
\usepackage{listings}
\usepackage{booktabs,multirow,tabularx}
\usepackage{amsmath}
\usepackage{amsfonts}
\usepackage{float} 
\usepackage{subfig} 
\usepackage{morefloats}
\usepackage{color}
\usepackage{multirow,bigdelim}
\usepackage{lineno}
\graphicspath{{./figures/}}
\usepackage[export]{adjustbox}
\usepackage{diagbox}

\newcommand{\harpoon}{\overset{\rightharpoonup}}

\newcommand{\ben}{\begin{enumerate}}
\newcommand{\een}{\end{enumerate}}
\newcommand{\bit}{\begin{itemize}}
\newcommand{\eit}{\end{itemize}}
\newcommand{\bc}{\begin{center}}
\newcommand{\ec}{\end{center}}
\newcommand{\bq}{\begin{equation}}
\newcommand{\eq}{\end{equation}}
\newcommand{\bqa}{\begin{eqnarray}}
\newcommand{\eqa}{\end{eqnarray}}

\def\ss{{\bigl.^3\hspace{-1mm}S^{[1]}_1}}
\def\sps{{\bigl.^1\hspace{-1mm}S^{[8]}_0}}
\def\so{{\bigl.^3\hspace{-1mm}S^{[8]}_1}}

\def\pj{{\bigl.^3\hspace{-1mm}P^{[8]}_J}}
\def\p0{{\bigl.^3\hspace{-1mm}P^{[8]}_0}}

\def\tpjs{{\bigl.^3\hspace{-1mm}P^{[1]}_J}}

\def\to{\rightarrow}

\def\ho{ {\sc \small HELAC-Onia}}

\def\Pythia{{\sc\small Pythia}}

\def\MG5aMC{{\sc \small MadGraph5\_aMC@NLO}}

\def\cCode#1{\begin{lstlisting}[mathescape,basicstyle=\small
\ttfamily,frame=leftline,aboveskip=4mm,belowskip=4mm,xleftmargin=20pt,framexleftmargin=10pt,
numbers=none,framerule=2pt,abovecaptionskip=0.0mm,belowcaptionskip=3.5mm #1]}
\def\chCode#1{\begin{lstlisting}[mathescape=true,basicstyle=\small
\ttfamily]}

\begin{document}

\title{$J/\psi$ meson production in association with an open charm hadron at the LHC: A reappraisal}

\author{Hua-Sheng Shao}
\affiliation{Laboratoire de Physique Th\'eorique et Hautes Energies (LPTHE), UMR 7589, Sorbonne Universit\'e et CNRS, 4 place Jussieu, 75252 Paris Cedex 05, France}

\date{\today}

\begin{abstract}
We critically (re)examine the associated production process of a $J/\psi$ meson plus an open charm hadron at the LHC in the proton-proton ($pp$) and proton-lead ($p{\rm Pb}$) collisions. Such a process is very intriguing in the sense of tailoring to explore the double parton structure of nucleons and to determine the geometry of partons in nuclei. In order to interpret the existing $pp$ data with the LHCb detector at the center-of-mass energy $\sqrt{s}=7$ TeV, we introduce two overlooked mechanisms for the double parton scattering (DPS) and single parton scattering (SPS) processes. Besides the conventional DPS mode, where the two mesons are produced almost independently in the two separate scattering subprocesses, we propose a novel DPS mechanism that the two constituent (heavy) quarks stemming from two hard scatterings can form into a composite particle, like the $J/\psi$ meson, during the hadronization phase. It yields a strong correlation in the final state from the two distinct scattering subprocesses per hadron-hadron collision. Such a mechanism should be ubiquitous for quarkonium associated production processes involving more-than-one pair of same-flavor heavy quarks. However, it turns out the corresponding contribution is small in $J/\psi+c\bar{c}$ hadroproduction. On the contrary, we point out that the resummation of the initial state logarithms due to gluon splitting into a charm quark pair is crucial to understand the LHCb measurement, which was overlooked in the literature. We perform a proper matching between the perturbative calculations in different initial-quark flavor number schemes, generically referring to the variable flavor number scheme. The new variable flavor number scheme calculation for the process strongly enhance the SPS cross sections, almost closing the discrepancies between theory and experiment. Finally, we present our predictions for the forthcoming LHCb measurement in $p{\rm Pb}$ collisions at $\sqrt{s_{NN}}=8.16$ TeV. Some interesting observables are exploited to set up the control regions of the DPS signal and to probe the impact-parameter-dependent parton densities in lead.
\end{abstract}

\maketitle


\section{Introduction}

The associated production of a $J/\psi$ meson plus an open charm hadron at the LHC is an intriguing process. It provides a mean to carry out studies on several QCD phenomena. First of all, very large yields of both charm (anti-)quark and charmonium give rise naturally to the emergence of two simultaneous parton scattering subprocesses, {\it aka} double parton scattering (DPS), in a single proton-proton collision due to the composite nature of the proton. In the conventional single parton scattering (SPS), the process has also been proposed to study the heavy quarkonium production mechanism~\cite{Artoisenet:2007xi,Li:2011yc,Shao:2018adj,Baranov:2006dh,Berezhnoy:1998aa,Lansberg:2019adr}, the intrinsic charm content of the proton~\cite{Brodsky:1980pb,Brodsky:2009cf}, and the factorization-breaking effect due to color transfer between the $J/\psi$ and the open charm in the mass threshold regime~\cite{Nayak:2007mb,Nayak:2007zb}. The measurement of such a process in heavy-ion collisions is further motivated to understand the spatial dependence of nuclear modification of parton flux in nuclei~\cite{Shao:2020acd}, given the observation of the large nuclear modification in the inclusive productions of $J/\psi$ and open charm (see, e.g., Ref.~\cite{Kusina:2017gkz} and references therein).

Despite the numerous theoretical motivations, the only measurement at the LHC so far was carried out with the LHCb detector based on a relatively low integrated luminosity $355$ pb$^{-1}$ of Run-I proton-proton ($pp$) data at $\sqrt{s}=7$ TeV in 2012~\cite{Aaij:2012dz}. Yet, the measurement is precise enough to be conclusive in the yields. It is claimed that the SPS stems from the gluon fusion $gg\to J/\psi+c\bar{c}$ (thereafter referred to as ``ggF") is quite negligible compared to the observed production rate. For instance, the measured cross section of $J/\psi+D^0$ and its charge conjugate mode $J/\psi+\bar{D}^0$ in the LHCb acceptance [$2.0<y_{J/\psi},y_C<4.0, P_{T,J/\psi}<12$ GeV and $3<p_{T,C}/{\rm GeV}<12$, where $C$ is either a charm hadron ($D^0, D^+,D_s^+,\Lambda_c^+$) or its charge conjugate] amounts to $161.0\pm 3.7\pm 12.2$ nb, while the corresponding ggF cross section is significantly smaller ranging from $3.7$ to $16$ nb. However, the DPS dominance picture seems to be in trouble in interpreting the shapes of the differential distributions, in particular for the transverse momentum $P_T$ of $J/\psi$, by assuming the standard zero correlation hypothesis between the $J/\psi$ and $D^0$ mesons. Under the simple assumption, the DPS cross section amounts to
\begin{eqnarray}
d\sigma_{J/\psi+D^0}^{{\rm DPS}_1}&=&\frac{d\sigma_{J/\psi}d\sigma_{D^0}}{\sigma_{{\rm eff},pp}},\label{eq:DPS1}
\end{eqnarray}
where we have denoted DPS as DPS$_1$ in order to differentiate another DPS mechanism that will be introduced in Sec.~\ref{sec:newDPS}. Presumably, by construction, the shape of such a DPS$_1$ distribution should be identical to that of the single inclusive (prompt) $J/\psi$ production, while the observed spectrum is much harder than the corresponding DPS$_1$ prediction (see Fig.11a in Ref.~\cite{Aaij:2012dz}).~\footnote{In the ggF SPS curves, we have also routinely included the other insignificant light quark-antiquark annihilation contributions.} In fact, the measured $P_{T,J/\psi}$ spectrum matches the ggF SPS result simulated with \ho\ 2.0~\cite{Shao:2012iz,Shao:2015vga} and  \Pythia\ 8.186~\cite{Sjostrand:2007gs} as reported in Fig.~\ref{fig:ggFDPS1a}. We have used the normalization of DPS$_1$ with $\sigma_{{\rm eff},pp}=15$ mb for the sum of both contributions (black solid line), which correctly reproduces the total yield, and the band stands for the missing higher-order QCD radiative corrections using the standard scale variation. Similar observation can be found in the invariant mass distribution of the two meson system in Fig.~\ref{fig:ggFDPS1b}. While there is no obvious reason that the zero correlation assumption in DPS$_1$ is strongly violated by the initial gluon-gluon correlation~\footnote{There are indeed several mechanisms on the market introducing initial parton-parton correlations in a proton (see, e.g., Ref.~\cite{Sjostrand:2017cdm} for the jointed interactions and rescattering). However, these contributions are expected to be subdominant in our interested domain. Moreover, the Bjorken $x$ (usually logarithmic) dependence in $\sigma_{{\rm eff},pp}$ is also not foreseen to dramatically change the picture here.} since this assumption works pretty well in other processes except the normalization encoded in $\sigma_{{\rm eff},pp}$, it is clear that a coherent physical picture is still missing in interpreting these LHCb data, and therefore prevents us from understanding the process and from using the process as a tool to study other interesting phenomenology.

\begin{figure}
  \centering
  \subfloat[Transverse momentum of $J/\psi$]{
    \includegraphics[width=0.45\textwidth,valign=c,draft=false]{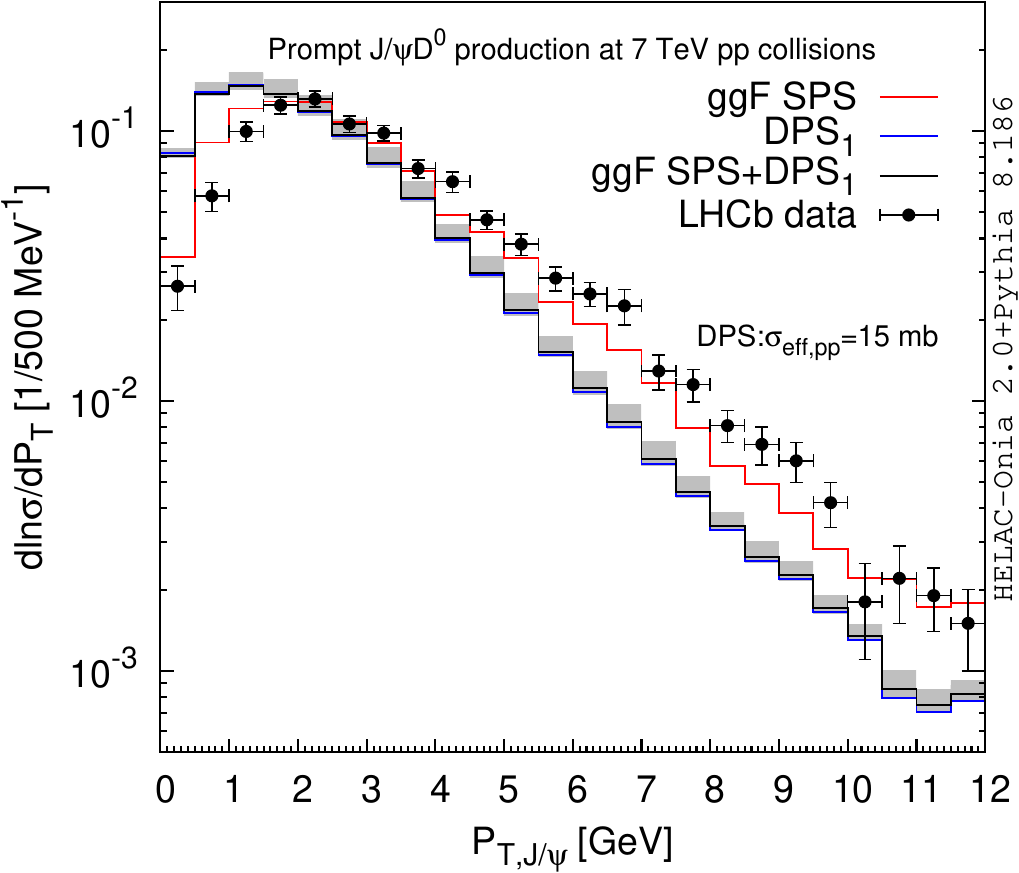}\label{fig:ggFDPS1a}}
  \subfloat[Invariant mass of $J/\psi$ and $D^0$]{\includegraphics[width=0.45\textwidth,valign=c,draft=false]{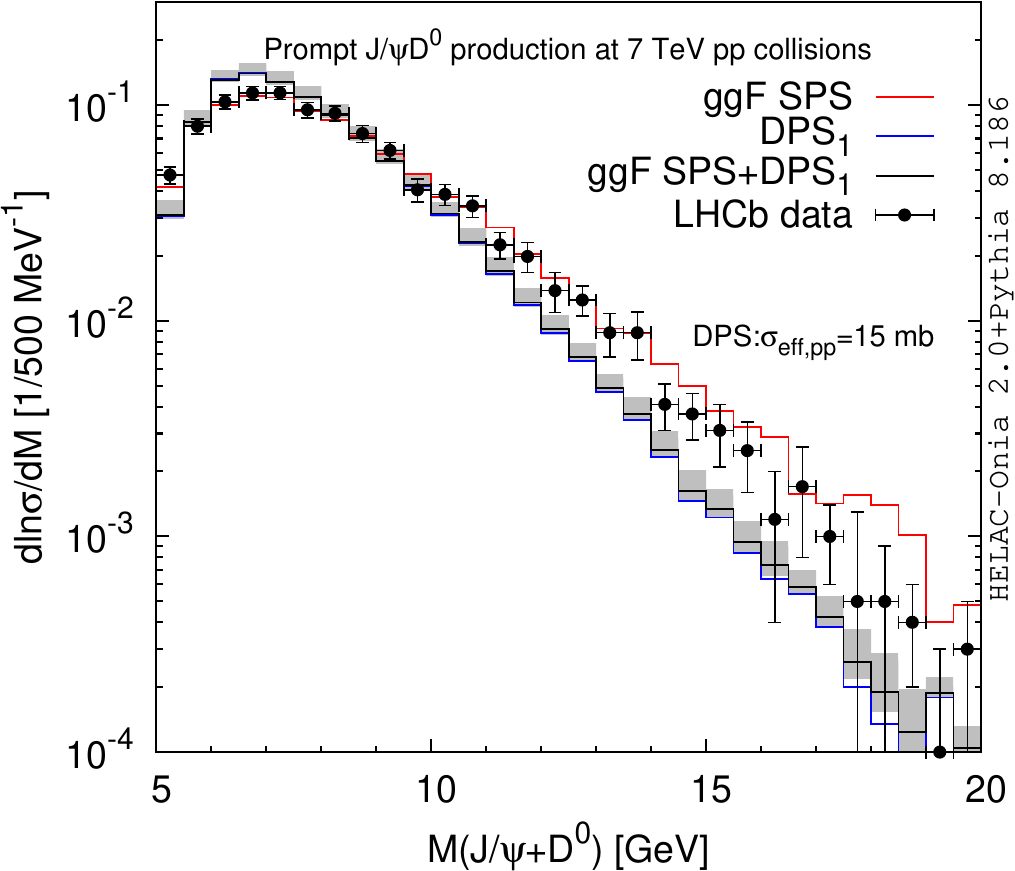}\label{fig:ggFDPS1b}}
  \caption{The shape comparisons between the LHCb data and theoretical calculations (ggF SPS and DPS$_1$) of (a) $P_{T,J/\psi}$ and (b) invariant mass $M(J/\psi+D^0)$. The differential cross sections have been divided by the corresponding integrated cross sections.
  \label{fig:ggFDPS1}}
\end{figure}

The main purpose of the present paper is to address the above mentioned issue in $pp$ collisions and to present our predictions for the forthcoming LHCb measurement of the same process in proton-lead ($p{\rm Pb}$) collisions at $\sqrt{s_{NN}}=8.16$ TeV. In addition, we would like to deduce the potential of constraining the impact-parameter-dependent parton densities in nuclei by measuring the process at the LHC. The remaining context is structured as follows. We introduce a new DPS mechanism in Sec.~\ref{sec:newDPS} and discuss the theoretical aspects of the initial (anti-)charm contributions and the variable flavor number scheme to improve our SPS predictions in Sec.~\ref{sec:vfns}. In Sec.~\ref{sec:ppres}, we perform an analysis on 7 TeV LHCb $pp$ data, where a theory-data comparison can be found. The predictions in $p{\rm Pb}$ collisions at $\sqrt{s_{NN}}=8.16$ TeV, in accompanying with a discussion on the potential of determining spatial-dependent nuclear parton distribution functions (nPDFs), are presented in Sec.~\ref{sec:pA}. The likelihood-base approach we have used to extract $\sigma_{{\rm eff},pp}$ is documented in the Appendix~\ref{app:likelihood}. Some additional plots for the theory-data comparison at 7 TeV $pp$ are collected in Appendix~\ref{app:diffcomp}.

\section{A new DPS mechanism: final state correlations\label{sec:newDPS}}

In contrast to the pointlike elementary particle production,  there could exist another new DPS mechanism for non-point charmonium if the charm quarks are abundantly produced at short distance. As opposed to the ordinary DPS$_1$ process shown in the left graph of Fig.~\ref{fig:feynmandiaga} with its expression given in Eq.(\ref{eq:DPS1}), the charm and anticharm quarks forming a $J/\psi$ meson can come from two different partonic scatterings, where a typical Feynman diagram is  sketched out at the right of Fig.~\ref{fig:feynmandiaga}. In the latter configuration, the final particles $J/\psi$ and open charm quark/hadron are strongly correlated. Such kind of channels are overlooked so far and should be in principle ubiquitous in one or multiple same-flavor quarkonia production (e.g. double $J/\psi$~\cite{Lansberg:2014swa} and triple $J/\psi$~\cite{Shao:2019qob}). On the other hand, they are absent in different flavor quarkonia production (e.g. $J/\psi+\Upsilon$~\cite{Shao:2016wor}) and a quarkonium in association with other elementary particles (e.g. $J/\psi+Z$~\cite{Lansberg:2016rcx} and $J/\psi+W^\pm$~\cite{Lansberg:2017chq}). We call such a contribution as DPS$_2$ in the following context.

\begin{figure}
  \centering
  \subfloat[DPS]{\fbox{
    \includegraphics[width=0.28\textwidth,valign=c,draft=false]{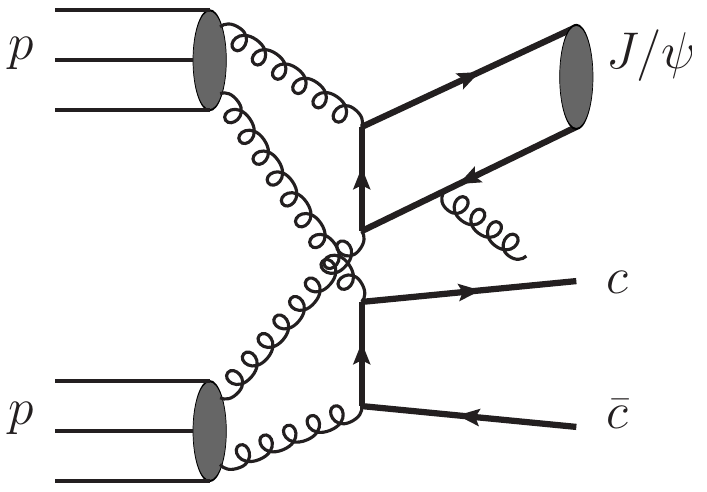}{\Huge $+$}
    \includegraphics[width=0.28\textwidth,valign=c,draft=false]{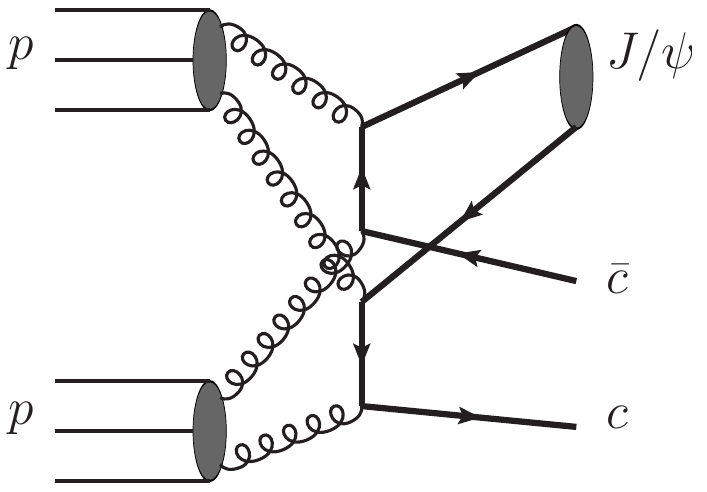}}\label{fig:feynmandiaga}}\\
  \subfloat[SPS]{\fbox{
    \includegraphics[width=0.28\textwidth,valign=c,draft=false]{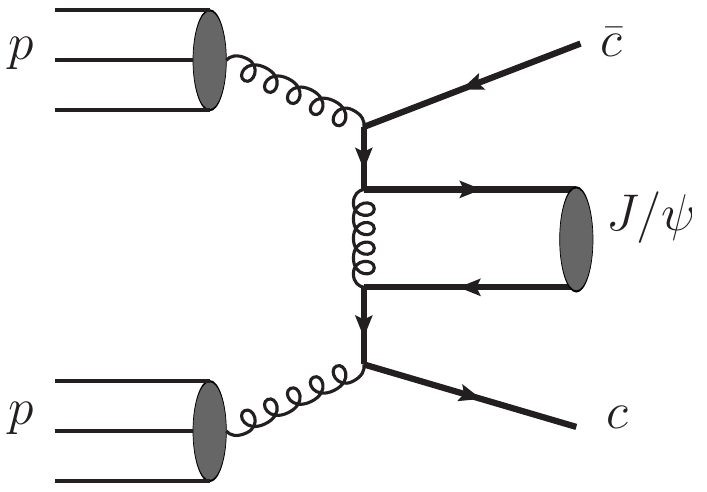}{\Huge $+$}
    \includegraphics[width=0.28\textwidth,valign=c,draft=false]{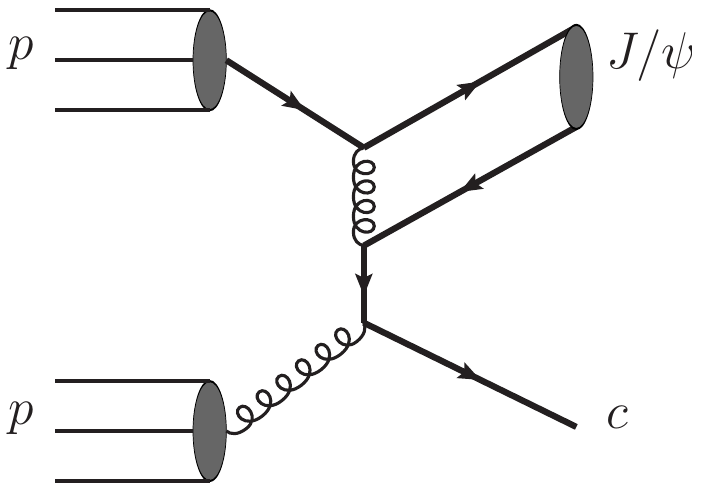}{\Huge $-$}
   \includegraphics[width=0.28\textwidth,valign=c,draft=false]{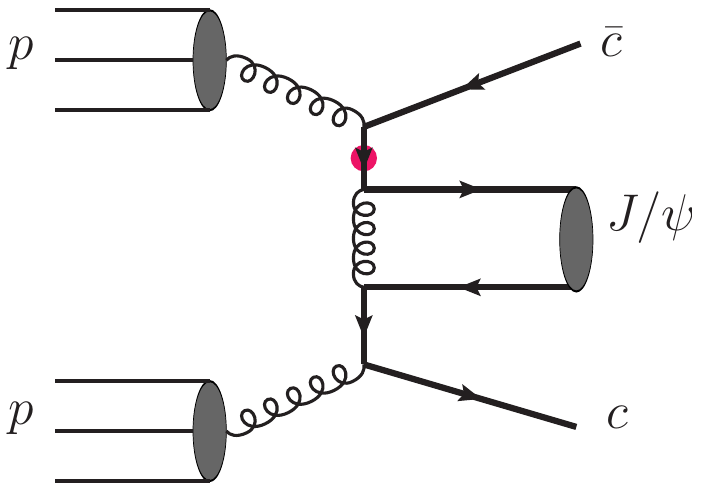}}\label{fig:feynmandiagb}}
  \caption{Representative Feynman diagrams for (a) DPS and
    (b) SPS production. In the third graph of SPS, the red bullet stands for the charm PDF counterterm defined in Eq.(\ref{eq:PDFCT}).
  \label{fig:feynmandiag}}
\end{figure}

By ignoring the correlation of the initial state, the expression of the differential cross section of DPS$_2$ in the collinear perturbative QCD factorization and in the nonrelativistic QCD approach~\cite{Bodwin:1994jh} can be written as
\begin{eqnarray}
d\sigma^{{\rm DPS}_{2}}_{pp\to J/\psi+c\bar{c}}&=&\frac{1}{\sigma_{{\rm eff},pp}}\sum_{i,j,k,l}{dx_1dx_1^{\prime}dx_2dx_2^{\prime}f_{i/p}(x_1)f_{j/p}(x_1^\prime)f_{k/p}(x_2)f_{l/p}(x_2^\prime)}\nonumber\\
&&\times \sum_{n}{\mathbb{P}_{12}^{(n)}{\left[d\hat{\sigma}(ik\to c(\frac{P_{J/\psi}}{2})\bar{c}(p_3))d\hat{\sigma}(jl\to c(p_2)\bar{c}(\frac{P_{J/\psi}}{2}))\right]}\langle\mathcal{O}^{J/\psi}_n\rangle}.\nonumber\\
\end{eqnarray}
where the projector $\mathbb{P}_{12}^{(n)}$ means to cast a charm quark from one partonic scattering and a charm antiquark from another partonic scattering into a given quantum number $n={\bigl.^{2s+1}\hspace{-1mm}L_J^{[c]}}$, $\langle\mathcal{O}^{J/\psi}_n\rangle$ is the long-distance matrix element (LDME), and the first sum runs over all possible initial partons from two beams with the corresponding parton distribution functions (PDFs) as $f_{i/p},f_{j/p},f_{k/p},f_{l/p}$.~\footnote{Without raising any ambiguity in the context, we will suppress the factorization scale dependence in these PDFs unless when necessary.} For example, for $n=\ss$ state, the precise expression is
\begin{eqnarray}
d\sigma^{{\rm DPS}_{2}}_{pp\to J/\psi(\ss)+c\bar{c}}&=&\frac{1}{\sigma_{{\rm eff},pp}}\sum_{i,j,k,l}{dx_1dx_1^{\prime}dx_2dx_2^{\prime}f_{i/p}(x_1)f_{j/p}(x_1^\prime)f_{k/p}(x_2)f_{l/p}(x_2^\prime)}\nonumber\\
&&\times\frac{1}{4\hat{s}_{ik}\hat{s}_{jl}}\overline{\sum}{\left|\mathcal{M}^{{\rm DPS}_{2}}_{ijkl\to J/\psi^{\lambda}(P_{J/\psi})+c(p_c)\bar{c}(p_{\bar c})}\right|^2}\nonumber\\
&&\times d\Phi(p_i,p_k\to \frac{P_{J/\psi}}{2}+q, p_{\bar{c}})d\Phi(p_j,p_l\to p_{c}, \frac{P_{J/\psi}}{2}-q)\nonumber\\
&&\frac{P^0_{J/\psi}}{2}\left(2\pi\right)^3\delta^{(3)}\left(\overrightarrow{q}\right)
\end{eqnarray}
where $\overline{\sum}$ means we have summed all the spin, color of the external particles and have taken the average of the quantum numbers of the initial partons, and $\hat{s}_{ik}=\left(p_i+p_k\right)^2,\hat{s}_{jl}=\left(p_j+p_l\right)^2$. The phase space measure is defined as
\begin{eqnarray}
d\Phi(p_1,p_2\to p_3,p_4)&\equiv & (2\pi)^4\delta^{(4)}\left(p_1+p_2-p_3-p_4\right)\frac{d^3\overrightarrow{p_3}}{(2\pi)^32p_3^0}\frac{d^3\overrightarrow{p_4}}{(2\pi)^32p_4^0}\nonumber\\
&=&\frac{x_1x_2}{2\hat{s}_{12}\left(4\pi\right)}dy_4dy_3dp_{T,3}^2\frac{d\phi_3}{2\pi}.
\end{eqnarray}
Moreover, the DPS$_2$ amplitude becomes
\begin{eqnarray}
\mathcal{M}^{{\rm DPS}_{2}}_{ijkl\to J/\psi^{\lambda}(P_{J/\psi})+c(p_c)\bar{c}(p_{\bar c})}&=&\sum_{s_1,s_2,a_1,a_2}{\frac{\delta^{a_1a_2}}{\sqrt{N_c}}\frac{N(\lambda|s_1,s_2)}{\sqrt{m_c}}\frac{R^{J/\psi}(0)}{\sqrt{4\pi}}}\nonumber\\
&&\times \mathcal{M}_{ik\to c_{a_1}^{s_1}(\frac{P_{J/\psi}}{2}) \bar{c}(p_{\bar c})}\mathcal{M}_{jl\to  c(p_{c}) \bar{c}_{a_2}^{s_2}(\frac{P_{J/\psi}}{2})},\label{eq:DPS2amp}
\end{eqnarray}
where $a_1,a_2$ are two color indices, $s_1,s_2$ are the (anti-)charm quark spin components, $\lambda$ is the helicity of $J/\psi$, $\delta^{a_1a_2}/\sqrt{N_c}$ is the color-singlet projector, $N(\lambda|s_1,s_2)\equiv\frac{1}{2\sqrt{2}m_c}\bar{v}(\frac{P_{J/\psi}}{2},s_2)\slashed{\varepsilon}^*_{\lambda}u(\frac{P_{J/\psi}}{2},s_1)$ is the spin projector in the nonrelativistic limit, and $R^{J/\psi}(0)$ is the wave function at the origin of $J/\psi$. It means in the zero correlation assumption, the DPS$_2$ amplitude is determined by the product of two $2\to 2$ $c\bar{c}$ production amplitudes.

It is convenient to express $k_1^\mu\equiv\left(\frac{P_{J/\psi}}{2}+q\right)^\mu,k_2^\mu\equiv\left(\frac{P_{J/\psi}}{2}-q\right)^\mu$. Then, $k_1+k_2=P_{J/\psi},\frac{k_1-k_2}{2}=q$. Because of $d^3\overrightarrow{k_1}d^3\overrightarrow{k_2}=d^3\overrightarrow{P_{J/\psi}}d^3\overrightarrow{q}$, we have
\begin{eqnarray}
\frac{d^3\overrightarrow{k_1}}{(2\pi)^32k_1^0}\frac{d^3\overrightarrow{k_2}}{(2\pi)^32k_2^0}\frac{P^0_{J/\psi}}{2}(2\pi)^3\delta^{(3)}\left(\overrightarrow{q}\right)&=&\frac{d^3\overrightarrow{P_{J/\psi}}}{(2\pi)^32P_{J/\psi}^0}.
\end{eqnarray}
Therefore, we have arrived at
\begin{eqnarray}
d\sigma^{{\rm DPS}_{2}}_{pp\to J/\psi(\ss)+c\bar{c}}&=&\frac{1}{\sigma_{{\rm eff},pp}}\sum_{i,j,k,l}{dx_1dx_1^{\prime}dx_2dx_2^{\prime}f_{i/p}(x_1)f_{j/p}(x_1^\prime)f_{k/p}(x_2)f_{l/p}(x_2^\prime)}\nonumber\\
&&\times\frac{1}{4\hat{s}_{ik}\hat{s}_{jl}}\overline{\sum}{\left|\mathcal{M}^{{\rm DPS}_{2}}_{ijkl\to J/\psi^{\lambda}(P_{J/\psi})+c(p_c)\bar{c}(p_{\bar c})}\right|^2}\nonumber\\
&&\times (2\pi)^8\delta^{(4)}\left(p_i+p_k-\frac{P_{J/\psi}}{2}-p_{\bar c}\right)\delta^{(4)}\left(p_j+p_l-\frac{P_{J/\psi}}{2}-p_{c}\right)\nonumber\\
&&\times \frac{d^3\overrightarrow{P_{J/\psi}}}{(2\pi)^32P_{J/\psi}^0}\frac{d^3\overrightarrow{p_c}}{(2\pi)^32p_c^0}\frac{d^3\overrightarrow{p_{\bar c}}}{(2\pi)^32p_{\bar c}^0}\nonumber\\
&=&\frac{1}{\sigma_{{\rm eff},pp}}\sum_{i,j,k,l}{f_{i/p}(x_1)f_{j/p}(x_1^\prime)f_{k/p}(x_2)f_{l/p}(x_2^\prime)}\nonumber\\
&&\times\frac{1}{4\hat{s}_{ik}\hat{s}_{jl}}\overline{\sum}{\left|\mathcal{M}^{{\rm DPS}_{2}}_{ijkl\to J/\psi^{\lambda}(P_{J/\psi})+c(p_c)\bar{c}(p_{\bar c})}\right|^2}\nonumber\\
&\times&\frac{x_1x_2x_1^\prime x_2^\prime}{4\hat{s}_{ik}\hat{s}_{jl}}dy_cdy_{\bar c}dy_{J/\psi}dP_{T,J/\psi}^2\frac{d\phi_{J/\psi}}{2\pi}\nonumber\\
&=&\frac{1}{\sigma_{{\rm eff},pp}}\sum_{i,j,k,l}{f_{i/p}(x_1)f_{j/p}(x_1^\prime)f_{k/p}(x_2)f_{l/p}(x_2^\prime)}\nonumber\\
&&\times\frac{1}{4\hat{s}_{ik}\hat{s}_{jl}}\overline{\sum}{\left|\mathcal{M}^{{\rm DPS}_{2}}_{ijkl\to J/\psi^{\lambda}(P_{J/\psi})+c(p_c)\bar{c}(p_{\bar c})}\right|^2}\nonumber\\
&\times&\frac{x_1x_2x_1^\prime x_2^\prime}{4\hat{s}_{ik}\hat{s}_{jl}}d\Delta y_{c\bar{c}}dy_{\bar c}dy_{J/\psi}dP_{T,J/\psi}^2\frac{d\phi_{J/\psi}}{2\pi}.\label{eq:XSDPS2}
\end{eqnarray}
The maximally allowed phase-space integration ranges are $\phi_{J/\psi}\in [0,2\pi), P_{T,J/\psi}\in [0, \frac{\sqrt{s-16m_c^2}}{2}],|y_{J/\psi}-y_0|\leq {\rm Arccosh}{\left(\frac{\sqrt{s}}{2M_{T,J/\psi}}\right)}$, $|\Delta y_{c\bar{c}}|\equiv |y_{c}-y_{\bar c}|\leq {\rm Arccosh}{\left(2({\rm cosh}(y_{J/\psi})-\frac{\sqrt{s}}{M_{T,J/\psi}})^2-1\right)}$ and $-\log{\left(2\frac{\sqrt{s}-M_{T,J/\psi}e^{-y_{J/\psi}}}{\left(1+e^{-\Delta y_{c\bar c}}\right)M_{T,J/\psi}}\right)} \leq y_{\bar c}\leq \log{\left(2\frac{\sqrt{s}-M_{T,J/\psi}e^{y_{J/\psi}}}{\left(1+e^{\Delta y_{c\bar c}}\right)M_{T,J/\psi}}\right)}$, where $y_0\equiv \frac{1}{2}\log{\frac{E_1}{E_2}}$ with $E_{1/2}$ being the two beam energies, $s=4E_1E_2$ and the transverse mass of $J/\psi$ as $M_{T,J/\psi}\equiv \sqrt{P_{T,J/\psi}^2+4m_c^2}$. Due to the momentum conservation, we notice the equality $p_{T,c}=p_{T,\bar{c}}=\frac{P_{T,J/\psi}}{2}$ at leading order, which is in contrast with DPS$_1$, where the momentum of $J/\psi$ is largely independent of the open (anti-)charm quark. At the LHC energies, due to the partonic luminosity, we will only consider the gluon initial state for both DPS mechanisms.

\section{Initial charm contribution and variable flavor number scheme\label{sec:vfns}}

Another important contribution ignored so far is the initial charm contribution in SPS production. Without considering the intrinsic source of charm quarks in the proton, $gg\to J/\psi+c\bar{c}$ has partially captured the  initial state logarithms of the type $\log{\frac{\mu_h^2}{m_c^2}}$ stemming from gluon splitting into a charm quark pair as shown in the first diagram of Fig.~\ref{fig:feynmandiagb}, where $\mu_h$ is the typical hard scale of the process. The resummation of these logarithms to all orders in $\alpha_s$ becomes essential when the logarithmic terms are much larger than other terms in the cross section.  In such a circumstance, charm (anti-)quark should be allowed in the initial state as depicted in the middle graph of Fig.~\ref{fig:feynmandiagb} (thereafter referred to as ``cgF"), and the QCD Dokshitzer-Gribov-Lipatov-Altarelli-Parisi evolution of PDF automatically resum these logarithms. In the LHCb acceptance, it was indeed (roughly) estimated in Table 1 of Ref.~\cite{Aaij:2012dz} that the contribution from sea charm quark is much larger than ggF. In particular, the forward LHCb detector only measures forward particles and therefore probes the relatively high $x$ domain of the forward beam. The initial state logarithms will be amplified with respect to the low $x$ domain as what has been found in bottom-quark-initiated processes~\cite{Maltoni:2012pa}. It is well known that the two approaches (ggF vs cgF) are compatible and complementary in phase space. The ggF calculation should be well applicable in the region close to the heavy quark threshold (i.e. $\mu_h\sim m_c$), while the cgF calculation is better in deciphering the cross section when $\mu_h\gg m_c$.

In order to correctly combine the two different calculations, a proper matching scheme must be used to subtract the overlap of the two, which is generally referring to variable flavor number scheme (VFNS). Following the pioneering work done by Aivazis, Collins, Olness and Tung (the so-called ACOT scheme)~\cite{Aivazis:1993pi}, the problem has been extensively studied in nonquarkonium processes in the literature~\cite{Buza:1996wv,Olness:1997yc,Thorne:1997ga,Cacciari:1998it,Kramer:2000hn,Tung:2001mv,Thorne:2006qt,
Forte:2010ta,Stavreva:2012bs,Kusina:2013slm,Forte:2015hba,Forte:2016sja,Krauss:2017wmx,Forte:2018ovl,Duhr:2020kzd}. Since we are only working at the lowest order for ggF and cgF processes, the matching procedure is rather simple. The only complication we have to  tackle on is that since the nonrelativistic nature of $J/\psi$, we cannot easily neglect the mass of the charm (anti-)quark in the cgF calculation. Instead, all charm quarks will be maintained as massive particles regardless where they are from. In such a situation, the differential cross section of cgF SPS $c(p_1) g(p_2)\to J/\psi(P_{J/\psi})c(p_c)$ is
\begin{eqnarray}
d\sigma_{cg\to J/\psi+c}^{\rm SPS}&=& \frac{1}{2\left(\hat{s}-m_c^2\right)}dx_1dx_2f_{c/p}(x_1,\mu_F^2)f_{g/p}(x_2,\mu_F^2)\overline{\sum}\left|\mathcal{M}_{cg\to J/\psi c}\right|^2d\Phi(p_1,p_2\to P_{J/\psi},p_c),
\end{eqnarray}
where we have taken $p_1=\left(\frac{\hat{s}+m_c^2}{2\sqrt{\hat{s}}},0,0,\frac{\hat{s}-m_c^2}{2\sqrt{\hat{s}}}\right),p_2=\left(\frac{\hat{s}-m_c^2}{2\sqrt{\hat{s}}},0,0,-\frac{\hat{s}-m_c^2}{2\sqrt{\hat{s}}}\right)$ in the initial partonic center-of-mass system with $\hat{s}=x_1x_2s$.~\footnote{In the center of the hadronic collision frame $P_{1,2}=\frac{\sqrt{s}}{2}\left(1,0,0,\pm1\right)$, we have $p_{i}=x_{i}P_{i}+\frac{m_i^2}{x_i s}P_{3-i}$ with $i=1,2$.} A similar formalism applies to gluon-charm scattering. The double counting between ggF and cgF can be attributed to  first $\alpha_s$ term of the charm PDF via
\begin{eqnarray}
f_{c/p}(x_1,\mu_F^2)&=&\tilde{f}_{c/p}^{(1)}(x_1,\mu_F^2)+\mathcal{O}(\alpha_s^2),
\end{eqnarray}
where
\begin{eqnarray}
\tilde{f}_{c/p}^{(1)}(x_1,\mu_F^2)&=&\frac{\alpha_s}{2\pi}\log{\left(\frac{\mu_F^2}{m_c^2}\right)}\int_{x_1}^{1}{\frac{dz}{z}P_{qg}(z)f_{g/p}\left(\frac{x_1}{z},\mu_F^2\right)}\label{eq:PDFCT}
\end{eqnarray}
with the well-known Altarelli-Parisi splitting function $P_{qg}(z)=\frac{1}{2}\left[z^2+\left(1-z\right)^2\right]$. The overlap counterterm [represented in the last diagram in Fig.~\ref{fig:feynmandiagb}] that must be subtracted is
\begin{eqnarray}
d\sigma_{{\rm CT}, cg\to J/\psi+c}^{\rm SPS}&=&\frac{1}{2\left(\hat{s}-m_c^2\right)}dx_1dx_2\tilde{f}_{c/p}^{(1)}(x_1,\mu_F^2)f_{g/p}(x_2,\mu_F^2)\overline{\sum}\left|\mathcal{M}_{cg\to J/\psi c}\right|^2d\Phi(p_1,p_2\to P_{J/\psi},p_c).\label{eq:cgCT}
\end{eqnarray}

Since we will be based on \Pythia, the gluon splitting into a charm quark pair must be written explicitly in the event files in order to ensure the correct backward evolution of the initial state shower based on the above equation. Because of the nonzero $m_c$, the momentum conservation and the on-shell condition cannot be guaranteed simultaneously in the splitting $g\to c\bar{c}$. A momentum reshuffling has to be imposed analogous to what has been advocated in Ref.~\cite{Frixione:2019fxg} for an entirely different purpose. In the laboratory frame, the momentum of the first initial gluon is assigned as $p_g=\left(\frac{x_1}{z}E_1,0,0,\frac{x_1}{z}E_1\right)$, and all the external momenta of $cg\to J/\psi c$ have been boosted to the same frame. In particular, a boost operation has been applied to the initial charm quark $p_1\to {\mathbb B}p_1$. Then, the momentum of anticharm is $p_{\bar{c}}=p_g-p_1$ thanks to the momentum conservation. However, the anticharm is not on-shell $p^2_{\bar{c}}\neq m_c^2$ without doing anything further. The momentum reshuffling is carried out as follows. We keep the invariant mass of the recoil system $J/\psi+c$ [$m_{\rm rec}^2=\left(P_{J/\psi}+p_c\right)^2$] invariant. The new initial gluon-gluon invariant mass square is $\hat{s}_{gg}=(p_g+p_2)^2=\frac{\hat{s}}{z}=\frac{x_1x_2s}{z}$. In the rest frame of the new partonic system, the energies of the final anticharm quark and the recoil system are
\begin{eqnarray}
\tilde{p}_{\bar{c}}^0&=&\frac{\hat{s}_{gg}+m_c^2-m_{\rm rec}^2}{2\sqrt{\hat{s}_{gg}}},\\
\tilde{p}^0_{\rm rec}&=&\frac{\hat{s}_{gg}-m_c^2+m_{\rm rec}^2}{2\sqrt{\hat{s}_{gg}}}.
\end{eqnarray}
The three-dimensional momenta preserve the directions of the original momenta with rescalings in order to satisfy the on-shell conditions
\begin{eqnarray}
\overrightarrow{\tilde{p}}_{\bar{c}}&=&\sqrt{\left(\tilde{p}_{\bar{c}}^0\right)^2-m_c^2}\frac{\overrightarrow{p}_{\bar{c}}}{\left|\overrightarrow{p}_{\bar{c}}\right|},\\
\overrightarrow{\tilde{p}}_{\rm rec}&=&\sqrt{\left(\tilde{p}_{\rm rec}^0\right)^2-m_{\rm rec}^2}\frac{\overrightarrow{p}_{\rm rec}}{\left|\overrightarrow{p}_{\rm rec}\right|},
\end{eqnarray}
where $\overrightarrow{p}_{\rm rec}\equiv \overrightarrow{P}_{J/\psi}+\overrightarrow{p}_{c}$. Because of $\overrightarrow{p}_{\rm rec}+\overrightarrow{p}_{\bar{c}}=\overrightarrow{0}$ in the initial gluon-gluon center-of-mass frame, 
$\overrightarrow{\tilde{p}}_{\bar{c}}+\overrightarrow{\tilde{p}}_{\rm rec}=\overrightarrow{0}$ is also guaranteed. Then, the new momenta of $J/\psi$ and $c$ in the same frame are
\begin{eqnarray}
\tilde{P}_{J/\psi}&=&{\mathbb B}_R^{-1}\left(\tilde{p}_{\rm rec}\right){\mathbb B}_R\left(p_{\rm rec}\right)P_{J/\psi},\\
\tilde{p}_{c}&=&{\mathbb B}_R^{-1}\left(\tilde{p}_{\rm rec}\right){\mathbb B}_R\left(p_{\rm rec}\right)p_{c},
\end{eqnarray}
where the boost operation ${\mathbb B}_R(p)$ represents a Lorentz boost which turns any four-dimensional momentum to the rest frame of $p$. In particular, we have ${\mathbb B}_R(p)p=\left(\sqrt{p^2},0,0,0\right)$. After the momentum reshuffling, all momenta of the external legs are boosted back to the laboratory frame.

Finally, the VFNS differential cross section is defined as
\begin{eqnarray}
d\sigma_{\rm VFNS}^{\rm SPS}&=&d\sigma_{gg\to J/\psi+c\bar{c}}^{\rm SPS}+\sum_{i=c,\bar{c}}{\left[\left(d\sigma_{ig\to J/\psi+i}^{\rm SPS}-d\sigma_{{\rm CT},ig\to J/\psi+i}^{\rm SPS}\right)+\left(d\sigma_{gi\to J/\psi+i}^{\rm SPS}-d\sigma_{{\rm CT},gi\to J/\psi+i}^{\rm SPS}\right)\right]}\nonumber\\
&=&d\sigma_{\rm ggF}^{\rm SPS}+d\sigma_{\rm cgF}^{\rm SPS}-d\sigma_{\rm CT}^{\rm SPS},\label{eq:VFNS}
\end{eqnarray}
where we have adopted
\begin{eqnarray}
d\sigma_{\rm ggF}^{\rm SPS}&\equiv&d\sigma_{gg\to J/\psi+c\bar{c}}^{\rm SPS},\label{eq:ggF}\\
d\sigma_{\rm cgF}^{\rm SPS}&\equiv&\sum_{i=c,\bar{c}}{\left(d\sigma_{ig\to J/\psi+i}^{\rm SPS}+d\sigma_{gi\to J/\psi+i}^{\rm SPS}\right)},\label{eq:cgF}\\
d\sigma_{\rm CT}^{\rm SPS}&\equiv&\sum_{i=c,\bar{c}}{\left(d\sigma_{{\rm CT},ig\to J/\psi+i}^{\rm SPS}+d\sigma_{{\rm CT},gi\to J/\psi+i}^{\rm SPS}\right)}\label{eq:CT}.
\end{eqnarray}
We, therefore, arrive at a similar master formula derived in Ref.~\cite{Aivazis:1993pi}.

A small inconsistence may occur due to the charm quark mass. In principle, one should take the exactly same charm mass in the computations of the cross sections and in the PDF. However, we cannot keep a single $m_c$ here, because we have to stick to the pole mass in the matrix element and in the phase space, while the $\overline{\rm MS}$ mass is usually adopted in PDF evolution. We do not bother such a small inconsistence in the paper in the view of much larger theoretical uncertainties from other sources.

\section{Proton-proton results at 7 TeV\label{sec:ppres}}

\subsection{Double parton scattering\label{sec:dpspp}}

We first compare the two DPS mechanisms (DPS$_1$ vs DPS$_2$) here in the LHCb $7$ TeV acceptance~\cite{Aaij:2012dz}. The estimate of the DPS$_1$ part is different from the DPS$_2$ part, where the computation of the latter has been detailed in Sec. \ref{sec:newDPS}. Since the two mesons are produced in two different partonic scatterings in DPS$_1$, we opt for a data-driven approach~\cite{Kom:2011bd,Lansberg:2014swa,Lansberg:2015lva,Massacrier:2015qba,Lansberg:2016deg} to estimate the matrix elements of the two hard scatterings. This is particularly motivated by the fact that we are still lacking a satisfactory description of inclusive $J/\psi$ hadroproduction at the LHC based on perturbative QCD. The data-driven approach is pure phenomenological, which does not provide any insight on the underlying production mechanisms of single-inclusive $J/\psi$ but allows us to alleviate the strong model-dependence at a large extent. Moreover, the precision of the experimental data is much better than the state-of-the-art perturbative QCD calculations for both the single-inclusive $J/\psi$ and open charm hadron production at the LHC. The high precision of these data can be inherently transfer to our data-driven approach. Finally, our approach is also very fast and therefore very economical from the point of view concerning the CPU expenses.

In our data-driven approach, the matrix elements of the single-inclusive prompt $J/\psi$ and open charm hadroproduction are characterized by an empirical formula,
\begin{eqnarray}
&&\overline{\sum}{\left|\mathcal{M}_{g(p_1)g(p_2)\to \mathcal{H}(P_{\mathcal{H}})+X(p_X)}\right|^2}=\nonumber\\
&&\frac{\lambda_{\mathcal{H}}^2\kappa_{\mathcal{H}} sx_1x_2}{M^2_{\mathcal{H}}}\exp{\left(-\kappa_{\mathcal{H}}\frac{{\rm min}\left(P_{T,\mathcal{H}}^2,\langle P_{T,\mathcal{H}}\rangle^2\right)}{M_{\mathcal{H}}^2}\right)}\left(1+\theta\left(P_{T,\mathcal{H}}^2-\langle P_{T,\mathcal{H}}\rangle^2\right)\frac{\kappa_{\mathcal{H}}}{n_{\mathcal{H}}}\frac{P_{T,\mathcal{H}}^2-\langle P_{T,\mathcal{H}}\rangle^2}{M_{\mathcal{H}}^2}\right)^{-n_{\mathcal{H}}},\label{eq:CBME}
\end{eqnarray}
where only the gluon-gluon initial state is retained due to its dominant partonic luminosity at the LHC, and $X=g$ when $\mathcal{H}=J/\psi$ and $X=\bar{C}$~\footnote{$\bar{C}$ stands for the charge conjugate particle of the open (anti-)charm hadron $C$.} when $\mathcal{H}=C$. The 4 free parameters $\lambda_{\mathcal{H}},\kappa_{\mathcal{H}},P_{T,\mathcal{H}},n_{\mathcal{H}}$ are determined from the $pp$ experimental data via a fit after convolution with a proton PDF,
\begin{eqnarray}
d\sigma_{pp\to \mathcal{H}+X}&=&dx_1dx_2f_{g/p}(x_1,\mu_F^2)f_{g/p}(x_2,\mu_F^2)\frac{1}{2\hat{s}_{12}}\overline{\sum}{\left|\mathcal{M}_{g(p_1)g(p_2)\to \mathcal{H}(P_{\mathcal{H}})+X(p_X)}\right|^2}d\Phi(p_1,p_2\to P_{\mathcal{H}}, p_X).
\end{eqnarray}
The fitted values of these parameters, along with their errors from the $\chi^2$ fit, for $\mathcal{H}=J/\psi,D^0,D^+,D_s^+$ and $\Lambda_c^+$ to the LHCb double differential data $d^2\sigma/dP_Tdy$~\cite{Aaij:2011jh,Aaij:2013yaa,Aaij:2013mga} can be found in Table~\ref{tab:fitCB} with the CT10NLO proton PDF~\cite{Lai:2010vv}, where the $\chi^2$ values and the numbers of fitted experimental data are also given in the table. The comparison between our fit results and the LHCb data is reported in Fig.~\ref{fig:ppfit}. The factorization scale $\mu_F$ entering in the PDF is fixed dynamically as the transverse mass of the particle $\mathcal{H}$. Thereby, the DPS$_1$ cross sections are
\begin{eqnarray}
d\sigma_{pp\to J/\psi+C}^{\rm DPS_1}&=&\frac{d\sigma_{pp\to J/\psi+X}d\sigma_{pp\to C+X}}{\sigma_{{\rm eff},pp}}.
\end{eqnarray}
Since no correlation is considered in DPS$_1$, we take an individual factorization scale for each scattering. Finally, because LHCb~\cite{Aaij:2012dz} has presented the measured values of the cross sections for prompt $J/\psi$ and open charm hadrons $C$ in the acceptance, we will normalize our DPS$_1$ cross sections according to these measured values.

\begin{table}[H]
\centering\renewcommand{\arraystretch}{1.2}
\begin{tabular}{|c|c|cccc|c|} 
\hline
$\mathcal{H}$ & data &  $\lambda_{\mathcal{H}}$ & $\kappa_{\mathcal{H}}$ & $\langle P_{T,\mathcal{H}}\rangle/{\rm GeV}$ & $n_{\mathcal{H}}$ & $\chi^2/N_{\rm data}$\\\hline
$J/\psi$ & LHCb~\cite{Aaij:2011jh,Aaij:2013yaa} & $0.285$ & $0.623$ & $3.68\pm 0.22$ & $2$ (fixed) & $328/136$\\
$D^0$ & LHCb~\cite{Aaij:2013mga} & $2.38$ & $1.62$ & $0.52$ & $2$ (fixed) & $62.6/38$\\
$D^+$ & LHCb~\cite{Aaij:2013mga} & $1.54\pm 0.25$ & $1.44$ & $0.72$ & $2$ (fixed) & $36.1/37$ \\
$D_s^+$ & LHCb~\cite{Aaij:2013mga} & $0.91$ & $0.55$ & $2.82$ & $2$ (fixed) & $23.9/28$ \\
$\Lambda_c^+$ & LHCb~\cite{Aaij:2013mga} & $1.53$ & $1.18$ & $2.00\pm0.18$ & $2$ (fixed) & $0.567/6$ \\\hline
\end{tabular}
\caption{\label{tab:fitCB} The fitted values of free parameters in Eq.(\ref{eq:CBME}) from $d^2\sigma/dP_Tdy$ of prompt $J/\psi$ and four open charm hadrons in  $pp$ collisions using CT10NLO, where we have fixed the value of $n_{\mathcal{H}}$. The uncertainties for the fitted values are from the $\chi^2$ fit, where those below the precent level are not shown. The last column lists the corresponding $\chi^2$ values and the number of data $N_{\rm data}$ in the fit.}
\end{table}

\begin{figure}
  \centering
   \subfloat[$J/\psi$@7 TeV]{\includegraphics[width=0.45\textwidth,valign=c,draft=false]{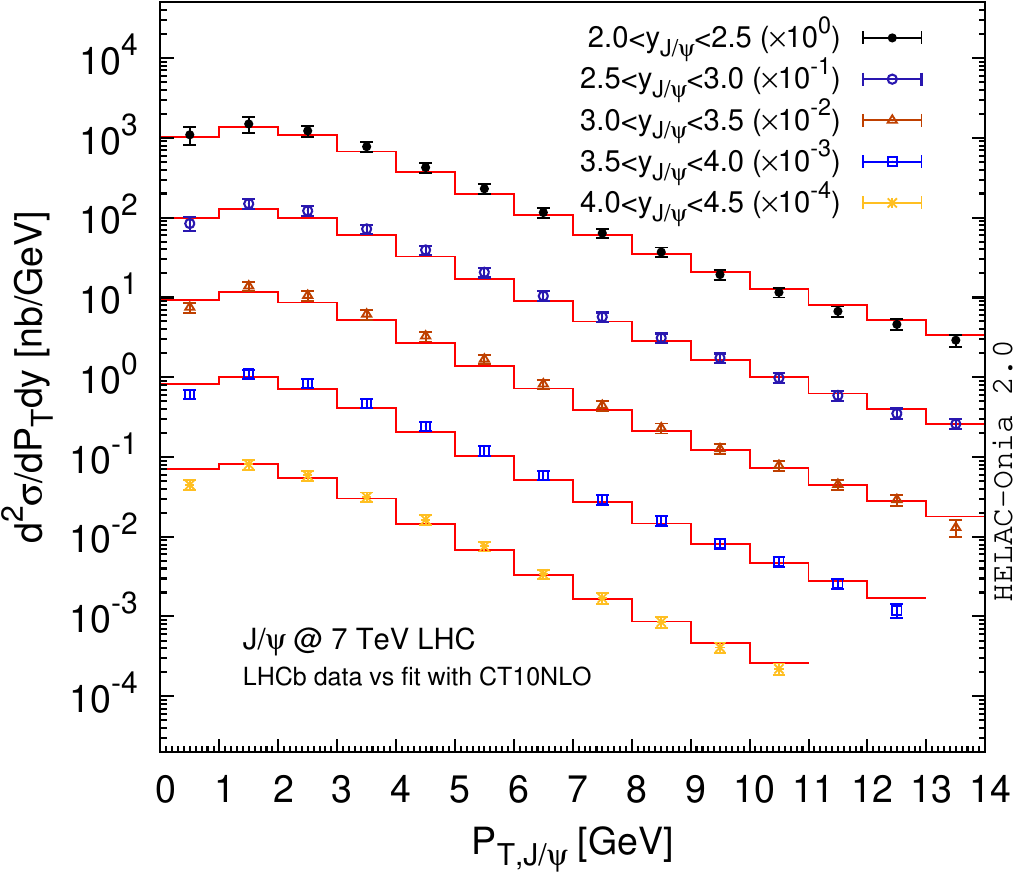}\label{fig:ppfita}}
  \subfloat[$J/\psi$@8 TeV]{
    \includegraphics[width=0.45\textwidth,valign=c,draft=false]{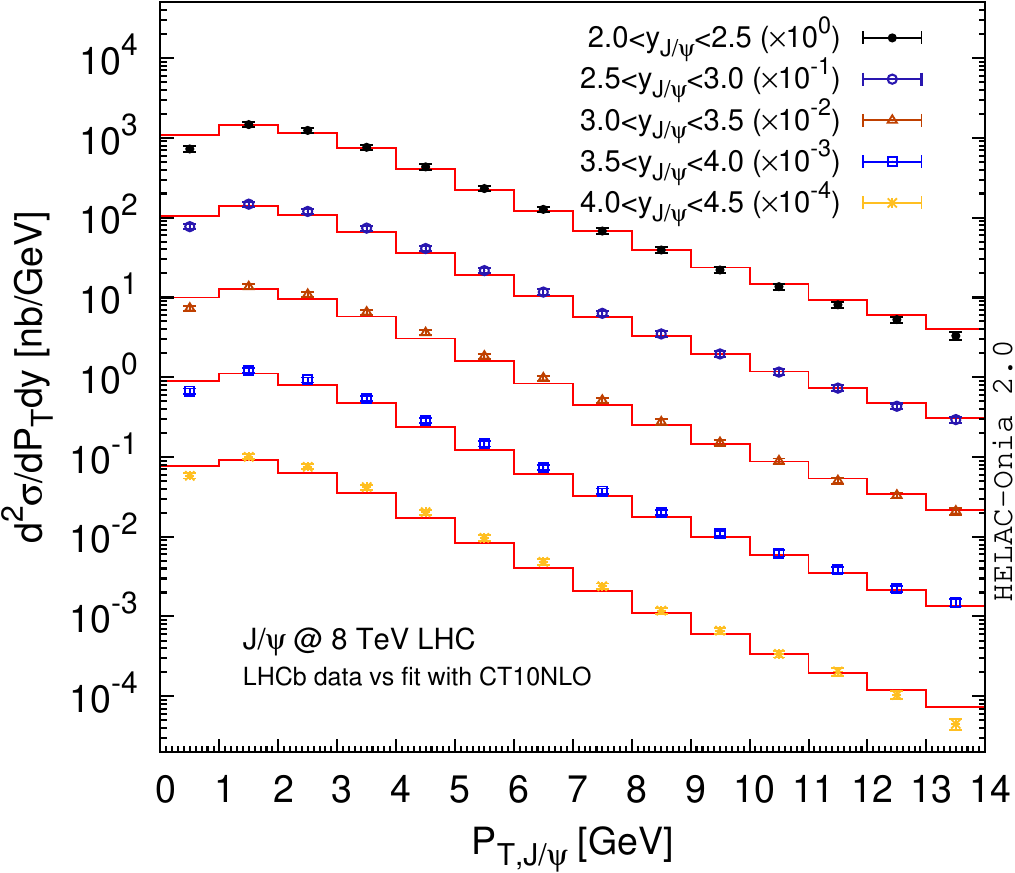}\label{fig:ppfitb}}\\
  \subfloat[$D^0$]{\includegraphics[width=0.45\textwidth,valign=c,draft=false]{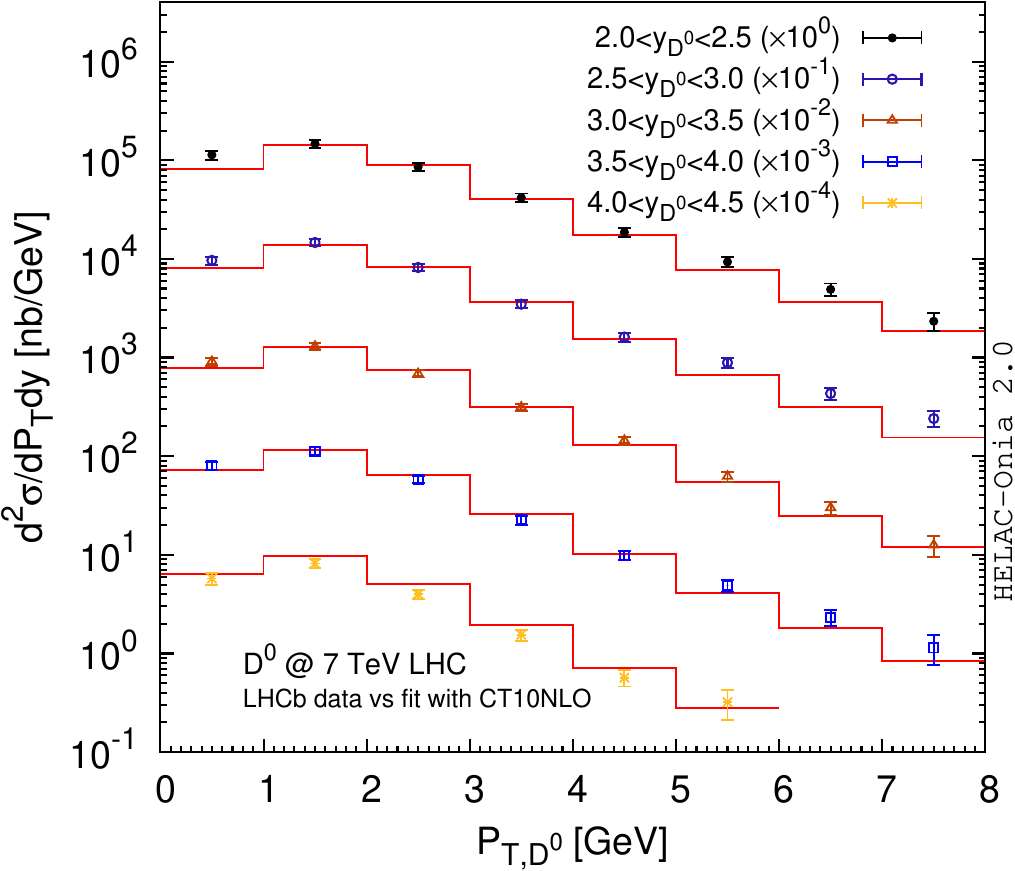}\label{fig:ppfitc}}
  \subfloat[$D^+$]{\includegraphics[width=0.45\textwidth,valign=c,draft=false]{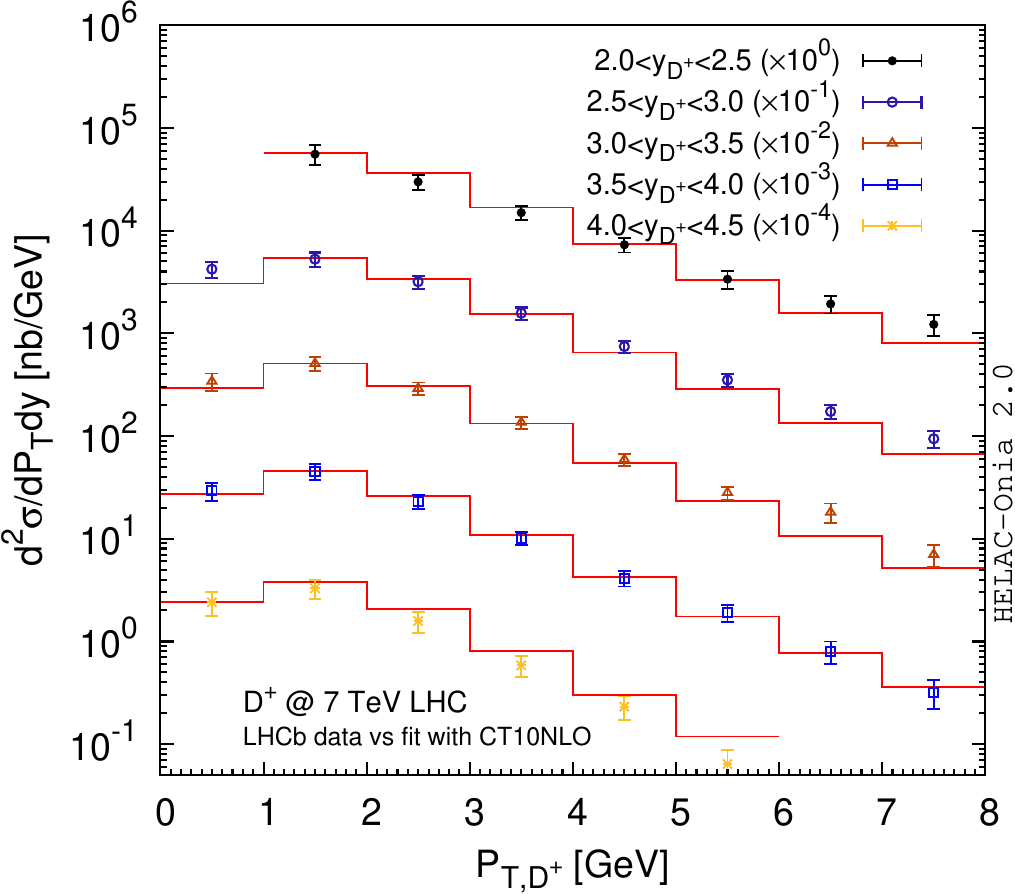}\label{fig:ppfitd}}\\
  \subfloat[$D_s^+$]{\includegraphics[width=0.45\textwidth,valign=c,draft=false]{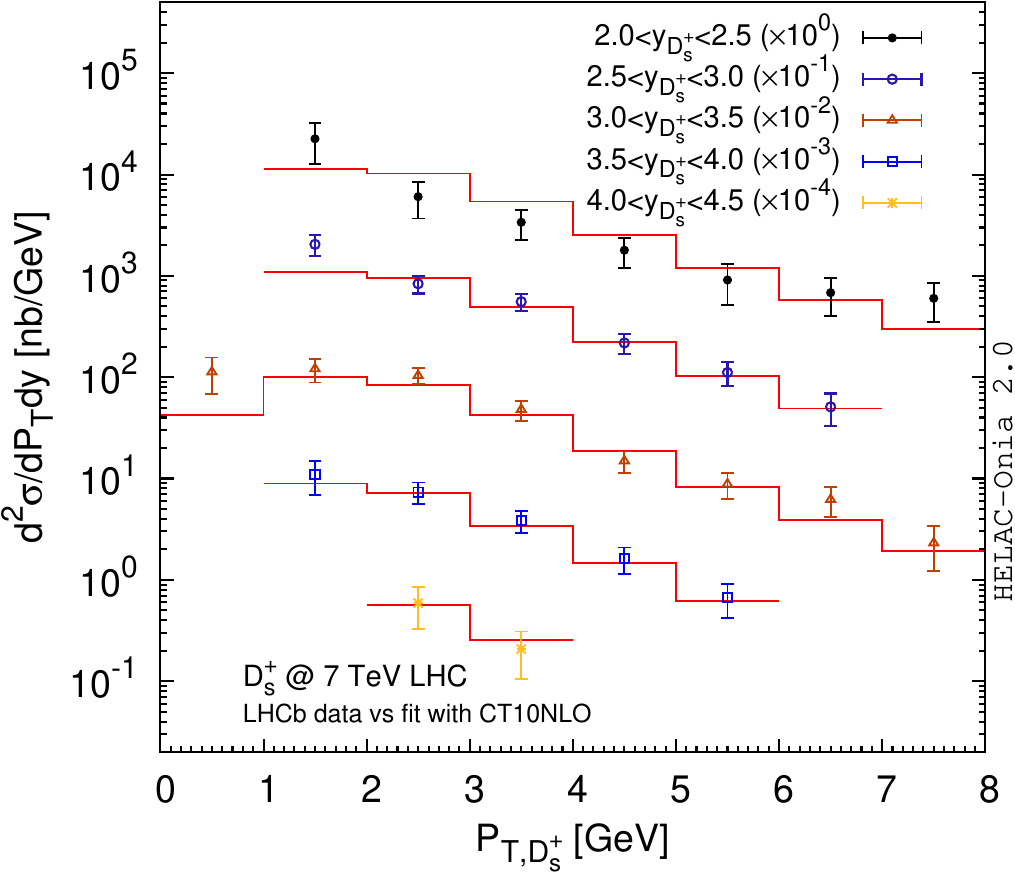}\label{fig:ppfite}}
   \subfloat[$\Lambda_c^+$]{\includegraphics[width=0.45\textwidth,valign=c,draft=false]{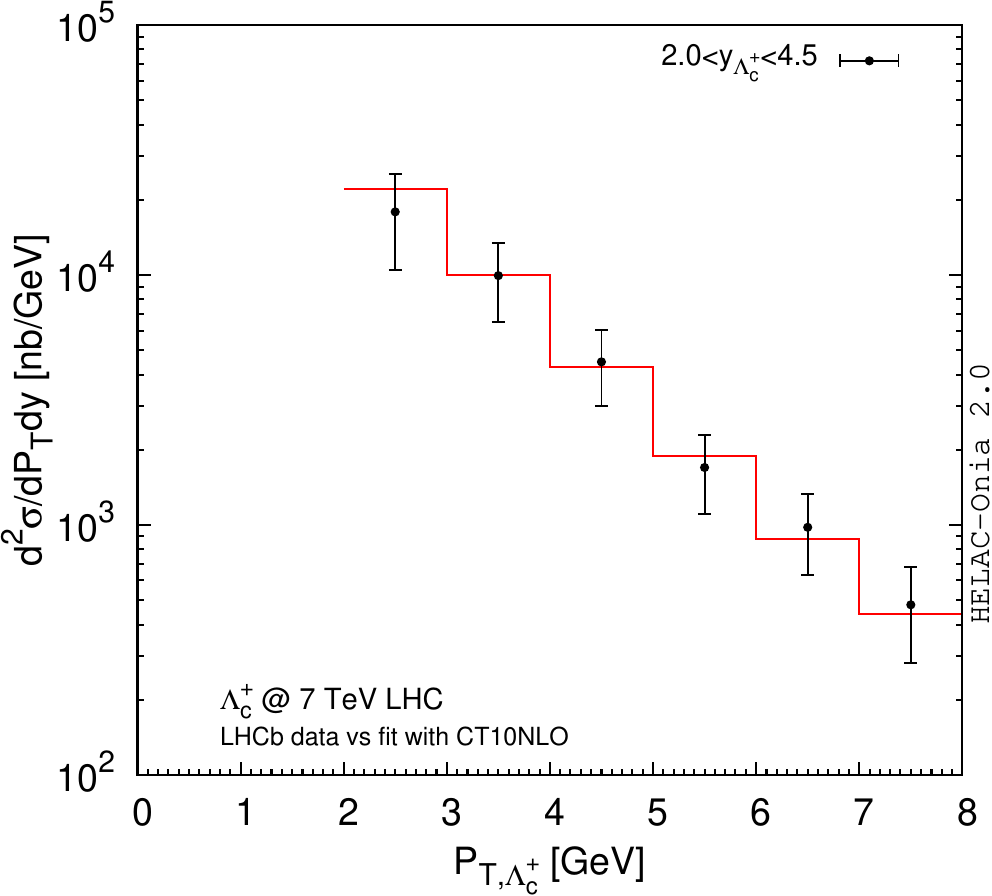}\label{fig:ppfitf}}
  \caption{Comparison of our fit results with the 7 TeV and 8 TeV prompt $J/\psi$ (a,b), $D^0$ (c), $D^+$ (d), $D_s^+$ (e), and $\Lambda_c^+$ (f) in $pp$ collisions at the LHC.\label{fig:ppfit}}
\end{figure}

Contrarily, because of the strong correlation in DPS$_2$, we use the perturbative QCD approach for evaluating the scattering amplitude along with Eq.(\ref{eq:DPS2amp}) (or similar variance for the color-octet channels), where the analytic helicity amplitudes for a heavy quark pair production are known in the literature (see, e.g., in Ref.~\cite{Chen:2017jvi}). The final differential cross sections of DPS$_2$ are obtained via Eq.(\ref{eq:XSDPS2}). The common dynamical scale $\mu_0$ for the central value of the factorization and renormalization scales is $H_T/2$, where $H_T$ is the scalar sum of the transverse mass of the final particles. The scale uncertainty for estimating the missing higher order QCD corrections is obtained by varying $\mu_R$ and $\mu_F$ independently around $\mu_0$ by a factor $2$. The leading order (LO) proton PDF CT10LO~\cite{Lai:2010vv} is adopted since the corresponding matrix elements are only LO accuracy here. For simplicity, we do not convolute any fragmentation function but multiply a global fragmentation fraction of $f(c\to C)$ [$f(c\to D^0)=0.565, f(c\to D^+)=0.246, f(c\to D_s^+)=0.080,f(c\to \Lambda_c^+)=0.094$]. It gives us the conservative upper values for the DPS$_2$ cross sections because the convolutions of fragmentation functions will soften the $p_T$ spectra of the open charm hadrons. The charm pole mass is taken as $m_c=1.5$ GeV, and the LDMEs for $J/\psi$ and feed-down ($\chi_{c}$ and $\psi(2S)$) are the Set 8 summarized in Table 3 of Ref.~\cite{Lansberg:2019fgm}, which is originally from Refs.~\cite{Han:2014jya,Shao:2014yta}. As an estimate of the order of the magnitude from DPS$_2$, we restrict ourselves to consider only three $c\bar{c}$ Fock states $\ss,\so$, and $\sps$ here. 

\begin{table}[H]
\centering\renewcommand{\arraystretch}{1.2}
\begin{tabular}{|c|cccc|} 
\hline
\diagbox[width=3.0cm,height=1.0cm]{Cross section}{Final state} & $J/\psi+D^0$ & $J/\psi+D^+$ & $J/\psi+D_s^+$ & $J/\psi+\Lambda_c^+$\\\hline
$\sigma^{\rm DPS_1}$ [$\frac{15~{\rm mb}}{\sigma_{{\rm eff},pp}}\cdot{\rm nb}$] & $159.2$ & $65.8$ & $25.9$ & $52.0$ \\
$\sigma^{\rm DPS_2}$ [$\frac{15~{\rm mb}}{\sigma_{{\rm eff},pp}}\cdot{\rm nb}$] & $(3.4_{-~2.8-1.0}^{+12.7+1.1})\cdot 10^{-1}$ & $(1.5_{-1.2-0.4}^{+5.5+0.5})\cdot 10^{-1}$ & $(4.8_{-~4.0-1.4}^{+17.9+1.6})\cdot 10^{-2}$ & $(5.6_{-~4.7-1.6}^{+21.1+1.8})\cdot 10^{-2}$ \\\hline
\end{tabular}
\caption{\label{tab:DPS1vsDPS2} The integrated cross sections of DPS$_1$ and DPS$_2$ within the LHCb acceptance at $7$ TeV~\cite{Aaij:2012dz}. The first quoted errors of $\sigma^{\rm DPS_2}$ are from the scale variation, and the second errors are from the PDF parameterization.}
\end{table}

The comparison of the DPS$_1$ and DPS$_2$ cross sections for the four different final states is reported in Table~\ref{tab:DPS1vsDPS2}. In DPS$_2$ cross sections, we have quoted two errors for each cross section. The first one is stemming from the scale uncertainty, and the second one is from the PDF uncertainty. The DPS$_2$ numbers are in anyway lower than the DPS$_1$ numbers by around 2 orders of magnitude. This is because in DPS$_2$, a $p_{T,C}>3$ GeV kinematic cut would also imply $P_{T,J/\psi}>6$ GeV and $p_{T,\bar{C}}>3$ GeV due to the strong correlation of the external momenta. In principle, the introduction of the initial primordial transverse momentum, because of the confinement and the uncertainty relation, can violate such a relation. However, we have explicitly checked that the kicks of the initial state in the transverse plane by a few GeV can maximally enhance the DPS$_2$ cross sections by a factor of $\sim 5$, which are still at least one order of the magnitude smaller than DPS$_1$. Such a conclusion will not change if we take other LDME sets in Ref.~\cite{Lansberg:2019fgm}. It is, however, not necessary to be still true that DPS$_2$ will be negligible compared to DPS$_1$ under other conditions. We will examine it again in proton-lead collisions at $\sqrt{s_{NN}}=8.16$ TeV in Sec.~\ref{sec:pA}, where there is no lower $p_T$ cut imposed.

\subsection{Single parton scattering\label{sec:SPSres}}

\subsubsection{Color-octet and feed-down contributions}

A brief survey of the possible impact of the color-octet contribution~\cite{Artoisenet:1900zz} and of $\chi_{c}$ feed-down~\cite{Li:2011yc} reveals that they only alter the dominant color-singlet $\ss$ predictions marginally until reaching to the tail of the $P_{T,J/\psi}$ spectra (e.g. $P_{T,J/\psi}>15$ GeV), which are outside the LHCb kinematic requirement. Here, we will assess their contributions with the diverse LDME fits on the market by considering ggF SPS. The simulation is carried out by the joint usage of \ho\ 2.0~\cite{Shao:2012iz,Shao:2015vga} and  \Pythia\ 8.186~\cite{Sjostrand:2007gs} with CT10LO as the proton PDF. In order to match what has been implemented in \Pythia\ 8, the charm quark mass $m_c$ is taken the half of the physical mass of the meson in the color-singlet channels (i.e., $\ss$ for $J/\psi$ and $\psi(2S)$, and $\tpjs$ for $\chi_{cJ}$), while a 100 MeV increment is applied to $m_c$ for the other S- and P-wave color-octet channels (i.e., $\sps,\so,\pj$ for $J/\psi$ and $\psi(2S)$, and $\so$ for $\chi_{cJ}$). Totally, we have considered $18$ Fock state channels. Each channel consists both gluon-gluon and quark-antiquark initial states. The central scale is same as DPS$_2$ case, i.e., $H_T/2$. The values of LDMEs from worldwide data fits are extremely disparate and inconsistent among different groups. The status has been well summarized in Sec. 5.2 of Ref.~\cite{Lansberg:2019fgm}. We take a global survey in order to minimize the possible bias introduced by a concrete LDME fit. A summary of nine LDME sets is reported in Table 3 of Ref.~\cite{Lansberg:2019fgm}, which are originally taken from Refs.~\cite{Sharma:2012dy,Braaten:1999qk,Kramer:2001hh,Sun:2012vc,Butenschoen:2011yh,Gong:2012ug,Shao:2014yta,Han:2014jya,Bodwin:2014gia}. We follow the same notation here and refrain from tabulating them again. We have reported the ratios of the prompt cross sections over the color-singlet direct production $J/\psi$~\footnote{We have fixed the LDME $\langle \mathcal{O}^{J/\psi}_{\ss}\rangle=1.16$ GeV$^3$ in the denominators regardless of the values of the same LDME used in the numerators of the ratios.} (i.e. no color-octet transition and no feed-down) in Table.~\ref{tab:COFD}. For a given LDME set, the ratios only mildly depend on the species of the open charm hadron. On the other hand, the dependencies of such ratios on LDMEs are more significant. Taking $J/\psi+D^0$ as an example, the ratio ranges from $1.06$ (Set 4) to $1.65$ (Set 2). This corroborates the similar conclusion drawn based on earlier (partial) studies: color-octet and feed-down contributions are not able to dramatically enhance the cross section.

\begin{table}[H]
\centering\renewcommand{\arraystretch}{1.2}
\begin{tabular}{|c|ccccccccc|} 
\hline
\diagbox[width=2.5cm,height=1.0cm]{Final state}{LDME Set} & 1 & 2 & 3 & 4 & 5 & 6 & 7 & 8 & 9\\\hline
$J/\psi+D^0$ & $1.49$ & $1.65$ & $1.48$ & $1.06$ & $1.20$ & $1.29$ & $1.56$ & $1.51$ & $1.43$ \\
$J/\psi+D^+$ & $1.53$ & $1.71$ & $1.51$ & $0.99$  & $1.17$ & $1.24$ & $1.60$ & $1.54$ &  $1.41$ \\
$J/\psi+D_s^+$ & $1.46$ & $1.60$ & $1.45$ & $1.14$  & $1.23$ & $1.34$ & $1.50$ & $1.47$ & $1.43$ \\
$J/\psi+\Lambda_c^+$ & $1.40$ & $1.53$ & $1.41$ & $1.25$ & $1.29$ & $1.38$ & $1.45$ & $1.42$ &  $1.51$ \\\hline
\end{tabular}
\caption{\label{tab:COFD} The integrated ggF SPS cross section ratios of the prompt $J/\psi$ over the color-singlet $J/\psi$ direct production. The LDME sets are from [Set 1: Sharma et al.~\cite{Sharma:2012dy}; Set 2: Braaten et al.~\cite{Braaten:1999qk}; Set 3: Kr{\"a}mer~\cite{Kramer:2001hh};  Set 4: Sun et al.~\cite{Sun:2012vc}; Set 5: Butensch{\"o}n et al.~\cite{Butenschoen:2011yh}; Set 6 : Gong et al.~\cite{Gong:2012ug}; Set 7: Shao et al.~\cite{Shao:2014yta}: Set 8: Han et al.~\cite{Han:2014jya}: Set 9: Bodwin et al.~\cite{Bodwin:2014gia}].}
\end{table}

\subsubsection{Fixed flavor number scheme versus variable flavor number scheme}

We compare the fixed flavor number scheme results and the variable flavor number scheme results in SPS. Again, for the sake of simplicity, we only consider the direct $J/\psi$ production via the leading $\ss$ channel since the inclusion of the color-octet and feed-down channels has been discussed in the previous subsection. Similarly to the setup in the last subsection, the LDME $\langle \mathcal{O}^{J/\psi}_{\ss}\rangle$ has been fixed as $1.16$ GeV$^3$ and the central scale $\mu_0=H_T/2$. The simulations are carried out by jointly using \ho\ 2.0~\cite{Shao:2012iz,Shao:2015vga} and  \Pythia\ 8.235~\cite{Sjostrand:2014zea} with primordial $k_T$ and underlying event enabled. At variance with what has been done before, we consider several proton PDF choices here. There are one LO PDF (CT10LO~\cite{Lai:2010vv}), four next-to-leading order (NLO) PDF sets (CT10NLO~\cite{Lai:2010vv}, CT14NLO~\cite{Dulat:2015mca}, MMHT14NLO~\cite{Harland-Lang:2014zoa}, and NNPDF3.1NLO~\cite{Ball:2017nwa}) and a next-to-next-to-leading order (NNLO) PDF set (CT14NNLO~\cite{Dulat:2015mca}). Although all of them are global-fitted PDFs based on variable flavor number schemes, there are a few significant differences among them. Notably, they differ by (i) perturbative orders in the scale evolution of PDF and $\alpha_s$ and in matrix elements to fit the PDF; (ii) PDF parameterizations, statistical fit methodologies, and uncertainty estimations; (iii) values of parameters (e.g., $\alpha_s(m_Z^2)$ and heavy quark masses) and input experimental data; (iv) the concrete implementation of variable flavor number schemes. None of them has been introduced the non-perturbative source of charm and bottom (anti-)quark densities.

\begin{table}[H]
\centering\renewcommand{\arraystretch}{1.2}
\begin{tabular}{|c|c|c|c|c|c|}
\hline
Final state & PDF & 3FS (ggF) & 4FS (cgF) & CT & VFNS \\\hline
\multirow{6}{*}{$J/\psi+D^0$} & CT10LO & $6.4^{+19.5+0.9}_{-4.9~-0.5}$ & $17.4^{+105.3+5.7}_{-15.8~-1.8}$ & $4.9^{+32.5+1.2}_{-4.6~-0.5}$ & $18.8^{+92.4+5.4}_{-16.1-1.8}$ \\\cline{2-6}
& CT10NLO & $6.5^{+18.6+0.9}_{-4.9~-0.5}$ & $17.2^{+92.0+5.6}_{-15.6-1.8}$ & $4.9^{+28.3+1.2}_{-4.5~-0.5}$ & $18.8^{+82.3+5.3}_{-16.0-1.8}$ \\\cline{2-6}
& CT14NLO & $6.6^{+18.7+1.0}_{-4.9~-0.4}$ & $17.9^{+95.1+6.9}_{-16.2-1.5}$ & $5.0^{+28.8+1.4}_{-4.6~-0.4}$ & $19.4_{-16.5-1.5}^{+84.9+6.5}$\\\cline{2-6}
& CT14NNLO & $5.8^{+17.6+0.8}_{-4.5~-0.3}$ & $14.4^{+80.3+5.4}_{-13.2-1.0}$ & $4.2^{+26.7+1.1}_{-3.9~-0.3}$ & $16.0_{-13.8-1.0}^{+71.2+5.1}$ \\\cline{2-6}
& MMHT14NLO & $6.7^{+19.0+0.4}_{-5.0~-0.3}$ & $18.6^{+101.4+2.4}_{-17.1~-1.9}$ & $5.1^{+29.3+0.5}_{-4.7~-0.4}$ & $20.2^{+91.0+2.2}_{-17.3-1.8}$ \\\cline{2-6}
& NNPDF3.1NLO & $5.9^{+17.4+0.5}_{-4.6~-0.5}$ & $12.1^{+72.2+2.0}_{-11.3-2.0}$ & $4.3^{+25.9+0.6}_{-4.0~-0.6}$ & $13.7^{+63.8+1.9}_{-11.9-1.9}$ \\\hline
\multirow{6}{*}{$J/\psi+D^+$} & CT10LO & $3.5^{+10.7+0.5}_{-2.6~-0.3}$ & $9.5^{+57.6+3.1}_{-8.6~-1.0}$ & $2.7^{+17.9+0.7}_{-2.5~-0.3}$ & $10.2^{+50.3+2.9}_{-8.7~-1.0}$ \\\cline{2-6}
& CT10NLO & $3.4^{+9.9~+0.5}_{-2.6~-0.3}$ & $9.4^{+50.1+3.0}_{-8.5~-1.0}$ & $2.6^{+15.5+0.7}_{-2.4~-0.3}$ & $10.2^{+44.5+2.9}_{-8.7~-1.0}$ \\\cline{2-6}
& CT14NLO & $3.5^{+10.1+0.5}_{-2.6~-0.2}$ & $9.7^{+51.6+3.8}_{-8.8~-0.8}$ & $2.7^{+15.7+0.8}_{-2.5~-0.2}$ & $10.5_{-9.0~-0.8}^{+46.0+3.5}$\\\cline{2-6}
& CT14NNLO & $3.1_{-2.4~-0.2}^{+9.4~+0.4}$ & $7.8^{+44.1+3.0}_{-7.2~-0.6}$ & $2.3_{-2.1~-0.1}^{+14.5+0.6}$ & $~8.6_{-7.5~-0.6}^{+39.0+2.8}$ \\\cline{2-6}
& MMHT14NLO & $3.6^{+10.3+0.2}_{-2.7~-0.2}$ & $10.2^{+55.6+1.3}_{-9.4-1.1}$ & $2.8^{+16.2+0.3}_{-2.6~-0.2}$ & $11.0^{+49.7+1.2}_{-9.4~-1.0}$ \\\cline{2-6}
& NNPDF3.1NLO & $3.2^{+9.6~+0.3}_{-2.5~-0.3}$ & $6.6^{+39.4+1.1}_{-6.2~-1.1}$ & $2.3^{+13.9+0.3}_{-2.1~-0.3}$ & $7.5^{+35.1+1.0}_{-6.5~-1.0}$ \\\hline
\multirow{6}{*}{$J/\psi+D_s^+$} & CT10LO & $1.2^{+3.7+0.2}_{-0.9-0.1}$ & $3.3^{+20.2+1.1}_{-3.0~-0.3}$ & $0.96^{+6.41+0.25}_{-0.89-0.10}$ & $3.5^{+17.5+1.0}_{-3.0~-0.3}$ \\\cline{2-6}
& CT10NLO & $1.2^{+3.4+0.2}_{-0.9-0.1}$ & $3.3^{+17.8+1.1}_{-3.0~-0.3}$ & $0.91^{+5.49+0.16}_{-0.85-0.09}$ & $3.5^{+15.7+1.0}_{-3.0~-0.3}$ \\\cline{2-6}
& CT14NLO & $1.2^{+3.5+0.2}_{-0.9-0.1}$ & $3.4^{+18.5+1.4}_{-3.1~-0.3}$ & $0.94^{+5.57+0.27}_{-0.88-0.07}$ & $3.7_{-3.2~-0.3}^{+16.5+1.3}$\\\cline{2-6}
& CT14NNLO & $1.1^{+3.2+0.2}_{-0.8-0.1}$  & $2.7^{+15.6+1.1}_{-2.5~-0.2}$ & $0.80^{+5.20+0.22}_{-0.75-0.05}$ & $3.0_{-2.6~-0.2}^{+13.7+1.0}$\\\cline{2-6}
& MMHT14NLO & $1.2^{+3.4+0.1}_{-0.9-0.1}$ & $3.6^{+19.7+0.5}_{-3.3~-0.4}$ & $0.98^{+5.67+0.10}_{-0.91-0.08}$ & $3.8^{+17.4+0.4}_{-3.3~-0.3}$ \\\cline{2-6}
& NNPDF3.1NLO & $1.1^{+3.3+0.1}_{-0.9-0.1}$ & $2.3^{+14.0+0.4}_{-2.2~-0.4}$ & $0.80^{+4.92+0.12}_{-0.75-0.12}$ & $2.6^{+12.4+0.4}_{-2.3~-0.4}$ \\\hline
\multirow{6}{*}{$J/\psi+\Lambda_c^+$} & CT10LO & $0.46^{+1.42+0.07}_{-0.35-0.02}$ & $1.2^{+7.0+0.5}_{-1.1-0.1}$ & $0.35^{+2.33+0.10}_{-0.33-0.02}$ & $1.3^{+6.1+0.4}_{-1.2-0.1}$ \\\cline{2-6}
& CT10NLO & $0.52^{+1.51+0.07}_{-0.39-0.04}$ & $1.5_{-1.3-0.2}^{+8.1+0.5}$ & $0.41^{+2.48+0.11}_{-0.38-0.04}$ & $1.6_{-1.4-0.2}^{+7.2+0.5}$ \\\cline{2-6}
& CT14NLO & $0.50^{+1.48+0.08}_{-0.38-0.03}$ & $1.5_{-1.4-0.1}^{+8.3+0.6}$ & $0.42^{+2.53+0.10}_{-0.39-0.03}$ & $1.6_{-1.4-0.1}^{+7.3+0.6}$\\\cline{2-6}
& CT14NNLO & $0.46^{+1.42+0.07}_{-0.35-0.02}$ & $1.2_{-1.1-0.1}^{+7.0+0.5}$ & $0.35^{+2.33+0.10}_{-0.33-0.02}$ & $1.3_{-1.2-0.1}^{+6.1+0.4}$\\\cline{2-6}
& MMHT14NLO & $0.48^{+1.38+0.03}_{-0.36-0.02}$ & $1.7^{+9.6+0.2}_{-1.5-0.2}$ & $0.43^{+2.52+0.04}_{-0.40-0.04}$ & $1.7^{+8.4+0.2}_{-1.5-0.2}$ \\\cline{2-6}
& NNPDF3.1NLO & $0.50^{+1.50+0.04}_{-0.39-0.04}$ & $1.0^{+6.3+0.2}_{-1.0-0.2}$ & $0.36^{+2.25+0.05}_{-0.34-0.05}$ & $1.2^{+5.6+0.2}_{-1.0-0.2}$ \\\hline
\end{tabular}
\caption{\label{tab:FFNSvsVFNS} The integrated cross sections (in unit of nb) of SPS of direct color-singlet $J/\psi$ plus prompt open charm hadron production within the LHCb acceptance at $7$ TeV~\cite{Aaij:2012dz}. The charge conjugate open charm hadrons have been included as well. The first quoted errors are from the 9-point scale variation and the second errors represent the $1\sigma$ PDF parameterization uncertainty.}
\end{table}

The integrated cross sections at $\sqrt{s}=7$ TeV within the LHCb acceptance for the four final states and by using six PDF sets are reported in Table.~\ref{tab:FFNSvsVFNS}. For the purpose of the comparison, we have shown ggF [cf. Eq.(\ref{eq:ggF})], cgF [cf. Eq.(\ref{eq:cgF})], CT [cf. Eq.(\ref{eq:CT})], and VFNS [cf. Eq.(\ref{eq:VFNS})] separately. Two theoretical errors are quoted for each configuration along with its central value. The first one is from the standard 9-point renormalization and factorization scale variation $\mu_{R/F}=\xi_{R/F}\mu_0, \xi_{R/F} \in \{1.0,0.5,2.0\}$. The second error is estimated from $68\%$ confidence level  (CL) PDF parameterization uncertainty.
In general, because we are working at the lowest order (i.e., LO) in the strong coupling $\alpha_s$ for the matrix elements and at rather low scales (a few GeV), the scale uncertainties anyway dominate the theoretical uncertainties. They shift the predicted cross sections up and down by factors of more than $3$ around the central values. The effect is more striking in processes with only three charm (anti-)quarks involved in the hard scatterings (i.e., cgF, CT) , which is anticipated from their even lower scales than ggF, where in the latter there are four heavy (anti-)quarks. Such theoretical errors can only be systematically reduced by including higher-order QCD terms in the $\alpha_s$ perturbative series. Charm quark hadroproduction up to NNLO reveals that the precise predictions with higher-order $\alpha_s$ corrections usually lie at the upper limits of the corresponding LO predictions. Such a statement can of course be altered by considering a different process and by choosing a different central scale. Given the absence of higher-order QCD calculations of the $J/\psi$ plus open charm process nowadays, we allow the missing higher-order corrections can be any values allowed by the scale variation, and the differential shapes can be distorted arbitrarily within the scale uncertainty bands. A remarkable observation from Table~\ref{tab:FFNSvsVFNS} is that the original ggF calculations in 3 flavor number scheme (3FS) proposed in the literature~\cite{Artoisenet:2007xi,Baranov:2006dh,Berezhnoy:1998aa} largely underestimate the SPS predictions of VFNS. VFNS enhances 3FS ggF central predictions by a factor $3$, while the upper limits due to the scale variation are more than $4$ times larger after including the (anti-)charm initial state. This implies the importance of resumming the initial collinear logarithms $\alpha_s^n\log^k{\frac{\mu_F^2}{m_c^2}}$ with $0<k\leq n$ in our interested kinematic regime. Hence, the statement of ``SPS can be negligible" should be revised, and it sheds light on resolving the tensions between the LHCb data and the theoretical predictions discussed in the Introduction. On the other hand, the PDF uncertainty is subdominant compared to the scale uncertainty. It introduces additional $\sim 15\%$ and $\sim 30\%$ theoretical errors for ggF and VFNS results. It is, however, interesting to notice that not all PDF set gives reasonable estimates of the true PDF uncertainty.

\subsection{$\sigma_{{\rm eff},pp}$ and theory-data comparison}

\subsubsection{Determination of $\sigma_{{\rm eff},pp}$}

Following the prescription detailed in Appendix~\ref{app:likelihood}, we redetermine $\sigma_{{\rm eff},pp}$ from the LHCb data via the likelihood-based approach. In order to maximize our prediction power and to minimize the possible bias, we only choose the integrated cross section, the normalized shapes of $P_{T,J/\psi}$ and invariant mass $M(J/\psi+D^0)$ distributions from the LHCb measurement of $J/\psi+D^0$ final state. We have checked that the integrated cross section alone does not give any meaningful constraint for the effective cross section $\sigma_{{\rm eff},pp}$, which is plagued with huge SPS theoretical uncertainties (especially the scale uncertainty) as shown in the Sec.~\ref{sec:SPSres}. Not all data in the aforementioned distributions will be adopted in the likelihood fit. We only select the $6$ $P_{T,J/\psi}\in [1.0,4.0]$ GeV data and $10$ $M(J/\psi+D^0) \in  [5.5,10.5]$ GeV data. The very low $P_{T,J/\psi}$ data ($P_{T,J/\psi}<1$ GeV) are excluded because of the imperfection in our DPS modeling in the regime (cf. Fig.~\ref{fig:ppfit}). For the same reason, the first near-threshold bin $5.0<M(J/\psi+D^0)/{\rm GeV}<5.5$ has been excluded as well. The reasons for not taking into account the data in the tails of the two spectra are twofold. The statistical Monte Carlo fluctuations in the theoretical curves start to be significant. In the $P_{T,J/\psi}$ distribution, we also encounter an unexpected large fluctuation in bin $4.0<P_{T,J/\psi}/{\rm GeV}<4.5$ of the LHCb data, which is beyond the size of the given experimental error. In total, after accounting for the integrated cross section, we used $17$ experimental data in our likelihood fit procedure to extract the value of $\sigma_{{\rm eff},pp}$. 

\begin{figure}
  \centering
    \includegraphics[width=0.65\textwidth,valign=c,draft=false]{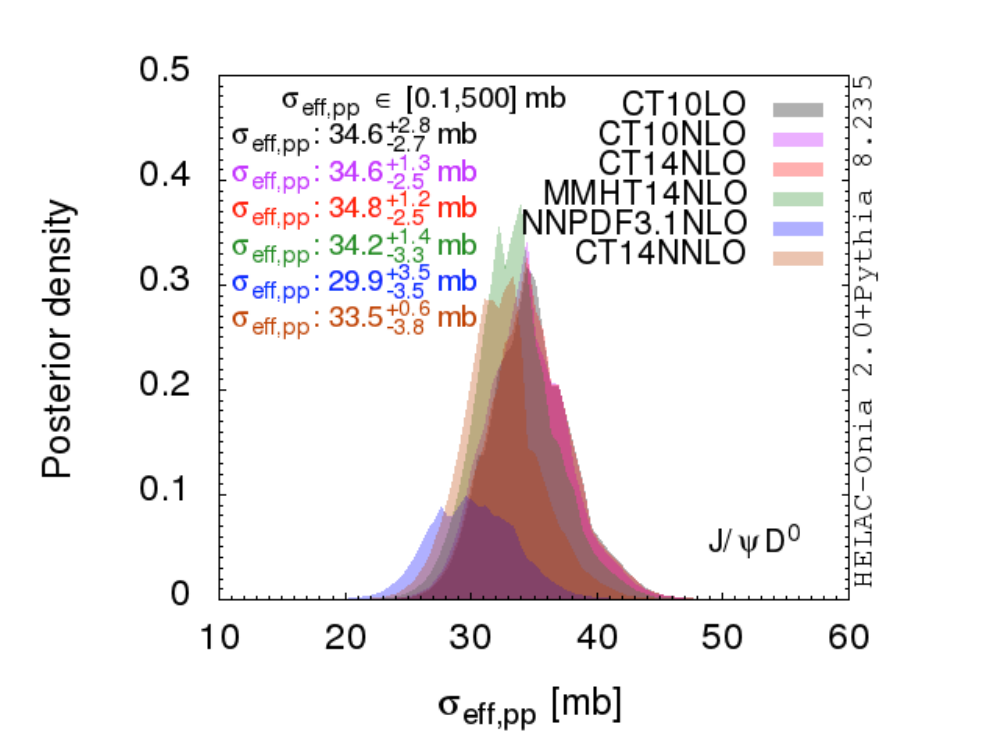}
  \caption{Posterior probability density of $\sigma_{{\rm eff},pp}$ for six different PDF sets used in VFNS SPS. The central values along with their $68\%$ CL ($1\sigma$) uncertainties are also shown in the figure legend.
  \label{fig:posterior}}
\end{figure}

\begin{figure}
  \centering
    \includegraphics[width=0.65\textwidth,valign=c,draft=false]{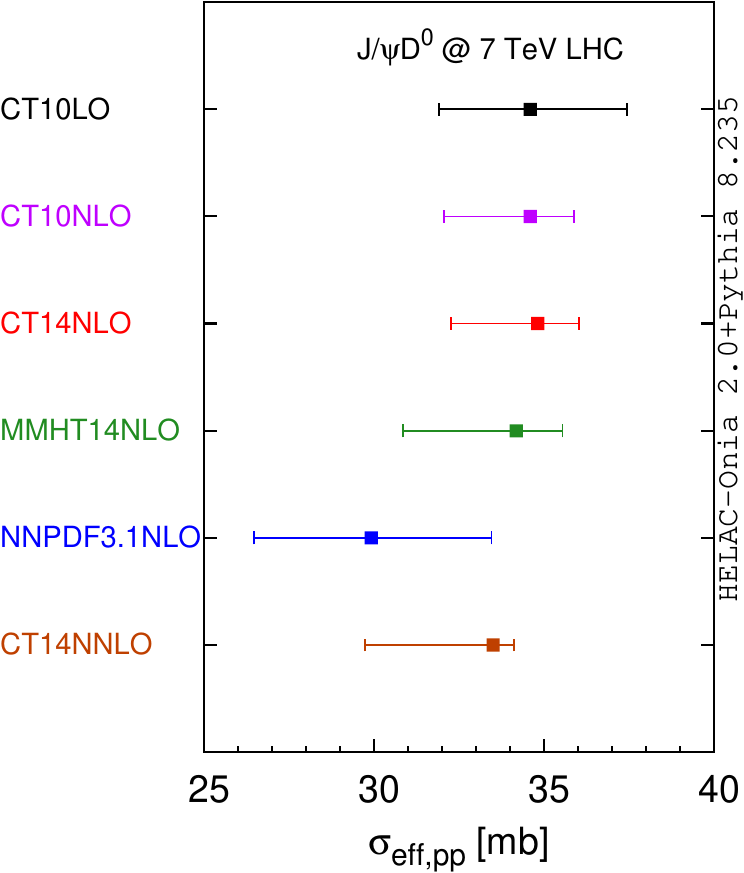}
  \caption{Constraints on $\sigma_{{\rm eff},pp}$ from 7 TeV LHCb $J/\psi+D^0$ data by using various PDFs in SPS. The central values are indicated as squares, and the $1\sigma$ errors are represented with the error bars.
  \label{fig:sigmaeff}}
\end{figure}

In Fig.~\ref{fig:posterior}, we show the marginalized constraints on $\sigma_{{\rm eff},pp}$ from the 7 TeV LHCb $J/\psi+D^0$ data. The posterior probability density is defined as $L_{17}(\sigma_{{\rm eff},pp})\left[\int_{0.1}^{500}{L_{17}(\sigma_{{\rm eff},pp})d\sigma_{{\rm eff},pp}}\right]^{-1}$, where we have used the global likelihood function Eq.(\ref{eq:globalLn}) in Appendix~\ref{app:likelihood}. We also assumed a uniform prior on $\sigma_{{\rm eff},pp}$ in the range $[0.1,500]$ mb. The final constraints on $\sigma_{{\rm eff},pp}$ are displayed in Fig.~\ref{fig:sigmaeff}. For VFNS SPS with various PDF sets as reported in Table~\ref{tab:FFNSvsVFNS}, the best-fitted values of $\sigma_{{\rm eff},pp}$ span from $29.9$ mb to $34.8$ mb. The $68\%$ quantiles of the effective cross section are reported both in the figure legends of Figs.~\ref{fig:posterior} and \ref{fig:sigmaeff}. The worst precision is from the SPS with the NNPDF3.1NLO PDF, but is still better than $12\%$. All of the inferred values of $\sigma_{{\rm eff},pp}$ from various PDF sets are consistent with each other, which are within $1$ standard deviation in Fig.~\ref{fig:sigmaeff}. The inclusion of VFNS SPS dramatically increases the $\sigma_{{\rm eff},pp}$ values by a factor of $2$ with respect to those assuming negligible SPS in Ref.~\cite{Aaij:2012dz}. As we will show in the later subsections, the theoretical calculation and the LHCb data will also be largely reconciled. The larger $\sigma_{{\rm eff},pp}$ values are also preferred by the LHCb double-D data~\cite{Aaij:2012dz} as pointed out in the VFNS analysis of Ref.~\cite{Helenius:2019uge}.

Given the PDF dependence is marginal with respect to the other sources of theoretical errors, in the following, we will only use CT14NLO in the SPS results. The final determination of $\sigma_{{\rm eff},pp}$ with this PDF is $34.8^{+1.2}_{-2.5}$ mb, posing the most precise (around $7\%$) constraint. Such a $\sigma_{{\rm eff},pp}$ value from $J/\psi+D^0$ will also be applied to other three final states $J/\psi+C (C=D^+,D_s^+,\Lambda^+_c)$, because they basically undergo the same hard scattering process.~\footnote{Because of the fiducial phase space cuts on the open charm hadrons, a slight dependence on the $C$ species may occur because of their different fragmentation functions.} With this setup, we start to have our theory-data comparison.

\subsubsection{Integrated cross sections}

After combining all the aforementioned information, we are able to refine the theoretical calculations, in particular the SPS postdictions. The comparison to the LHCb measurement can be found both in Table~\ref{tab:XSpp} and in Fig.~\ref{fig:XSpp}. We call the coverage of the prompt over color-singlet $J/\psi$ ratios listed in Table~\ref{tab:COFD} as ``LDME uncertainty". The central values are rescaled with the ratio numbers of the LDME Set 5, which should be taken with a grain of salt. For the sake of being conservative,  we take the envelope of the 9-point scale variation in 9 LDME sets as our combined scale+LDME uncertainty estimate. PDF uncertainty will be either given separately (e.g. in Table~\ref{tab:XSpp}) or summed in the quadrature way with other theoretical uncertainties. As pointed out in the last subsection, we use $\sigma_{{\rm eff},pp}=34.8$ mb in the following context in DPS estimates. Since our DPS results are more-or-less data driven (modulo $\sigma_{{\rm eff},pp}$), we do not associate any theoretical uncertainty to them. As announced, the scale uncertainty is the dominant theoretical error, which can be clearly seen in Fig.~\ref{fig:XSpp}. A higher-order QCD calculation, at least NLO, is demanding in the near future to improve the accuracy. Nevertheless, assuming that the scale variation has successfully captured the correct size of the missing higher-order $\alpha_s$ terms, the VFNS SPS can be as sizeable as the experimental data. Therefore, from the viewpoint of the integrated cross section, in principle, no DPS is needed. However, the current prominent uncertainty cannot forbid a significant DPS contribution. We will also give a remark on the open charm baryon $\Lambda_c^+$ here. The baryon $\Lambda_c^+$ has been shown at the LHC that the traditional single-parton fragmentation function is insufficient to describe the single-inclusive data~\cite{Acharya:2017kfy}, which calls for novel mechanisms (e.g., the coalescence mechanism~\cite{Plumari:2017ntm}). Hence, one should be careful to interpret the $J/\psi+\Lambda_c^+$ data here, since we only include the fragmentation contribution in our SPS simulations.

\begin{table}[H]
\centering\renewcommand{\arraystretch}{1.2}
\begin{tabular}{c|ccc|c} 
\hline\hline
Final state & VFNS SPS & DPS & VFNS SPS+DPS & LHCb data~\cite{Aaij:2012dz} \\\hline\hline
$J/\psi+D^0$ & $23.3^{+148.9+7.8}_{-20.3~-1.8}$ & $68.6$ & $92.0^{+148.9+7.8}_{-20.3~-1.8}$ & $161.0\pm3.7\pm12.2$ \\\hline
$J/\psi+D^+$ & $12.3^{+84.3+4.1}_{-10.8-1.0}$ & $28.4$ & $40.6^{+84.3+4.1}_{-10.8-1.0}$ & $56.6\pm1.7\pm5.9$\\\hline
$J/\psi+D_s^+$ & $4.6^{+27.7+1.6}_{-4.0~-0.4}$ & $11.2$ & $15.7^{+27.7+1.6}_{-4.0~-0.4}$ & $30.5\pm2.6\pm3.4$\\\hline
$J/\psi+\Lambda_c^+$ & $2.1^{+11.5+0.7}_{-1.8~-0.2}$ & $22.4$ & $24.5^{+11.5+0.7}_{-1.8~-0.2}$ & $43.2\pm7.0\pm12.0$\\\hline\hline
\end{tabular}
\caption{\label{tab:XSpp} The integrated cross sections (in unit of nb) of $J/\psi+C$ [$C: D^0,D^+,D_s^+,\Lambda_c^+$] within the LHCb acceptance at $7$ TeV~\cite{Aaij:2012dz}. The quoted theoretical errors in the second and fourth columns represent scale+LDME and PDF uncertainties, respectively. The first (second) uncertainty in the LHCb data column is statistical (systematic).}
\end{table}

\begin{figure}
  \centering
    \includegraphics[width=0.65\textwidth,valign=c,draft=false]{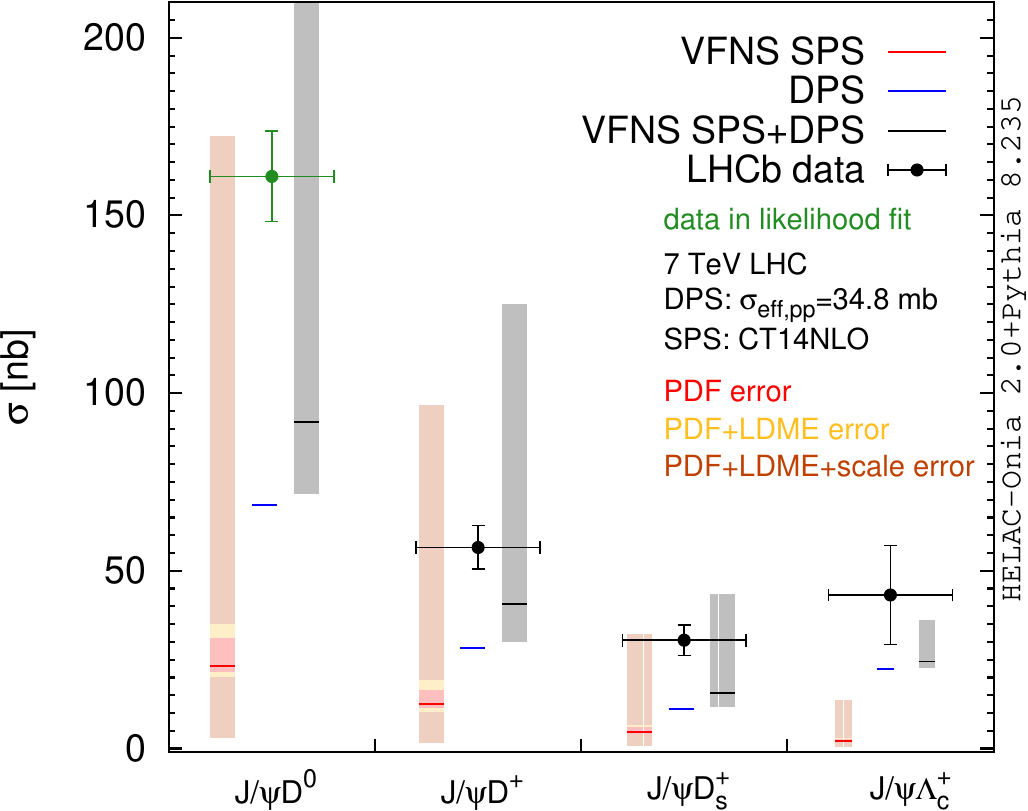}
  \caption{The comparison of theoretical calculations (VFNS SPS, DPS and VFNS SPS+DPS) and the LHCb measurement~\cite{Aaij:2012dz} of the integrated cross sections for prompt $J/\psi$ and open charm hadron associated production at $\sqrt{s}=7$ TeV $pp$ collisions.
  \label{fig:XSpp}}
\end{figure}

\subsubsection{Differential distributions}

We now turn to the differential distributions, which are originally in tensions with the DPS-dominance hypothesis. Given the similarity among different open charm hadron species, we will only focus on $J/\psi+D^0$ here, while the comparison for other hadron species can be found in Appendix~\ref{app:diffcomp}. Following what has been given in the LHCb paper~\cite{Aaij:2012dz}, the comparison is only performed at the normalized distribution (shape) level, i.e., the sum of all bins gives unity. This will help to cancel global and correlated systematical experimental errors, which are present in the absolute differential distributions. On the contrary, we view the scale uncertainties in SPS are not bin-by-bin correlated. In fact, the missing higher order quantum radiative corrections may distort the shapes within the given uncertainty bands. For other theoretical uncertainties, they can be treated in the full correlation way. However, they are subdominant and unimportant from this perspective.

\begin{figure}
  \centering
  \subfloat[Transverse momentum of $J/\psi$]{
    \includegraphics[width=0.45\textwidth,valign=c,draft=false]{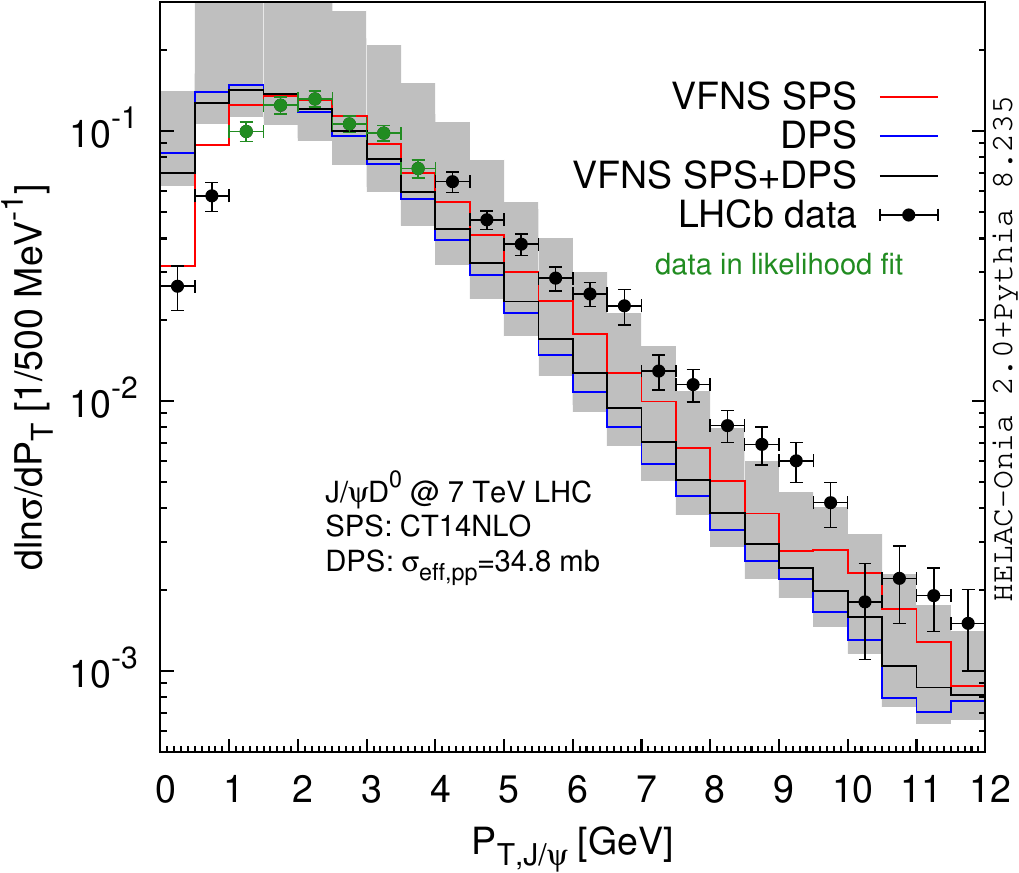}\label{fig:ppD0a}}
  \subfloat[Invariant mass of $J/\psi$ and $D^0$]{\includegraphics[width=0.45\textwidth,valign=c,draft=false]{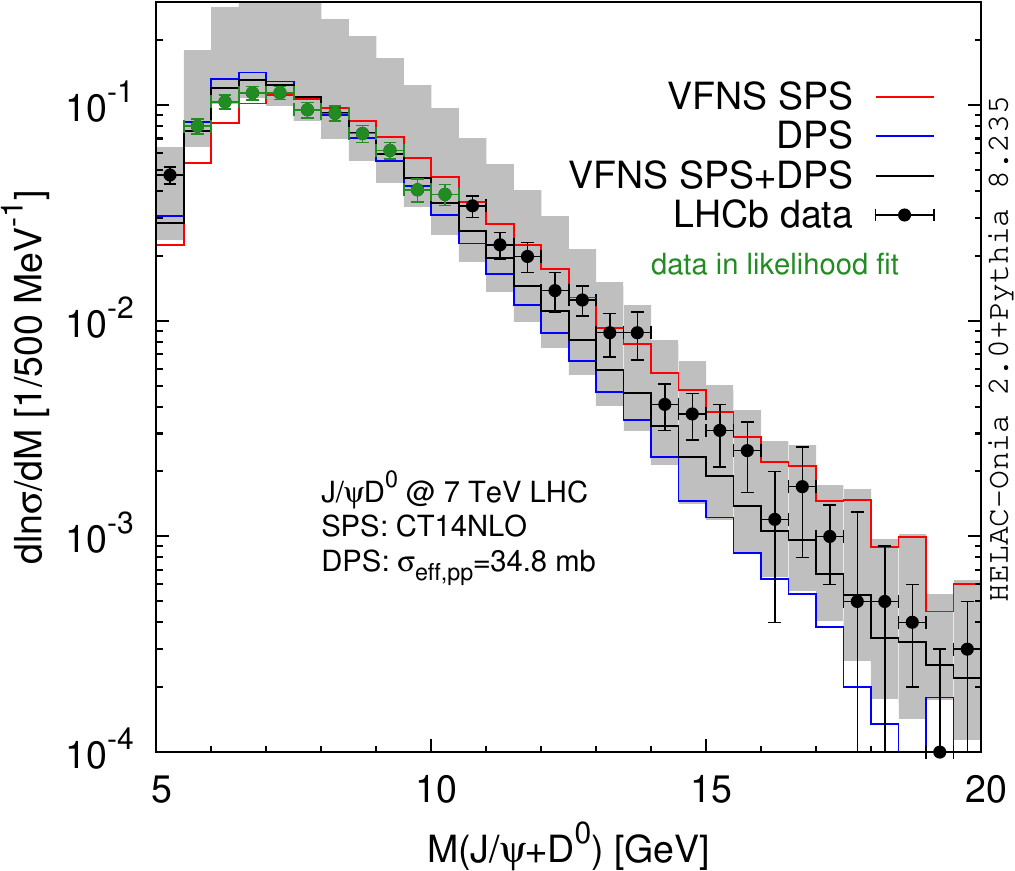}\label{fig:ppD0b}}\\
  \subfloat[Invariant mass of $J/\psi$ and $D^0$]{\includegraphics[width=0.45\textwidth,valign=c,draft=false]{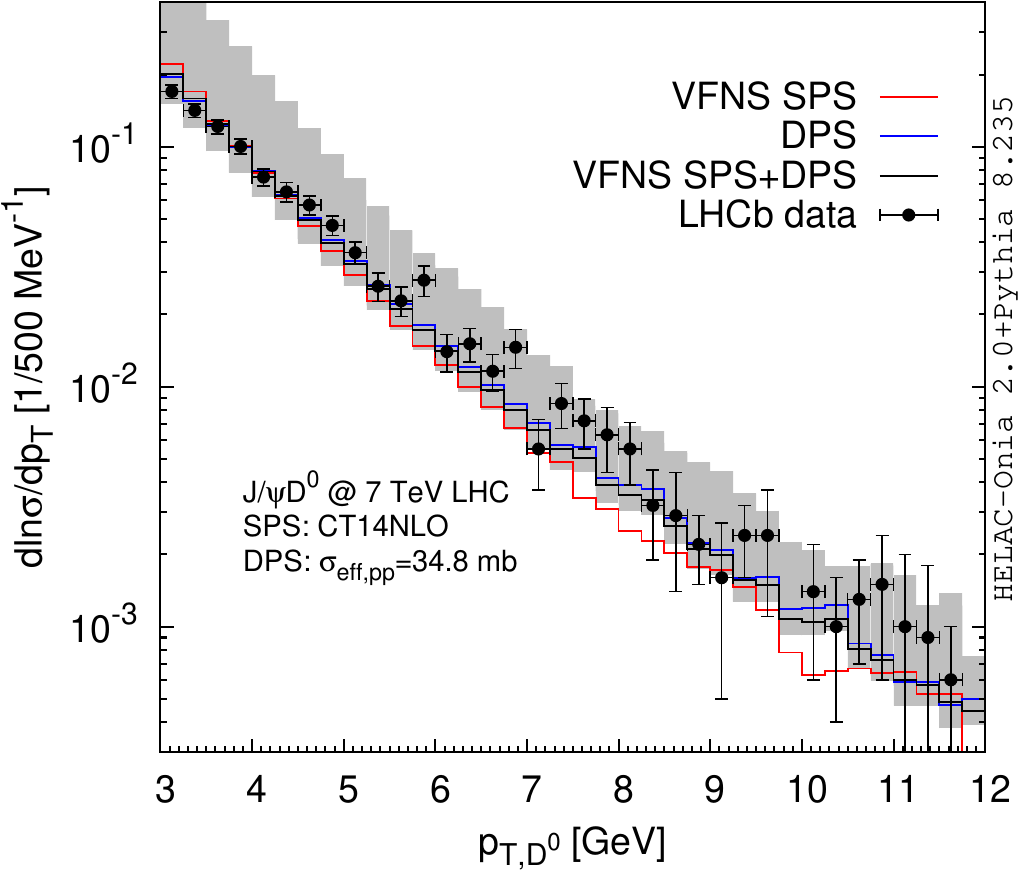}\label{fig:ppD0c}}
  \subfloat[Rapidity gap between $J/\psi$ and $D^0$]{\includegraphics[width=0.45\textwidth,valign=c,draft=false]{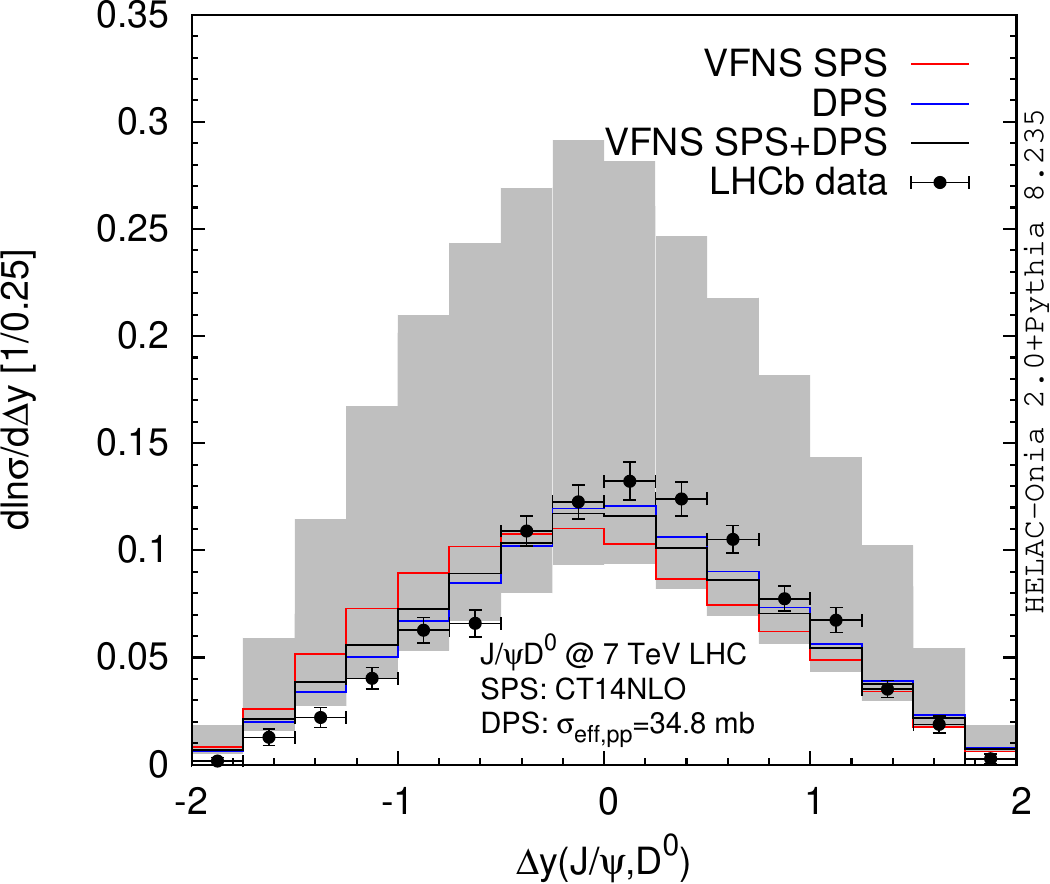}\label{fig:ppD0d}}\\
   \subfloat[Azimuthal angle difference between $J/\psi$ and $D^0$]{\includegraphics[width=0.45\textwidth,valign=c,draft=false]{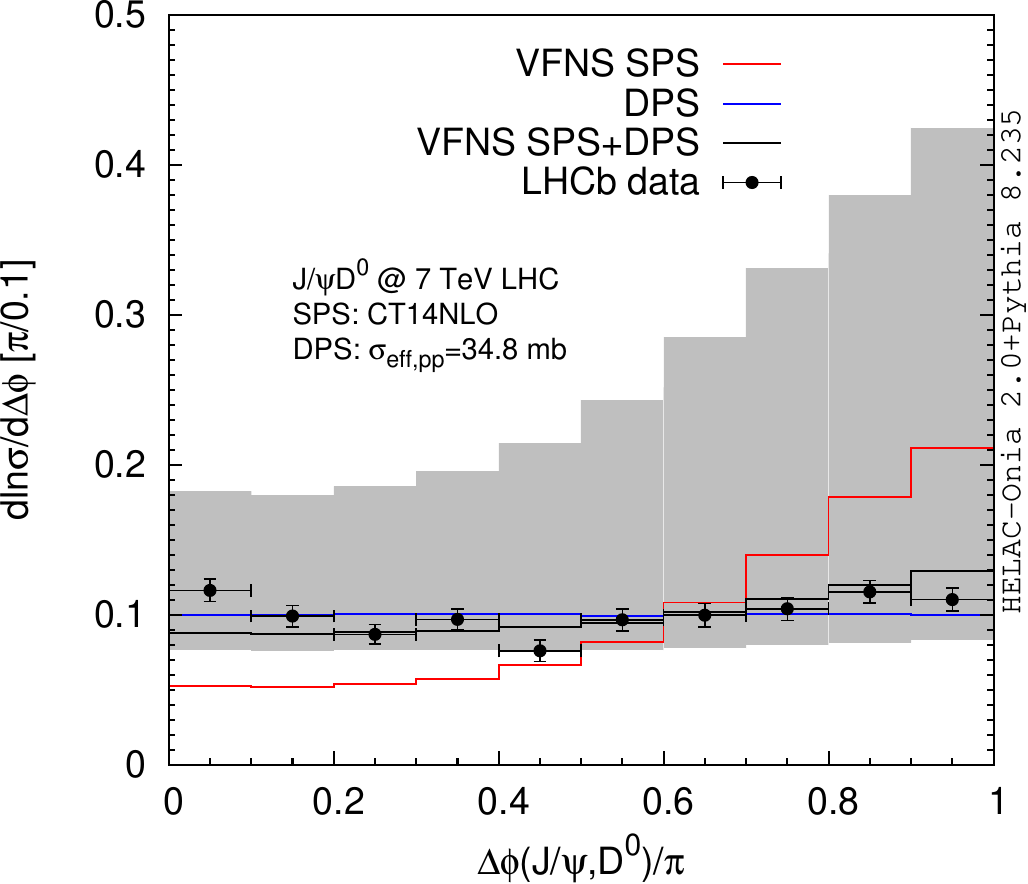}\label{fig:ppD0e}}
  \caption{The differential shape comparison between the LHCb $J/\psi+D^0$ data and the corresponding theoretical calculations. They are (a) $P_{T,J/\psi}$, (b) invariant mass $M(J/\psi+D^0)$, (c) $p_{T,D^0}$, (d) $\Delta y(J/\psi,D^0)$, and (e) $\Delta \phi(J/\psi,D^0)$. The differential cross sections have been divided by the corresponding integrated cross sections.
  \label{fig:ppD0}}
\end{figure}

The theory and data comparison for the five distributions is presented in Fig.~\ref{fig:ppD0}. The distributions are the transverse momenta of $J/\psi$ [Fig.~\ref{fig:ppD0a}] and $D^0$ [Fig.~\ref{fig:ppD0c}] mesons, the invariant mass of the two mesons [Fig.~\ref{fig:ppD0b}], the rapidity gap $\Delta y(J/\psi,D^0)=y_{J/\psi}-y_{D^0}$ [Fig.~\ref{fig:ppD0d}], and the azimuthal angle difference $\Delta \phi(J/\psi,D^0)=\left|\phi_{J/\psi}-\phi_{D^0}\right|$ [Fig.~\ref{fig:ppD0e}] between the meson pair. We have marked the data those enter into our likelihood fit in determining $\sigma_{{\rm eff},pp}$ in green. In general, the previous tensions are greatly alleviated. The reasonable agreements have been achieved in most of the distributions, except the first two bins in the $P_{T,J/\psi}$ distribution. The disagreement in the first two bins is anticipated due to the limitation of our DPS modeling. Although there are a few random fluctuations in the experimental data, almost all of them lie within the grey bands.

The spectra of $P_{T,J/\psi}$ and $M(J/\psi+D^0)$ are quite distinct between SPS and DPS. This fact reflects that $J/\psi$ undergoes the single-charm fragmentation process in $J/\psi+c$ production at high $P_T$, while several competitive mechanisms are present in DPS $J/\psi$ production, which is equivalent to the single inclusive $J/\psi$ process. The LHCb data favour the hard spectra close to the SPS ones, implying the necessities of the SPS events in interpreting the data.  On the other hand, SPS curves are quite similar to the DPS ones in the $p_{T,D^0}$ and $\Delta y(J/\psi,D^0)$ distributions. In the $p_{T,D^0}$ distribution, $D^0$ are generated via the charm quark fragmentation in both SPS and DPS. Its shape, hence, has no discrimination power to separate the two contributions. In the rapidity gap $\Delta y(J/\psi,D^0)$ distribution, DPS shape turns out to be very symmetric around $\Delta y(J/\psi,D^0)=0$ because the two mesons are mainly produced in two independent scattering subprocesses, while due to the correlation in SPS, the $D^0$ meson is slightly more forward than the $J/\psi$ meson. However, such an asymmetric behavior in the SPS distribution is still quite mild. The azimuthal angle difference $\Delta \phi(J/\psi,D^0)$, in principle, could be useful to disentangle the SPS and the DPS mechanisms, because DPS contribution is flat due to the dominance of DPS$_1$, while the SPS has no reason to be flat due to the correlation of the final particles. However, as pointed out several times before, the nonflat $\Delta \phi(J/\psi,D^0)$ distribution can be diluted by the introduction of the initial $k_T$ smearing. The smearing effect can make the original nontrivial structures in $\Delta \phi(J/\psi,D^0)$ fading out. In our simulation, such an effect is modeled by both the initial state radiation and the primordial $k_T$ in the beam remnants implemented in \Pythia8. However, we should remind the reader that the primordial $k_T$ is purely phenomenological and suffers from the uncertainty in model dependence. With our simulation, it seems that the shape of the SPS is inadequate to describe the experimental data of $\Delta \phi(J/\psi,D^0)$, while the corresponding DPS shape matches the data in a better way. The SPS contribution peaks at the away side $\Delta \phi(J/\psi,D^0)=\pi$, i.e., back-to-back in the transverse plane. The sum of the SPS and the DPS has a right shape, which indicates that the DPS process should not be ignored as well.

%
%
%

\section{Predictions for proton-lead collisions at $\sqrt{s_{NN}}=8.16$ TeV\label{sec:pA}}

After understanding the $pp$ data, we are in the position to present our theoretical predictions for the upcoming LHCb measurement in proton-lead ($p{\rm Pb}$) collisions at the average nucleon-nucleon center-of-mass energy $\sqrt{s_{NN}}=8.16$ TeV.~\footnote{A collection of the theoretical predictions for a bunch of other inclusive observables at the same energy can be found in Ref.~\cite{Albacete:2017qng}.} As advocated in Ref.~\cite{Shao:2020acd}, such a measurement will be very useful, because it will, for the first time, reveal both the nucleus geometrical enhancement effect~\cite{Strikman:2001gz} and the impact-parameter dependence of the nuclear modification on the incoming parton flux. In this context, we assume the only dominant cold nuclear matter effect has been encoded in the universal nPDFs. Such a hypothesis has been effectively justified for (nonexcited) heavy flavor production at the LHC~\cite{Kusina:2017gkz,Lansberg:2016deg}, where the nPDF effect is quite significant and therefore must be taken into account. 

In order to quantify the spatial dependence of nPDFs, we assume such a dependence is only related to the local nuclear thickness function $T_A(\harpoon{b})$, where $\harpoon{b}$ is the two-dimensional impact parameter vector in the colliding transverse plane. The nuclear modification $R_{k}^{A}(x,\harpoon{b})$ for a parton $k=g,q,\bar{q}$ at the position $\harpoon{b}$ is expressed as
\begin{eqnarray}
R_{k}^{A}(x,\harpoon{b})-1&=&\left(R_{k}^{A}(x)-1\right)G\left(\frac{T_A(\harpoon{b})}{T_A(\harpoon{0})}\right),\label{eq:bnPDF}
\end{eqnarray}
where $R_{k}^{A}(x)$ is simply the ratio of nPDF over free-nucleon PDF for the parton $k$ at the Bjorken $x$. $G()$ could be any function that satisfies the condition
\begin{eqnarray}
\int{T_A(\harpoon{b})G\left(\frac{T_A(\harpoon{b})}{T_A(\harpoon{0})}\right)d^2\harpoon{b}}&=&A.
\end{eqnarray}
We choose a test function $G\left(\frac{T_A(\harpoon{b})}{T_A(\harpoon{0})}\right)\propto \left(\frac{T_A(\harpoon{b})}{T_A(\harpoon{0})}\right)^a$ and opt for the simple hard-sphere form for the thickness function $T_A$.~\footnote{The possible refinements of using other $T_A$ forms as well as by accounting for the neutron-skin effect can be found in Ref.~\cite{Shao:2020acd}. These effects are, however, expected to be minor.} After incorporating both the nuclear-collision geometry and the spatial-dependent nPDFs, we have the following DPS$_i$ differential cross section in proton-nucleus collisions~\cite{Shao:2020acd}:
\begin{eqnarray}
d\sigma_{pA\to J/\psi+C}^{{\rm DPS}_i}&=&Ad\sigma_{J/\psi+C,11}^{{\rm DPS}_i}\left[\frac{3^{1-2a}(a+3)^{2a}}{2a+3}+\frac{\sigma_{{\rm eff},pp}}{\pi R_A^2}\left(A-1\right)\frac{9^{1-a}(a+3)^{2a}}{4(a+2)}\right]\nonumber\\
&&+A\left(d\sigma_{J/\psi+C,10}^{{\rm DPS}_i}+d\sigma_{J/\psi+C,01}^{{\rm DPS}_i}\right)\left[1-\frac{3^{1-2a}(a+3)^{2a}}{2a+3}+\frac{\sigma_{{\rm eff},pp}}{\pi R_A^2}\left(A-1\right)\left(\frac{3^{2-a}(a+3)^a}{2(a+4)}-\frac{9^{1-a}(a+3)^{2a}}{4(a+2)}\right)\right]\nonumber\\
&&+Ad\sigma_{J/\psi+C,00}^{{\rm DPS}_i}\left[-1+\frac{3^{1-2a}(a+3)^{2a}}{2a+3}+\frac{\sigma_{{\rm eff},pp}}{\pi R_A^2}\left(A-1\right)\left(\frac{9}{8}+\frac{9^{1-a}(a+3)^{2a}}{4(a+2)}-\frac{3^{2-a}(a+3)^a}{(a+4)}\right)\right],\label{eq:pADPS}
\end{eqnarray}
where $d\sigma_{J/\psi+C,kl}^{{\rm DPS}_i}$ is the DPS$_i$ (differential) cross section in (free or bounded) nucleon-nucleon collisions. The convoluted PDFs in $d\sigma_{J/\psi+C,kl}^{{\rm DPS}_i}$, for the two initial partons from the ion beam, are two (spatial-averaged) nPDFs when $kl=11$, one nPDF and one free-nucleon PDF  when $kl=10$ or $01$, and two free-nucleon PDFs when $kl=00$, respectively. It means $d\sigma_{J/\psi+C,00}^{{\rm DPS}_i}=d\sigma_{pp\to J/\psi+C}^{{\rm DPS}_i}$ without considering the possible isospin effect (e.g., the gluon initial state in our case). In this context, we take EPPS16~\cite{Eskola:2016oht} with the LHC $J/\psi$ or $D^0$ constraint derived in Ref.~\cite{Kusina:2017gkz} as our nPDF set~\footnote{The constraints from $J/\psi$ and $D^0$ data are quite similar, and they pose the strongest limits in Ref.~\cite{Kusina:2017gkz}.} and CT14NLO as the corresponding proton PDF. According to our $pp$ fit, $\sigma_{{\rm eff},pp}$ is fixed to $34.8$ mb, and $A=208$ and $R_A=6.624$ fm for ${\rm Pb}$. In the remainder of the paper, without losing generality, all the kinematics will be defined in the center-of-mass frame of the averaged per nucleon-nucleon collision.

\subsection{Cross sections}

We have reported the breakdown of cross sections from three different sources, VFNS SPS, DPS$_1$, and DPS$_2$, within the LHCb detector acceptance~\cite{Zhang:2020} in Table~\ref{tab:XSpPb}, where the power $a$ appearing in the $G()$ is fixed to the widespread value $a=1$, in which the nuclear modification is proportional to the thickness function. The forward (backward) region imposes both the $J/\psi$ meson and the open (anti-)charm hadron $C$ being in the rapidity interval $1.7<y_{J/\psi},y_{C}<3.7$ ($-4.7<y_{J/\psi},y_{C}<-2.7$).~\footnote{Following the ongoing LHCb analysis, we have also imposed the two additional cuts $P_{T,J/\psi}, p_{T,C}<12$ GeV on the generated events.} Several theoretical uncertainties are associated to various contributions. In the category of VFNS SPS, the first quoted errors are from the 9-point scale variation and the LDME uncertainty, while the second errors stand for both the nPDF and proton PDF parameterization uncertainties. The two errors in DPS$_2$ numbers are the scale uncertainty and the reweighted nPDF uncertainty, where, as mentioned in Sec.~\ref{sec:dpspp}, we have only included three Fock states ($\ss,\so$ and $\sps$) and used LDMEs of Set 8 in Table 3 in Ref.~\cite{Lansberg:2019fgm}. Moreover, only $68\%$ nPDF uncertainties are associated to the DPS$_1$ cross sections. Similar to the finding in the 7 TeV $pp$ case, DPS$_2$ contributions are at least 20 times smaller than DPS$_1$ even taking into account the large scale uncertainty. On the other hand, the upper limits of the VFNS SPS predictions are comparable with the DPS$_1$ components. Since there is no lower $p_T$ cut, we are probing even lower scales than the 7 TeV $pp$ case. The scale uncertainty in VFNS SPS is extremely large. Such a situation is inevitable before carrying out a computation including higher-order QCD corrections. For the most abundant process $J/\psi+D^0$, the $p{\rm Pb}$ cross sections are predicted to be around $0.5-1.0$ mb in the both fiducial regions. Given the similarities of the results in various open charm hadron species, we will only concentrate on $J/\psi+D^0$ plus its charge conjugate mode in the following.

\begin{table}[H]
\centering\renewcommand{\arraystretch}{1.2}
\begin{tabular}{c|c|c|c|c|c|c} 
\hline\hline
\multirow{3}{*}{Final state} & \multicolumn{3}{c|}{$1.7<y_{J/\psi},y_{C}<3.7$} & \multicolumn{3}{c}{$-4.7<y_{J/\psi},y_{C}<-2.7$}\\\cline{2-7}
& \multirow{2}{*}{VFNS SPS} & \multicolumn{2}{c|}{DPS} & \multirow{2}{*}{VFNS SPS} & \multicolumn{2}{c}{DPS}\\\cline{3-4}\cline{6-7}
& & DPS$_1$ & DPS$_2$ & & DPS$_1$ & DPS$_2$\\\hline\hline
$J/\psi+D^0$ & $(3.9^{+47.9+2.0}_{-3.1~-0.8})\cdot 10^{-2}$ & $0.36\pm0.06$ & $(1.3^{+16.5}_{-1.2}\pm 0.2)\cdot 10^{-3}$ & $(4.2^{+43.9+4.1}_{-4.1~-0.5})\cdot 10^{-2}$ & $0.47\pm0.04$ & $(1.5^{+17.7}_{-1.4}\pm 0.1)\cdot 10^{-3}$\\\hline
$J/\psi+D^+$ & $(2.0^{+26.1+1.0}_{-1.6~-0.4})\cdot 10^{-2}$ & $0.14\pm0.02$ & $(5.6^{+71.8}_{-5.3}\pm 0.8)\cdot 10^{-4}$ & $(2.2^{+24.0+2.1}_{-2.1~-0.2})\cdot 10^{-2}$ & $0.18\pm0.01$ & $(6.7^{+77.2}_{-6.3}\pm 0.6)\cdot 10^{-4}$\\\hline
$J/\psi+D_s^+$ & $(6.9^{+78.8+3.5}_{-5.3~-1.4})\cdot 10^{-3}$ & $0.08\pm0.01$ & $(1.8^{+23.3}_{-1.7}\pm 0.2)\cdot 10^{-4}$ & $(7.4^{+71.7+7.1}_{-7.0~-0.8})\cdot 10^{-3}$ & $0.11\pm0.01$ & $(2.2^{+25.1}_{-2.1}\pm 0.2)\cdot 10^{-4}$\\\hline
$J/\psi+\Lambda_c^+$ & $(2.8^{+28.8+1.4}_{-2.1~-0.6})\cdot 10^{-3}$ & $0.11\pm0.02$ & $(2.1^{+27.4}_{-2.0}\pm 0.3)\cdot 10^{-4}$ & $(3.0^{+26.5+2.9}_{-2.9~-0.3})\cdot 10^{-3}$ & $0.15\pm0.01$ & $(2.5^{+29.5}_{-2.4}\pm 0.2)\cdot 10^{-4}$\\\hline\hline
\end{tabular}
\caption{\label{tab:XSpPb} The forward and backward cross sections (in unit of mb) of $J/\psi+C$ [$C: D^0,D^+,D_s^+,\Lambda_c^+$] in proton-lead collisions at $\sqrt{s_{NN}}=8.16$ TeV. See the text for the description of the quoted theoretical errors and the parameter setup.}
\end{table}

The potential in constraining impact-parameter-dependent gluon nPDF from the process is very intriguing. The cross sections of $J/\psi+D^0$ are displayed in the functions of $a\in [0.0,4.0]$ in Fig.~\ref{fig:pPbD0XS}. As obvious, the VFNS SPS (red bands) are independent of the $a$ value.~\footnote{The two different errors listed in the VFNS SPS and DPS$_2$ columns of Tab.~\ref{tab:XSpPb} have been summed in the quadrature way in Fig.~\ref{fig:pPbD0XS} as well as in the following differential distributions.} The $a$ dependencies of the two DPS contributions (DPS$_1$ vs DPS$_2$) are more significant in the forward region than in the backward region. This is simply because the nuclear modification in the latter is much smaller. Figure~\ref{fig:pPbD0XSa} tells us if the impact-parameter dependence of the gluon nuclear modification is strong enough ($a>3$), we can easily observe the abundant $J/\psi+D^0$ events in the forward detection. In turn, this parameter space can be excluded if the measured cross section is smaller than $1$ mb. The challenging case is of course when the $|\harpoon{b}|$ dependence in $R_{g}^{A}(x,\harpoon{b})$ is weaker, in other words when $a\leq 3$ in our simplest parameterization of $G()$.

\begin{figure}
  \centering
  \subfloat[Forward]{
    \includegraphics[width=0.45\textwidth,valign=c,draft=false]{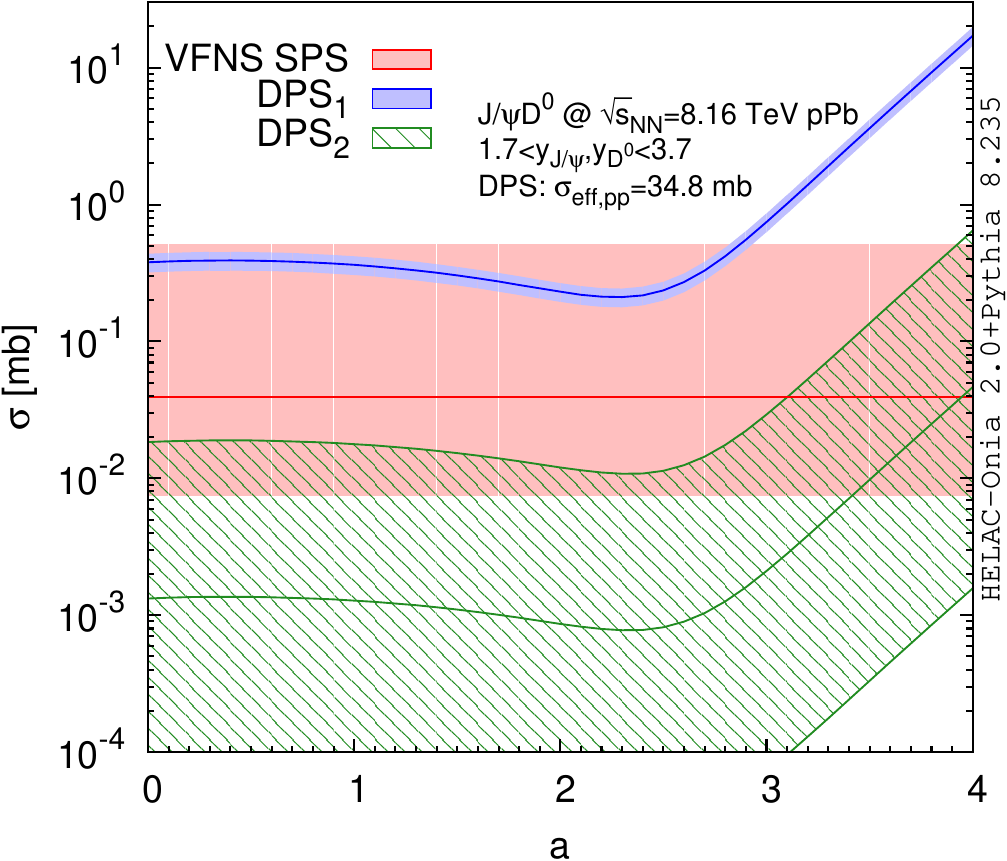}\label{fig:pPbD0XSa}}
  \subfloat[Backward]{\includegraphics[width=0.45\textwidth,valign=c,draft=false]{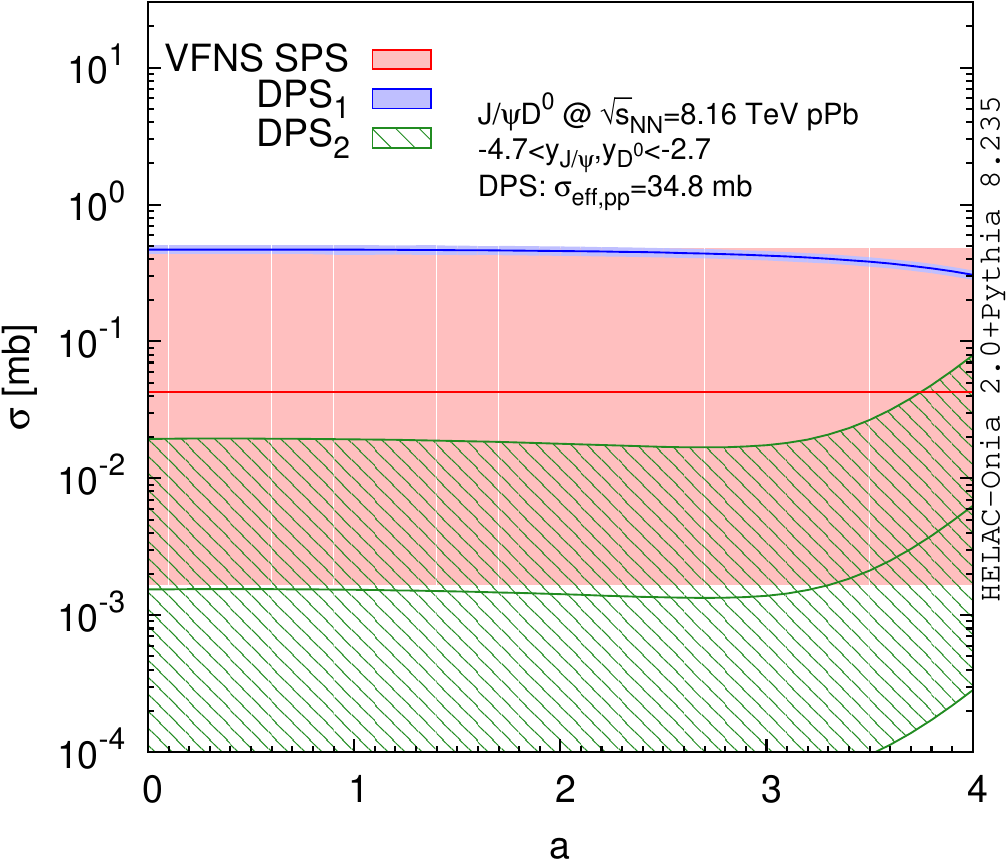}\label{fig:pPbD0XSb}}
  \caption{The ``$a$" dependencies of the integrated $J/\psi+D^0$ cross sections from VFNS SPS, DPS$_1$, and DPS$_2$ in proton-lead collisions at $\sqrt{s_{NN}}=8.16$ TeV, where $a$ is the power appearing in $G\left(\frac{T_A(\harpoon{b})}{T_A(\harpoon{0})}\right)\propto \left(\frac{T_A(\harpoon{b})}{T_A(\harpoon{0})}\right)^a$ that quantifies the impact-parameter dependence of the nuclear modification via Eq.(\ref{eq:bnPDF}). There are predictions both in the forward (left) and in the backward (right) rapidity intervals of the LHCb detector.
  \label{fig:pPbD0XS}}
\end{figure}

Aside from the integrated cross sections, the differential distributions can provide additional useful information. The invariant mass $M$ distributions of the $J/\psi+D^0$ system can be found in Fig.~\ref{fig:pPbD0dM}. The left (right) panel is for the forward (backward) rapidity region. The two DPS components have been summed, in which DPS$_1$ always dominants. As our showcases, we consider five different $a$ values in our $G()$ parameterization. They are $a=0$ (zero spatial dependence, green filled circle), $1$ (most widely used assumption, blue empty circle), $2$ (purple empty square), $3$ (brown filled square), and $4$ (orange). The same layout will be applied for the other distributions too. In order to make the plots less busy, only one DPS uncertainty band $a=1$ is associated to the DPS predictions in each plot. In the forward region, the probing $x$ in nPDF resides in the shadowing region, which has the significant nuclear modification effects. On the contrary, the backward region is exploring the $x$ regime close to the transition from the shadowing to the antishadowing. Hence, the nuclear modification is smaller. Likewise, the observation in the total cross sections, the DPS $a\leq 3$ distributions are quite similar and close to each other, while the very different features can be observed when $a=4$. In the forward case [Fig.~\ref{fig:pPbD0dMa}],  the $a=4$ DPS distribution is strongly enhanced uniformly in all bins with respect to the $a=3$ distribution. On the other hand, the backward $a=4$ curve in Fig.~\ref{fig:pPbD0dMb} features a more peculiar shape. In particular, the DPS prediction becomes negative if $5.5<M(J/\psi+D^0)/{\rm GeV}<8.5$. This can be understood from Eq.(\ref{eq:pADPS}) with $a=4$, which is reduced to 
\begin{eqnarray}
d\sigma_{p{\rm Pb}\to J/\psi+C}^{{\rm DPS}_i}&\overset{a\to 4}{=}&372763\left[1.09452d\sigma_{J/\psi+C,11}^{{\rm DPS}_i}-1.04534\left(d\sigma_{J/\psi+C,10}^{{\rm DPS}_i}+d\sigma_{J/\psi+C,01}^{{\rm DPS}_i}\right)+d\sigma_{J/\psi+C,00}^{{\rm DPS}_i}\right].\label{eq:DPSpPba4}
\end{eqnarray}
The differential cross section would be negative when the second piece in the brackets is larger than the other two terms. Such a pathological behavior, however, could be very sensitive to the fine tuning among the different input parameters. Moreover, although the SPS predictions are plagued with sizeable theoretical uncertainties, the DPS cross section, whatever the value of $a$, starts to overshoot the VFNS SPS cross section in large $M(J/\psi+D^0)$ in the backward region, especially when the mass is from $10-18$ GeV. It benefits from the reduced scale uncertainty in the VFNS SPS cross section at large scale. Such a specific kinematic region provides a useful mean to improve the purity of the DPS signal. 

\begin{figure}
  \centering
  \subfloat[Forward]{
    \includegraphics[width=0.45\textwidth,valign=c,draft=false]{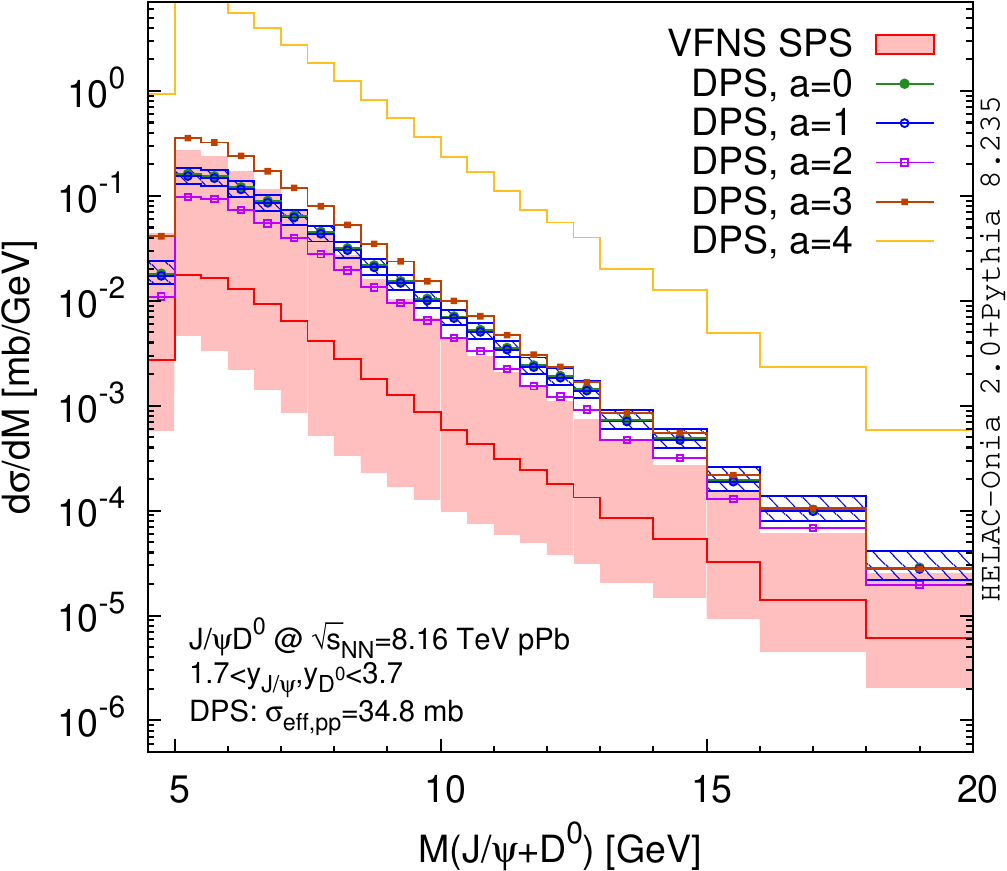}\label{fig:pPbD0dMa}}
  \subfloat[Backward]{\includegraphics[width=0.45\textwidth,valign=c,draft=false]{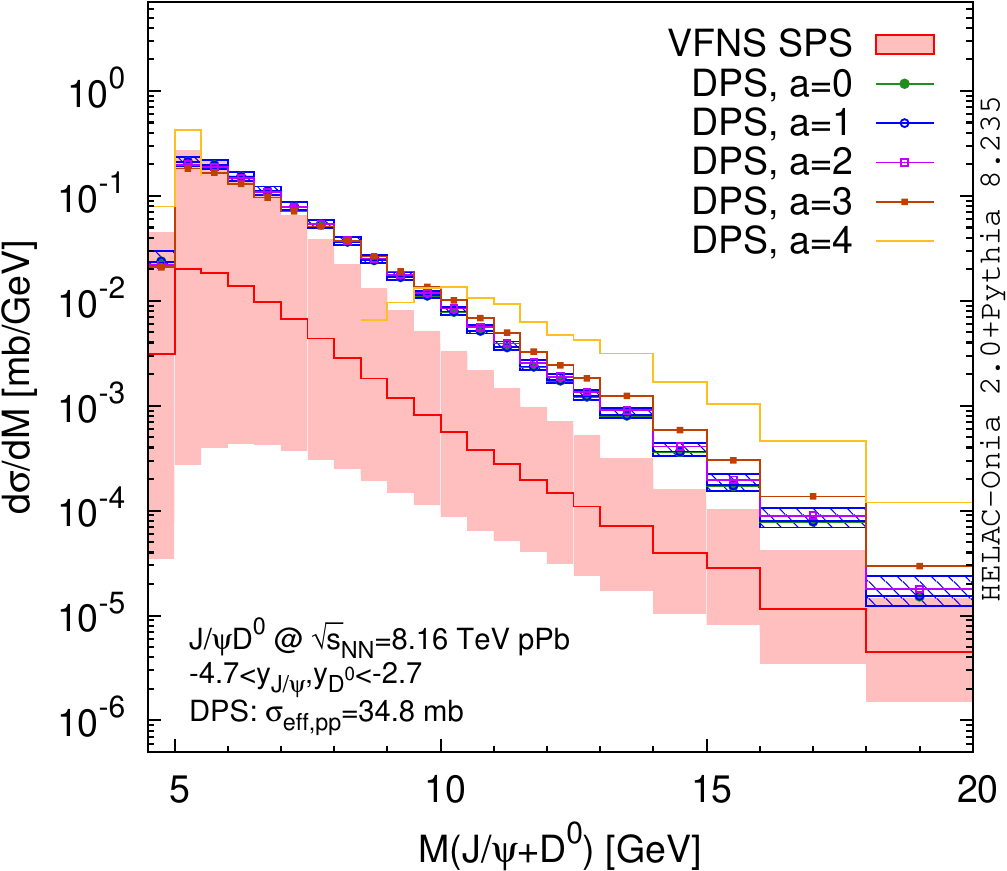}\label{fig:pPbD0dMb}}
  \caption{The distributions in the invariant mass $M$ of the $J/\psi+D^0$ system from VFNS SPS, DPS (the sum of DPS$_1$ and DPS$_2$) in proton-lead collisions at $\sqrt{s_{NN}}=8.16$ TeV. Several $a$ choices in the DPS distributions as showcases. Both forward (left) and backward (right) regions are considered.
  \label{fig:pPbD0dM}}
\end{figure}

Another kind of interesting distributions is the transverse momentum $P_T$ of the two meson pair, as shown in Fig.~\ref{fig:pPbD0dPT}. Due to the importance of initial (anti-)charm, the $P_T(J/\psi+D^0)$ spectra of VFNS SPS are soft, because the produced $J/\psi$ meson and the (anti-)charm quark are back-to-back in the LO 4FS calculation. Such a picture would not change even with higher order QCD corrections, since the two mesons are still likely along the beam pipe axis following the fragmentation of the (anti-)charm quark in the new topology appearing in the real emission diagrams. As opposite, the two mesons in the DPS events are almost independently produced. Hence, the DPS curves in Fig.~\ref{fig:pPbD0dPT} are generally harder than the VFNS SPS curves. A lower $P_T(J/\psi+D^0)$ cut can substantially enhance the DPS fraction in the events in the both interested rapidity intervals. $P_T(J/\psi+D^0)$ could be used to define the control regions of the DPS events on the experimental side. Similar to the invariant mass distributions, the strongest impact-parameter hypothesis with $a=4$ yields a peculiar dip structure in Fig.~\ref{fig:pPbD0dPTb}, which again can be attributed to the delicate cancellation among the three pieces in Eq.(\ref{eq:DPSpPba4}). Figure~\ref{fig:pPbD0dPTa} tells us that there is a quasiuniform enhancement of the $a=4$ curve compared to the $a\leq 3$ in the forward region.

\begin{figure}
  \centering
  \subfloat[Forward]{
    \includegraphics[width=0.45\textwidth,valign=c,draft=false]{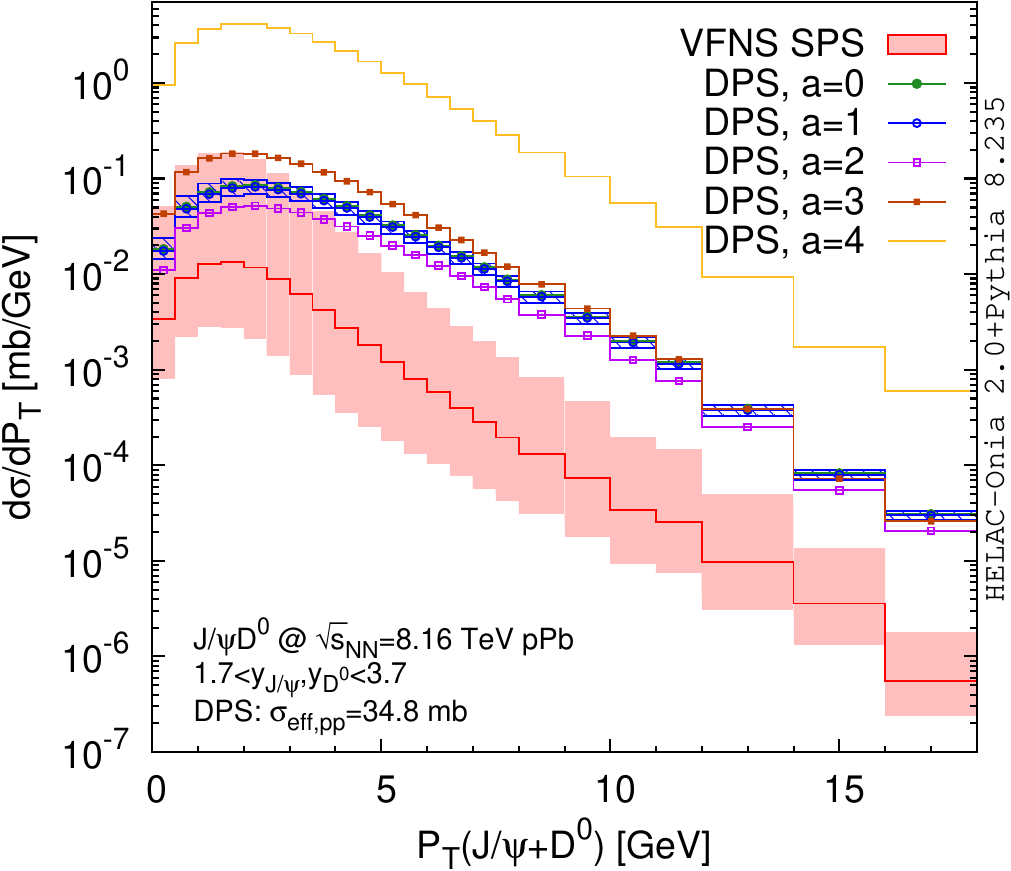}\label{fig:pPbD0dPTa}}
  \subfloat[Backward]{\includegraphics[width=0.45\textwidth,valign=c,draft=false]{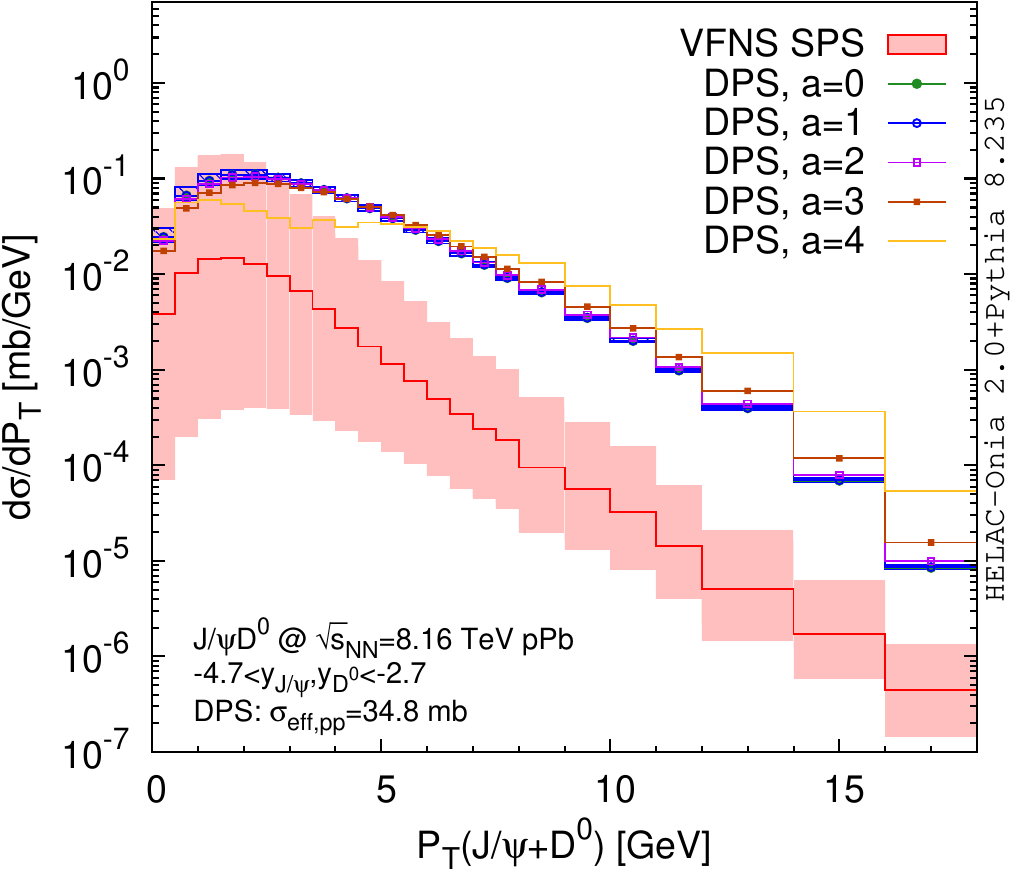}\label{fig:pPbD0dPTb}}
  \caption{Same as Fig.~\ref{fig:pPbD0dM} but for the transverse momentum $P_T$ of the $J/\psi+D^0$ system.
  \label{fig:pPbD0dPT}}
\end{figure}

The rapidity gap $\Delta y$ between the two detected mesons is widely used to differentiate the DPS and SPS contributions in the double $J/\psi$ production process. Here, we examine the same quantity for the $J/\psi+D^0$ associated production process. The corresponding distributions at the two rapidities can be found in Fig.~\ref{fig:pPbD0dy}. In order to fit in the frame, we have multiplied a factor $0.05$ to the DPS $a=4$ curve in Fig.~\ref{fig:pPbD0dya}. Because the two tagged mesons are not identical, these distributions are not necessary to be symmetric around $\Delta y(J/\psi,D^0)=0$. This has indeed been observed in the VFNS SPS distributions. In both Figs.~\ref{fig:pPbD0dya} and \ref{fig:pPbD0dyb}, our simulation shows that $J/\psi$ tends to be more central in rapidity than the $D^0$ meson. Due to the zero correlation of the two mesons in the DPS$_1$ events, the DPS distributions are (very close to be) symmetric around $\Delta y(J/\psi,D^0)=0$. A small asymmetric feature in Fig.~\ref{fig:pPbD0dyb} simply reflects some delicate details of our DPS$_1$ template, which is driven by the single-inclusive particle production data in $pp$ collisions. This asymmetry is amplified further in the DPS $a=4$ curve. The aforementioned cancellation results in a very narrow $\Delta y(J/\psi,D^0)$ distribution when $a=4$ and the two mesons are backward.

\begin{figure}
  \centering
  \subfloat[Forward]{
    \includegraphics[width=0.45\textwidth,valign=c,draft=false]{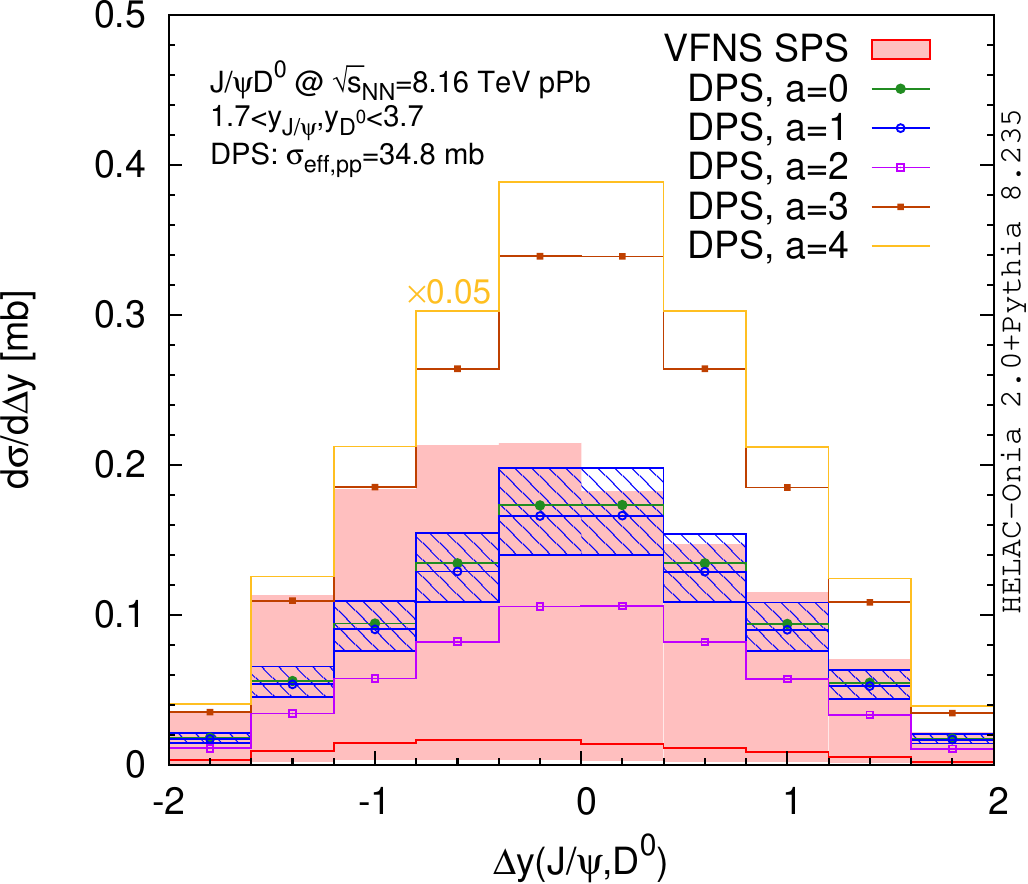}\label{fig:pPbD0dya}}
  \subfloat[Backward]{\includegraphics[width=0.45\textwidth,valign=c,draft=false]{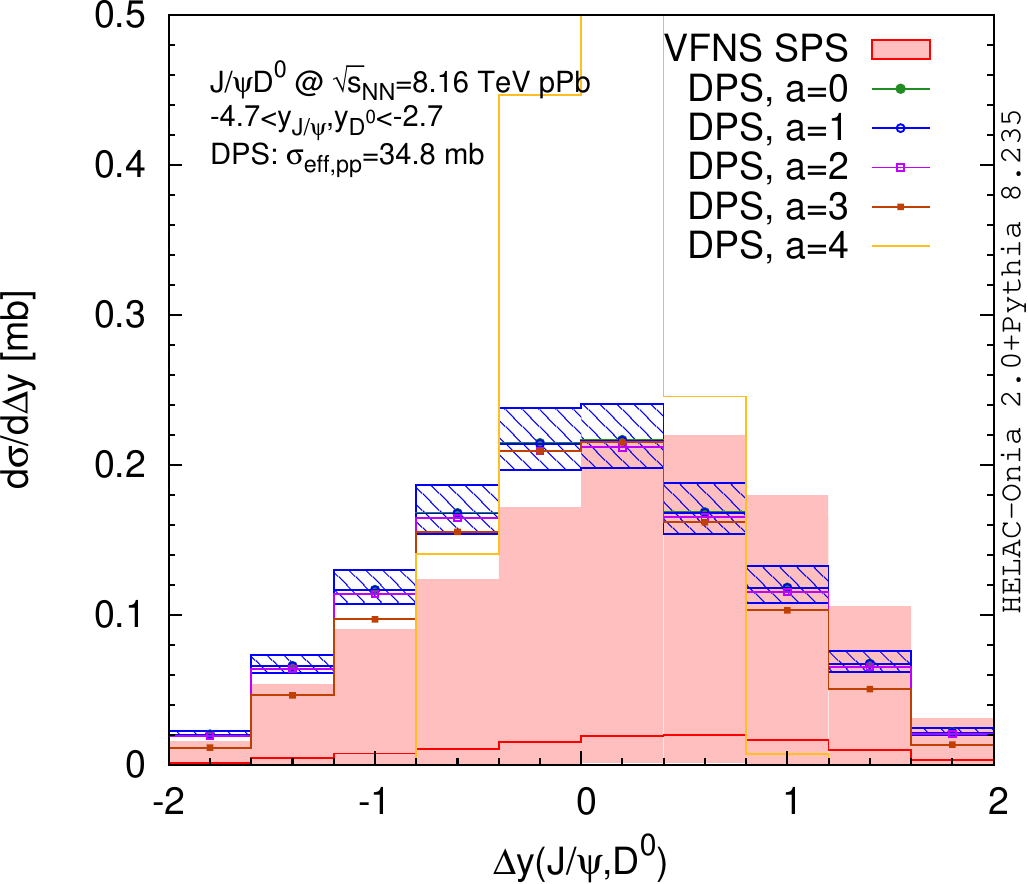}\label{fig:pPbD0dyb}}
  \caption{Same as Fig.~\ref{fig:pPbD0dM} but for the rapidity gap $\Delta y$ between the $J/\psi$ and  the $D^0$ mesons. A factor $0.05$ has been multiplied to the $a=4$ DPS curve in the forward case (left) in order to make it visible in the frame.
  \label{fig:pPbD0dy}}
\end{figure}

The last distributions, which we are interested in, are the azimuthal angle difference $\Delta \phi$ between $J/\psi$ and $D^0$. They are exhibited in Fig.~\ref{fig:pPbD0dphi}. The $\Delta \phi(J/\psi,D^0)$ observable is known to be very sensitive to the transverse momentum of the incoming partons from the intrinsic nonperturbative source and the perturbative initial shower. These effects have been taken into account in the VFNS SPS simulation via \Pythia8 and in the template of DPS$_1$, while they are absent in the LO calculation of DPS$_2$. The peak structures at $\Delta \phi(J/\psi,D^0)=\pi$ in Fig.~\ref{fig:pPbD0dphi} are from the DPS$_2$ events. We have examined that such peaks quickly fade out if we allow nonzero $k_T$ of the initial partons. After taking into account such an initial-$k_T$ smearing effect, the final DPS events are uniformly distributed in all bins, while the VFNS SPS events tend to populate at the back-to-back region $\Delta \phi(J/\psi,D^0)\simeq\pi$. However, the distinction between the two is still not remarkable enough given the current theory precision.

\begin{figure}
  \centering
  \subfloat[Forward]{
    \includegraphics[width=0.45\textwidth,valign=c,draft=false]{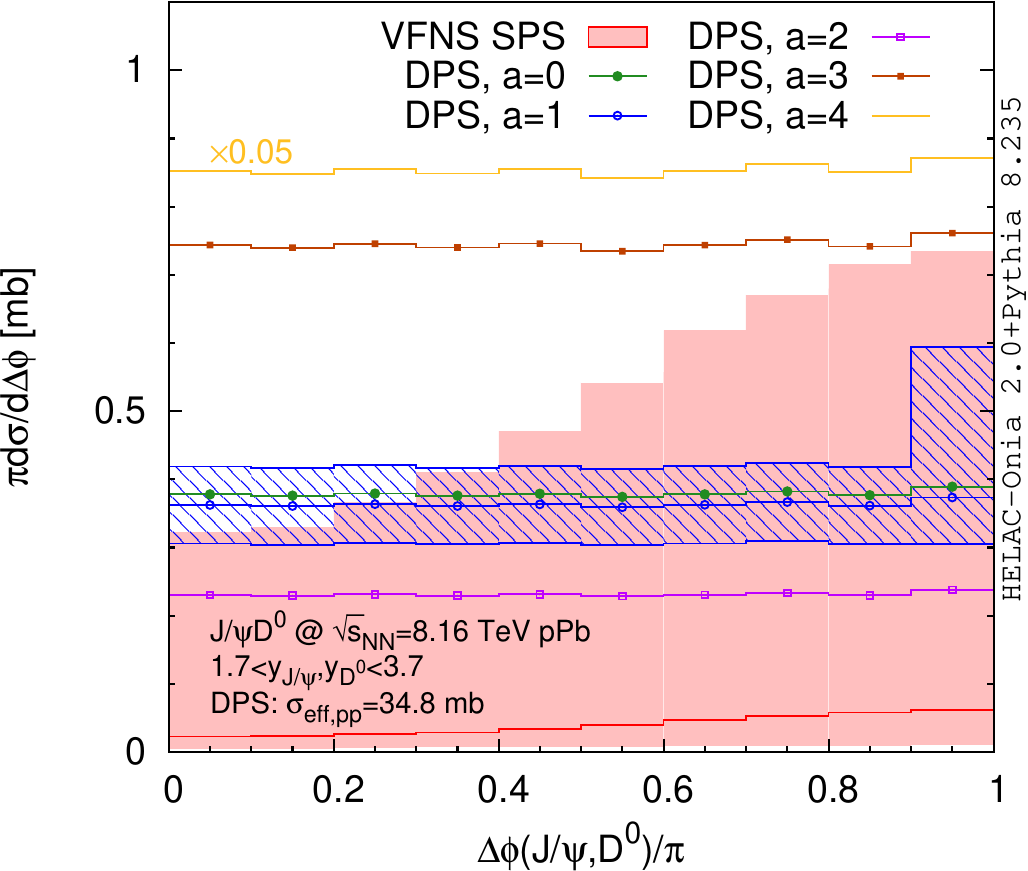}\label{fig:pPbD0dphia}}
  \subfloat[Backward]{\includegraphics[width=0.45\textwidth,valign=c,draft=false]{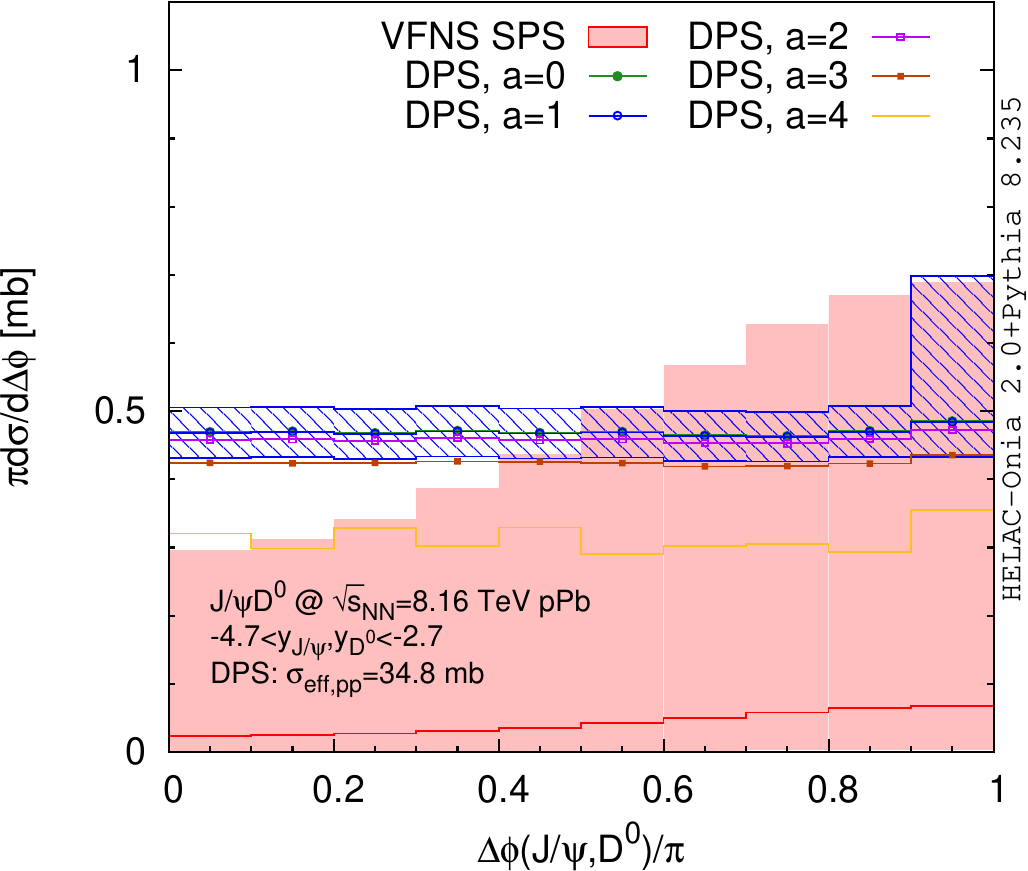}\label{fig:pPbD0dphib}}
  \caption{Same as Fig.~\ref{fig:pPbD0dy} but for the azimuthal angle difference between the $J/\psi$ and  the $D^0$ mesons.
  \label{fig:pPbD0dphi}}
\end{figure}

\subsection{Nuclear modification factors}

A traditionally useful observable in proton-nucleus collisions is the nuclear modification factor, which quantifies the relative modification of the cross section in $pA$ collisions with respect to its counterpart in $pp$ collisions. For our interested process, it is defined as
\begin{eqnarray}
R_{pA\to J/\psi+C}&=&\frac{d\sigma_{pA\to J/\psi+C}}{Ad\sigma_{pp\to J/\psi+C}}.
\end{eqnarray}
For the two DPS cross sections, according to Eq.(\ref{eq:pADPS}), we can write down the nuclear modification factor as
\begin{eqnarray}
R_{pA\to J/\psi+C}^{{\rm DPS}_i}&=&R_{J/\psi+C,11}^{{\rm DPS}_i}\left[\frac{3^{1-2a}(a+3)^{2a}}{2a+3}+\frac{\sigma_{{\rm eff},pp}}{\pi R_A^2}\left(A-1\right)\frac{9^{1-a}(a+3)^{2a}}{4(a+2)}\right]\nonumber\\
&&+\left(R_{J/\psi+C,10}^{{\rm DPS}_i}+R_{J/\psi+C,01}^{{\rm DPS}_i}\right)\left[1-\frac{3^{1-2a}(a+3)^{2a}}{2a+3}+\frac{\sigma_{{\rm eff},pp}}{\pi R_A^2}\left(A-1\right)\left(\frac{3^{2-a}(a+3)^a}{2(a+4)}-\frac{9^{1-a}(a+3)^{2a}}{4(a+2)}\right)\right]\nonumber\\
&&+\left[-1+\frac{3^{1-2a}(a+3)^{2a}}{2a+3}+\frac{\sigma_{{\rm eff},pp}}{\pi R_A^2}\left(A-1\right)\left(\frac{9}{8}+\frac{9^{1-a}(a+3)^{2a}}{4(a+2)}-\frac{3^{2-a}(a+3)^a}{(a+4)}\right)\right],\label{eq:RpADPS}
\end{eqnarray}
with the notations $R_{J/\psi+C,kl}^{{\rm DPS}_i}\equiv\frac{d\sigma_{J/\psi+C,kl}^{{\rm DPS}_i}}{\sigma_{J/\psi+C,00}^{{\rm DPS}_i}}$.~\footnote{Due to $J/\psi$ and $C$ mesons are produced independently in DPS$_1$, the further simplifications $R_{J/\psi+C,11}^{{\rm DPS}_1}=R_{pA\to J/\psi}R_{pA\to C},R_{J/\psi+C,10}^{{\rm DPS}_1}=R_{pA\to J/\psi}$ and $R_{J/\psi+C,01}^{{\rm DPS}_1}=R_{pA\to C}$ are possible.} In order to have the concrete impression on the relative importance of each piece in the above equation, we can take $a\in [0,1,2,3,4]$ explicitly. Then, $R_{pA\to J/\psi+C}^{{\rm DPS}_i}$ becomes
\begin{eqnarray}
R_{pA\to J/\psi+C}^{{\rm DPS}_i}&=&\left\{
\begin{array}{ll}
R_{J/\psi+C,11}^{{\rm DPS}_i}\left(1+\frac{9}{8}r_{{\rm eff},A}\right)
&\mbox{if $a=0$} \\
R_{J/\psi+C,11}^{{\rm DPS}_i}\left(\frac{16}{15}+\frac{4}{3}r_{{\rm eff},A}\right)-\left(R_{J/\psi+C,10}^{{\rm DPS}_i}+R_{J/\psi+C,01}^{{\rm DPS}_i}\right)\left(\frac{1}{15}+\frac{2}{15}r_{{\rm eff},A}\right)+\left(\frac{1}{15}+\frac{7}{120}r_{{\rm eff},A}\right) & \mbox{if $a=1$}\\
R_{J/\psi+C,11}^{{\rm DPS}_i}\left(\frac{625}{189}+\frac{625}{144}r_{{\rm eff},A}\right)-\left(R_{J/\psi+C,10}^{{\rm DPS}_i}+R_{J/\psi+C,01}^{{\rm DPS}_i}\right)\left(\frac{436}{189}+\frac{325}{144}r_{{\rm eff},A}\right)+\left(\frac{187}{144}+\frac{436}{189}r_{{\rm eff},A}\right) & \mbox{if $a=2$}\\
R_{J/\psi+C,11}^{{\rm DPS}_i}\left(\frac{64}{3}+\frac{144}{5}r_{{\rm eff},A}\right)-\left(R_{J/\psi+C,10}^{{\rm DPS}_i}+R_{J/\psi+C,01}^{{\rm DPS}_i}\right)\left(\frac{61}{3}+\frac{828}{35}r_{{\rm eff},A}\right)+\left(\frac{61}{3}+\frac{5499}{280}r_{{\rm eff},A}\right) & \mbox{if $a=3$}\\
239\left[R_{J/\psi+C,11}^{{\rm DPS}_i}\left(1+1.4r_{{\rm eff},A}\right)-\left(R_{J/\psi+C,10}^{{\rm DPS}_i}+R_{J/\psi+C,01}^{{\rm DPS}_i}\right)\left(1+1.3r_{{\rm eff},A}\right)+\left(1+1.2r_{{\rm eff},A}\right)\right] & \mbox{if $a=4$}\\
\end{array}
\right..\nonumber\\
\end{eqnarray}
We have used the abbreviation $r_{{\rm eff},A}\equiv(A-1)\frac{\sigma_{{\rm eff},pp}}{\pi R_A^2}$. For lead Pb, $r_{{\rm eff},A}\simeq 5.23\left(\frac{\sigma_{{\rm eff},pp}}{34.8~{\rm mb}}\right)$. It means that the terms linear in $r_{{\rm eff},A}$ are more important in the nuclear modification factor $R_{pA\to J/\psi+C}^{{\rm DPS}_i}$ unless $\sigma_{{\rm eff},pp}$ is smaller than $7$ mb. Thus, the DPS $R_{p{\rm Pb}}$ should be more sensitive to the value of $\sigma_{{\rm eff},pp}$ than the absolute cross sections in $p{\rm Pb}$ collisions. In the following context, we will also consider the uncertainty in $\sigma_{{\rm eff},pp}=34.8^{+1.2}_{-2.5}$ mb in evaluating the DPS $R_{p{\rm Pb}}$ values. We want to point out that the nuclear modification factor $R_{p{\rm Pb}}$ also provides us a new handle to simultaneously extract $\sigma_{{\rm eff},pp}$ and $G()$. A caveat is that since the fiducial volumes of the 7 TeV $pp$ and 8.16 TeV $p{\rm Pb}$ data are not completely identical, in principle, the identification of $\sigma_{{\rm eff},pp}$ in the two measurements may not be fully justified if a slight kinematic dependence of $\sigma_{{\rm eff},pp}$ is allowed for instance. Therefore, a dedicated $pp$ measurement under the same or at least similar condition would help to reduce such an uncertainty.

The $R_{p{\rm Pb}}$ values in the two LHCb fiducial regions are reported in Table~\ref{tab:RpPb}. We only consider $J/\psi+D^0$ in this section since other three final states share the very similar nuclear modification factors. $R_{p{\rm Pb}}$ predictions from VFNS SPS are shown in the second row, while those from DPS, as well as its breakdown into DPS$_1$ and DPS$_2$ parts, can be found in the last, third, and fourth rows respectively. The $R_{p{\rm Pb}}$ of VFNS SPS is smaller than unity in the forward rapidity interval $1.7<y_{J/\psi},y_{D^0}<3.7$. Such a fact is anticipated because we are probing the shadowing region of nPDF. Alternatively, the parton nuclear modification encoded in nPDF is very small when $-4.7<y_{J/\psi},y_{D^0}<-2.7$, which yields the SPS $R_{p{\rm Pb}}$ very close to unity. Several theoretical uncertainties are quoted. The first errors, including those in the parenthesis, are from the scale variation. The usual 7- and 9-point (in the parenthesis) scale variation~\footnote{The 7-point scale variation means the renormalization and factorization scales vary as $\mu_{R/F}=\xi_{R/F}\mu_0,\xi_{R/F}\in \{1.0,0.5,2.0\}$, but the two points $(\xi_R,\xi_F)=(2.0,0.5)$ or $(0.5,2.0)$ have been excluded.} errors are very large when $1.7<y_{J/\psi},y_{D^0}<3.7$. The big scale uncertainty stems from the low factorization scale $\mu_F=0.5\mu_0$ in $R_{k}^{{\rm Pb}}(x,\mu_F^2)$ with $k=g,c,\bar{c}$, which could be close to the charm quark mass threshold. In the later analysis of the differential $R_{p{\rm Pb}}$, we will explicitly show that such a scale dependence actually rapidly diminishes when the hard scale increases. In the backward region, because $R_{k}^{{\rm Pb}}(x,\mu_F^2)\sim 1$, the scale uncertainty largely cancels in the ratio and it becomes marginal. The second quoted errors are from the $R_{k}^{{\rm Pb}}(x)$ parameterization at the initial scale. They have been greatly constrained by the inclusive heavy-flavor data in Ref.~\cite{Kusina:2017gkz}. We have taken the envelope of the three reweighted EPPS16 grids, which have utilized the single heavy-flavor meson production data in three different factorization scale choices. In addition, we have assumed that there is no correlation of the hard scales between the SPS $J/\psi+D^0$ production and the single inclusive meson production processes. The final nPDF uncertainty is $18\%$ for $1.7<y_{J/\psi},y_{D^0}<3.7$ and $6\%$ for $-4.7<y_{J/\psi},y_{D^0}<-2.7$.

\begin{table}[H]
\centering\renewcommand{\arraystretch}{1.2}
\begin{tabular}{|c|c|c|c|} 
\hline
\multicolumn{2}{|c|}{Rapidity interval} & $1.7<y_{J/\psi},y_{D^0}<3.7$ & $-4.7<y_{J/\psi},y_{D^0}<-2.7$\\\hline
\multicolumn{2}{|c|}{VFNS SPS} & $0.66^{+0.94(1.83)+0.12}_{-0.00(0.00)-0.11}$ & $1.01^{+0.02(0.02)+0.06}_{-0.00(0.00)-0.06}$\\\hline
\multirow{3}{*}{DPS} & DPS$_1$ & $3.10^{+0.09}_{-0.19}\pm0.48$ & $6.65^{+0.20}_{-0.41}\pm0.52$ \\\cline{2-4}
 & DPS$_2$ & $3.41^{+0.10}_{-0.21}\pm 0.46$ & $5.90^{+0.17}_{-0.36}\pm 0.56$\\\cline{2-4}
 & Sum & $3.10^{+0.14+0.09}_{-0.01-0.19}\pm 0.48$ & $6.65^{+0.25+0.20}_{-0.02-0.41}\pm 0.52$\\\hline
\end{tabular}
\caption{\label{tab:RpPb} The forward and backward $R_{p{\rm Pb}}$ of $J/\psi+D^0$ in $p{\rm Pb}$ collisions at $\sqrt{s_{NN}}=8.16$ TeV. See the text for the description of the quoted theoretical errors and the parameter setup.}
\end{table}

We have only shown the DPS $R_{p{\rm Pb}}$'s with $a=1$ in Table~\ref{tab:RpPb}, while their $a$ dependencies in a wider range $a\in [0,4]$ can be found in Fig.~\ref{fig:RpPbD0}. Due to the geometrical enhancement effect from those terms proportional to $r_{{\rm eff},A}$, the DPS $R_{p{\rm Pb}}$'s are generically much larger than unity whatever the value of $a$, as long as $\sigma_{{\rm eff},pp}$ is not too small. With our nominal value of $\sigma_{{\rm eff},pp}$ from the $pp$ data, we have $R^{{\rm DPS}}_{p{\rm Pb}}\sim 3$ for the forward mesons and $\sim 6$ for the backward mesons when $a\leq 3$. A factor $2$ suppression between the two rapidity intervals is attributed to the double nuclear parton suppression $R_{k}^{{\rm Pb}}$ in $R^{{\rm DPS}}_{J/\psi+D^0,11}$. In the breakdown rows (DPS$_1$ and DPS$_2$) of Table~\ref{tab:RpPb}, only the $\sigma_{{\rm eff},pp}$ (first errors) and nPDF (second errors) uncertainties are considered. Other commonly present errors both in $p{\rm Pb}$ and in $pp$ can be taken as zero because of the cancellations in the ratio. As opposite, the cancellation is not guaranteed when we sum the two components of DPS concerning the scale variation in DPS$_2$. Therefore, the last row of  Table~\ref{tab:RpPb} shows three different errors. The first ones are because of the 9-point scale variation in DPS$_2$, and the second (third) ones are from the uncertainty in $\sigma_{{\rm eff},pp}$ (in nPDF). Unlike VFNS SPS, the dominant uncertainty in DPS is due to the nPDF parameterization. Such an uncertainty changes the central value by $15\%$ (forward) and $8\%$ (backward). When $a>3$, $R^{{\rm DPS}}_{p{\rm Pb}}$ increases dramatically in the forward region [see Fig.~\ref{fig:RpPbD0a}]. It, therefore, provides a smoking gun signal, if one is able to observe an anomalously large nuclear modification factor with an experiment. Finally, we comment on an additional feature in the layout of the figures in this subsection. There are two bands associated to each VFNS prediction: one dark-red band and one light-red band. The dark (light) bands are the combined theoretical uncertainty with the 9-point (7-point) scale variation. The differences between the two will be quite significant at low scales but will not be even visible if we go to higher scales.

\begin{figure}
  \centering
  \subfloat[Forward]{
    \includegraphics[width=0.45\textwidth,valign=c,draft=false]{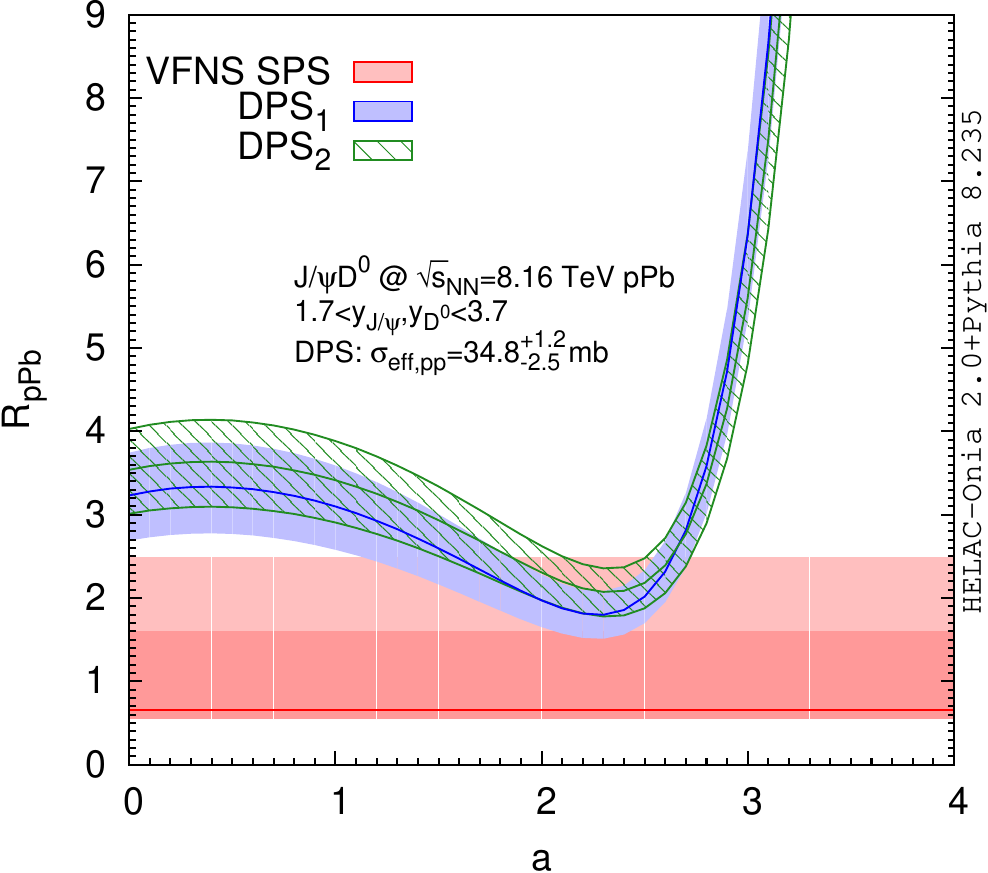}\label{fig:RpPbD0a}}
  \subfloat[Backward]{\includegraphics[width=0.45\textwidth,valign=c,draft=false]{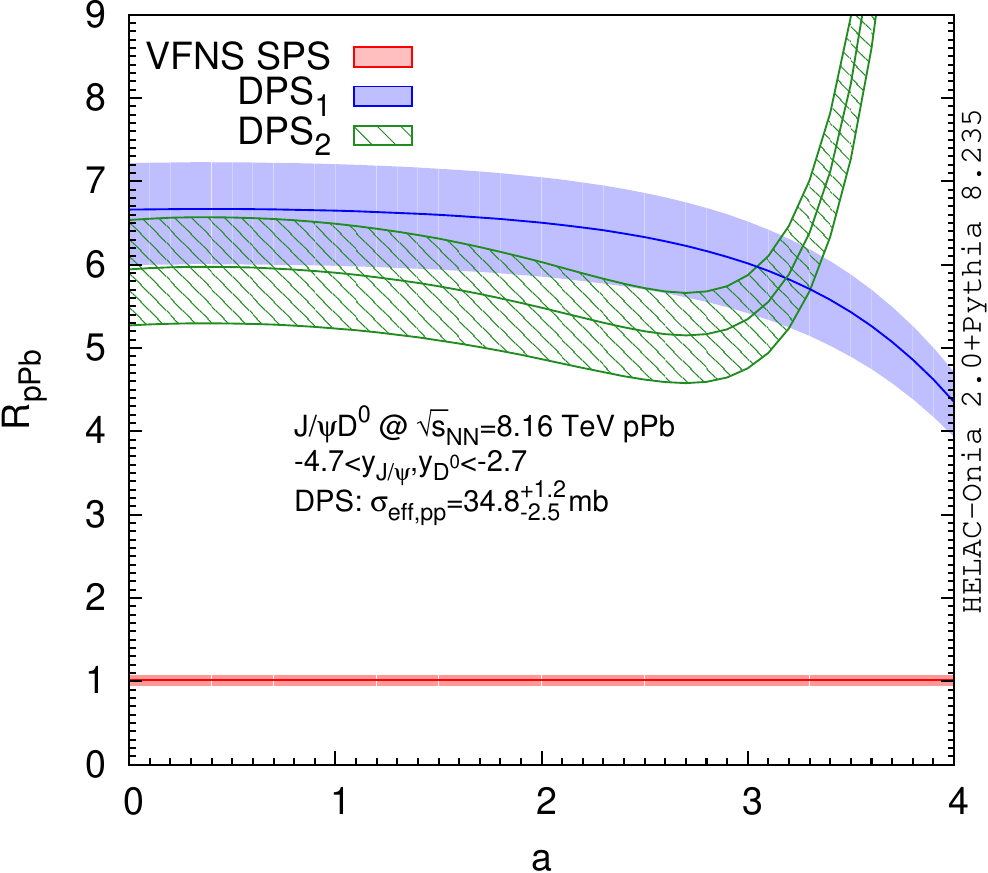}\label{fig:RpPbD0b}}
  \caption{Same as Fig.~\ref{fig:pPbD0XS} but for $R_{p{\rm Pb}}$ of $J/\psi+D^0$.
  \label{fig:RpPbD0}}
\end{figure}

We now turn to the discussions of the kinematical distributions of $R_{p{\rm Pb}}$. Following the absolute cross section discussions, we consider four different observables: the invariant mass $M(J/\psi+D^0)$ (Fig.~\ref{fig:RpPbD0dM}), the transverse momentum $P_T(J/\psi+D^0)$ (Fig.~\ref{fig:RpPbD0dPT}), the rapidity (Fig.~\ref{fig:RpPbD0dy}), and azimuthal angle (Fig.~\ref{fig:RpPbD0dphi}) differences between the two mesons. In order to improve the visibility, some additional factors, shown in the plots, have been multiplied to the DPS $a=4$ curves.

\begin{figure}
  \centering
  \subfloat[Forward]{
    \includegraphics[width=0.45\textwidth,valign=c,draft=false]{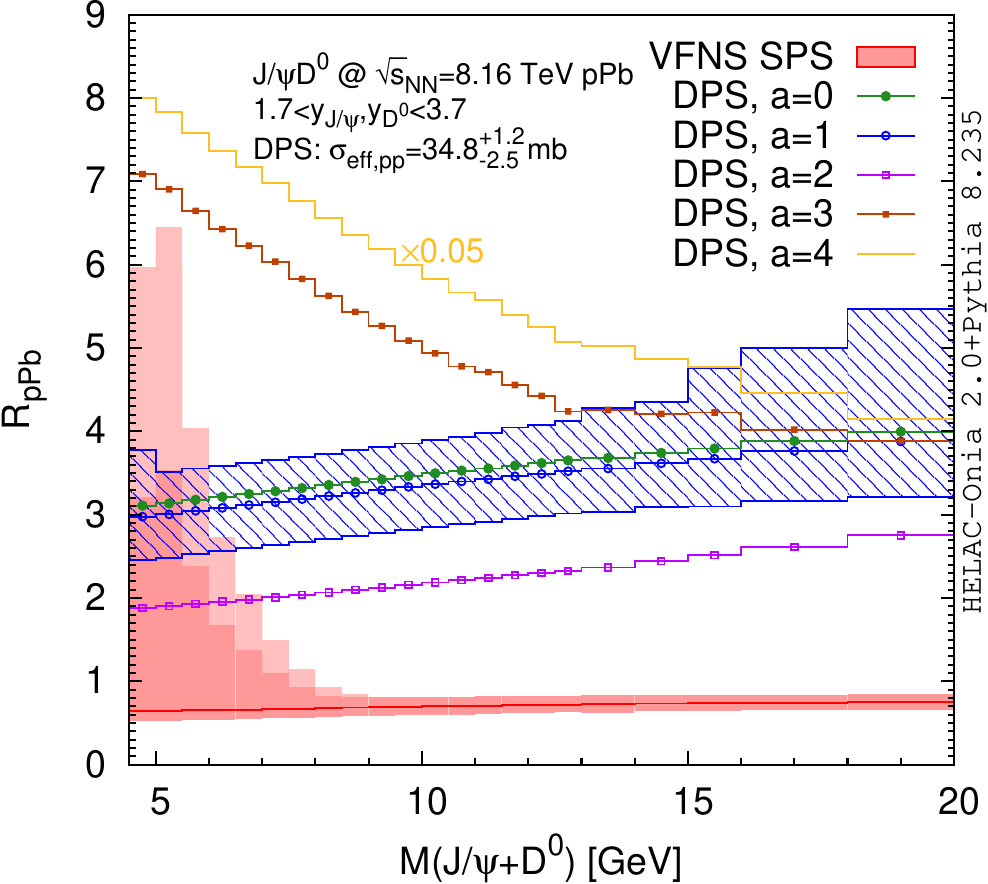}\label{fig:RpPbD0dMa}}
  \subfloat[Backward]{\includegraphics[width=0.45\textwidth,valign=c,draft=false]{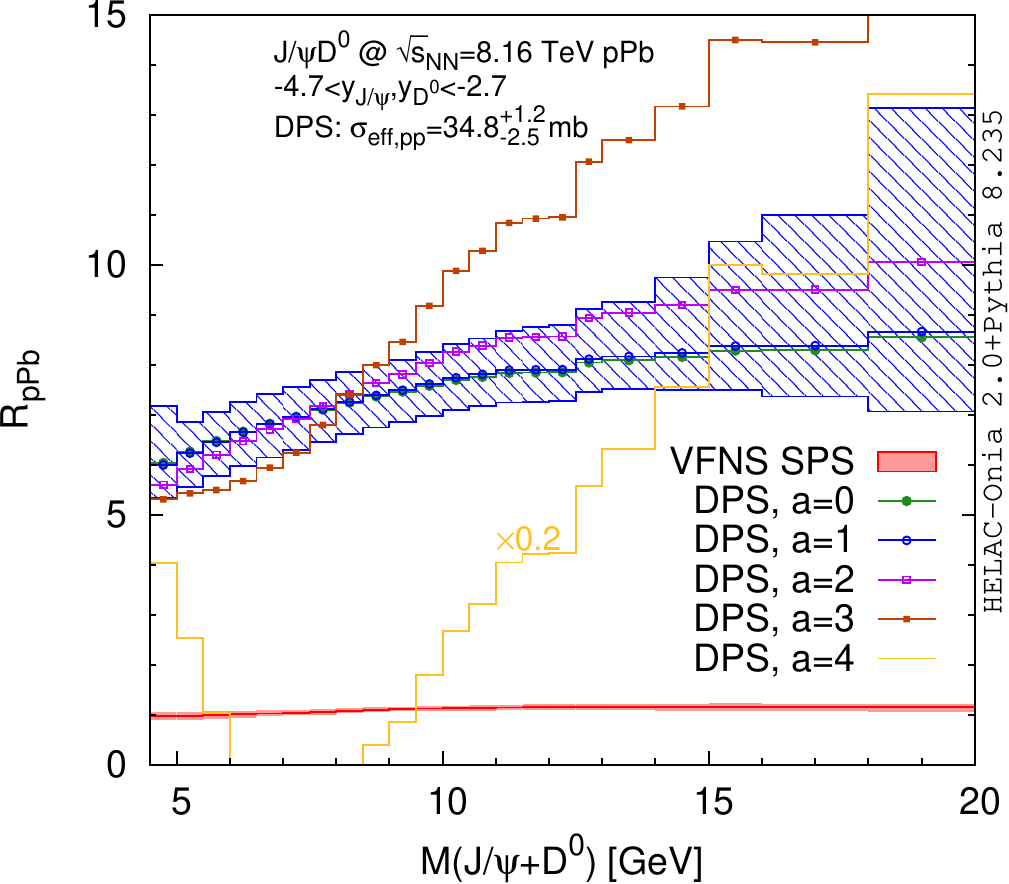}\label{fig:RpPbD0dMb}}
  \caption{The $R_{p{\rm Pb}}$ distributions versus the invariant mass $M$ of the $J/\psi+D^0$ system in proton-lead collisions at $\sqrt{s_{NN}}=8.16$ TeV. Five $a$ values are displayed in the DPS distributions. A factor $0.05$ ($0.2$) has been multiplied to the $a=4$ DPS curve in the left (right) panel in order to keep its visibility in the frame.
  \label{fig:RpPbD0dM}}
\end{figure}

\begin{figure}
  \centering
  \subfloat[Forward]{
    \includegraphics[width=0.45\textwidth,valign=c,draft=false]{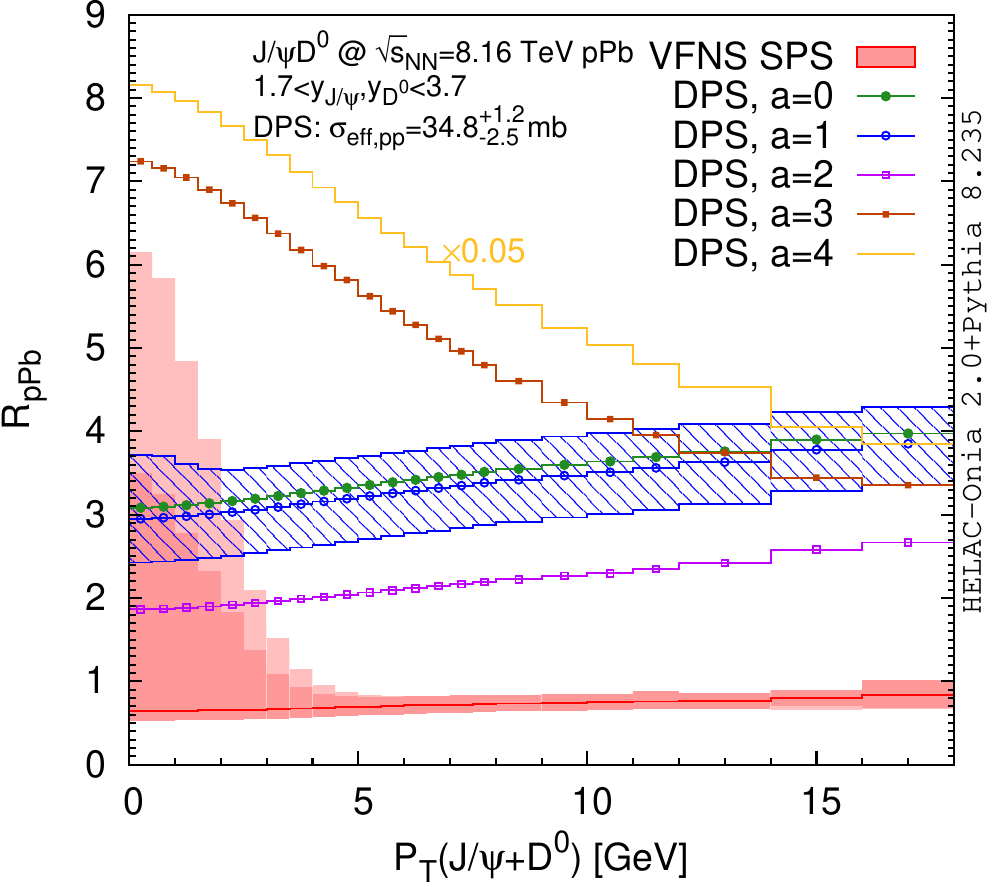}\label{fig:RpPbD0dPTa}}
  \subfloat[Backward]{\includegraphics[width=0.45\textwidth,valign=c,draft=false]{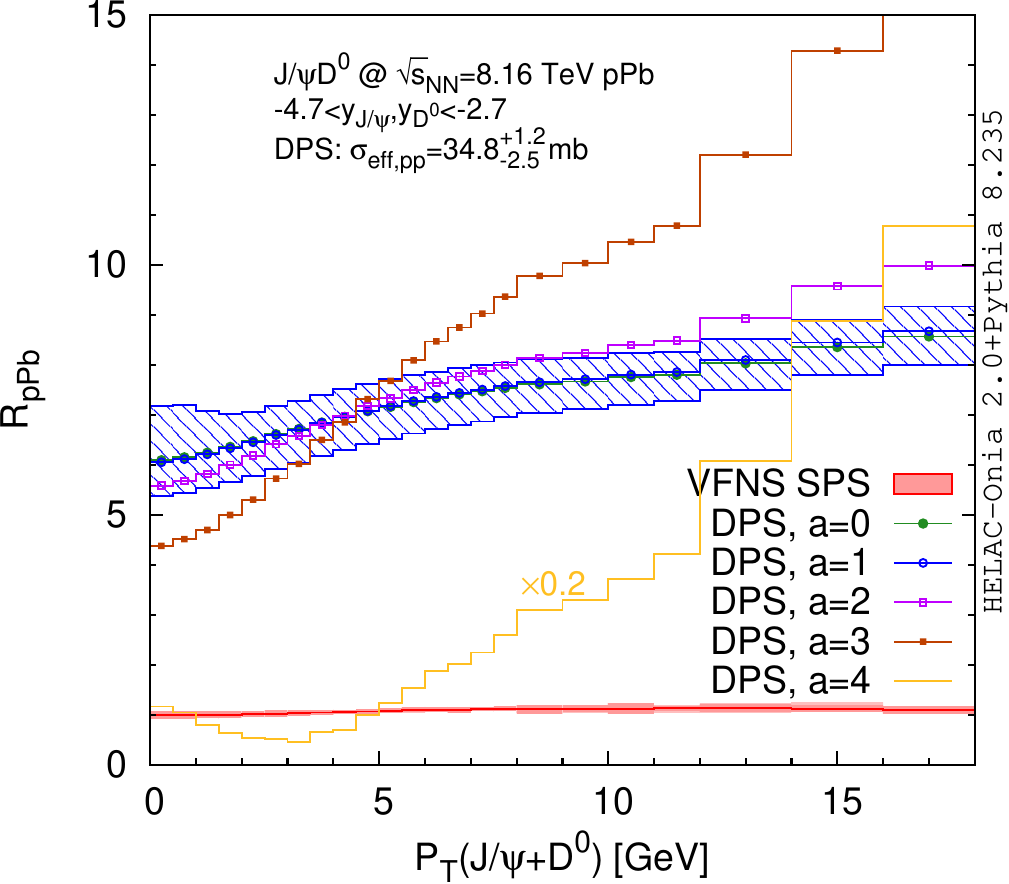}\label{fig:RpPbD0dPTb}}
  \caption{Same as Fig.~\ref{fig:RpPbD0dM} but for the transverse momentum $P_T$ of the $J/\psi+D^0$ system.
  \label{fig:RpPbD0dPT}}
\end{figure}

The $M(J/\psi+D^0)$ and $P_T(J/\psi+D^0)$ distributions share the similar features. In the left panels of Figs.~\ref{fig:RpPbD0dM} and \ref{fig:RpPbD0dPT}, the theoretical uncertainties in VFNS SPS are huge at the low scale regions, and they quickly reduce when the scale increases. The central $R_{p{\rm Pb}}^{{\rm SPS}}$ values trend to unity in the tails of the distributions. It simply reflects the fact that the nuclear modification effect diminishes in the tails. Such a diminishment also drives all the DPS curves in Figs.~\ref{fig:RpPbD0dMa} and \ref{fig:RpPbD0dPTa} to the pure geometrical enhancement $1+\frac{9}{8}r_{{\rm eff},A}$ asymptotically. On the other hand, in the backward region, the longitudinal momentum fraction $x$ of the initial parton in lead shifts from the shadowing-antishadowing transition point ($R_{k}^{{\rm Pb}}(x,\mu_F^2)\sim 1$) to the antishadowing regime ($R_{k}^{{\rm Pb}}(x,\mu_F^2)>1$) as $M(J/\psi+D^0)$ or $P_T(J/\psi+D^0)$ increases. Such an effect compensates by the scale evolution, where the latter reduces the size of the absolute modification $|R_{k}^{{\rm Pb}}(x,\mu_F^2)-1|$. The VFNS SPS curves, which are not populated by the geometrical effect, are slightly above unity in the tails of the two distributions. The same increasing trend can be found in the DPS curves in Figs.~\ref{fig:RpPbD0dMb} and \ref{fig:RpPbD0dPTb}. These kinematical regions are potentially interesting to explore the impact-parameter-dependent antishadowing regime of nPDF.

\begin{figure}
  \centering
  \subfloat[Forward]{
    \includegraphics[width=0.45\textwidth,valign=c,draft=false]{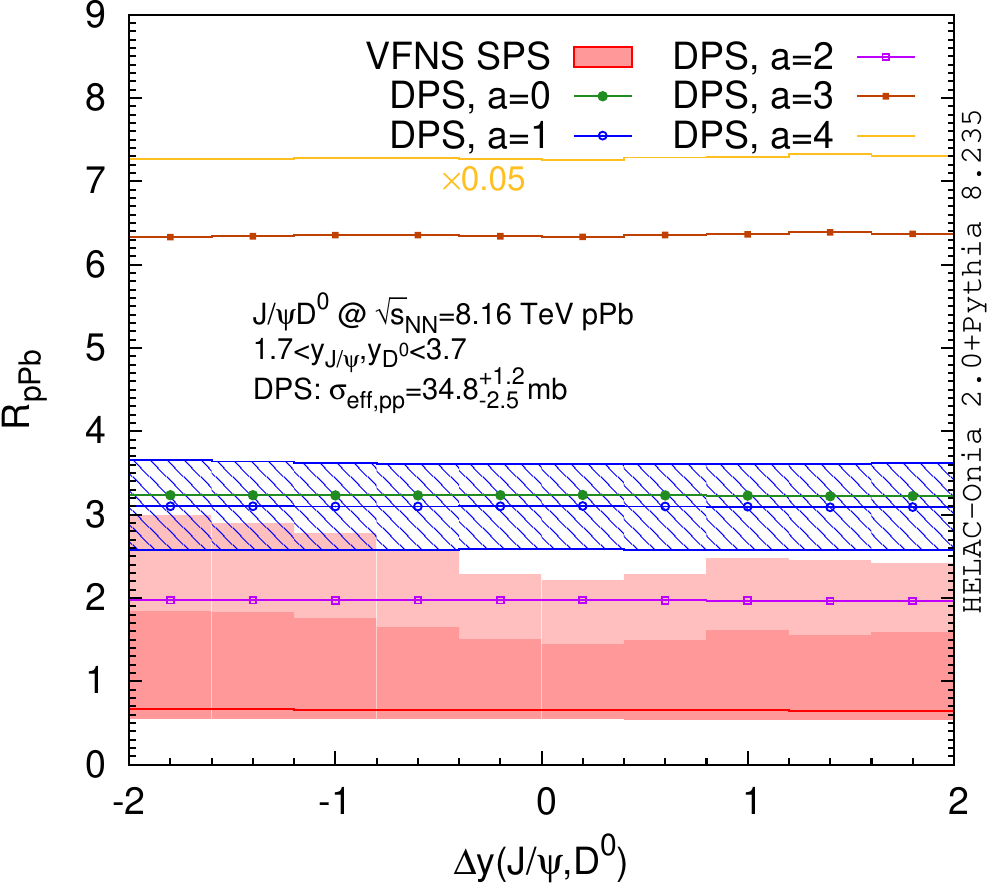}\label{fig:RpPbD0dya}}
  \subfloat[Backward]{\includegraphics[width=0.45\textwidth,valign=c,draft=false]{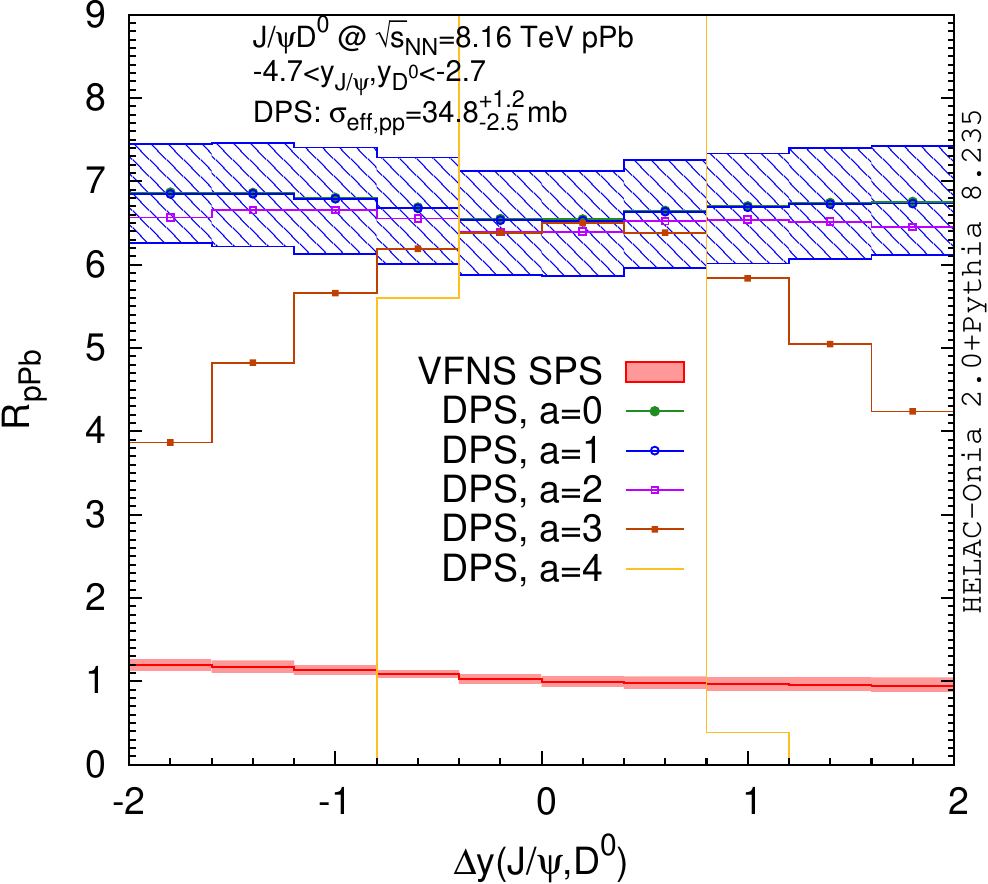}\label{fig:RpPbD0dyb}}
  \caption{Same as Fig.~\ref{fig:RpPbD0dM} but for the rapidity gap $\Delta y$ between $J/\psi$ and  $D^0$. No additional factor is multiplied to the $a=4$ DPS curve in the right panel.
  \label{fig:RpPbD0dy}}
\end{figure}

\begin{figure}
  \centering
  \subfloat[Forward]{
    \includegraphics[width=0.45\textwidth,valign=c,draft=false]{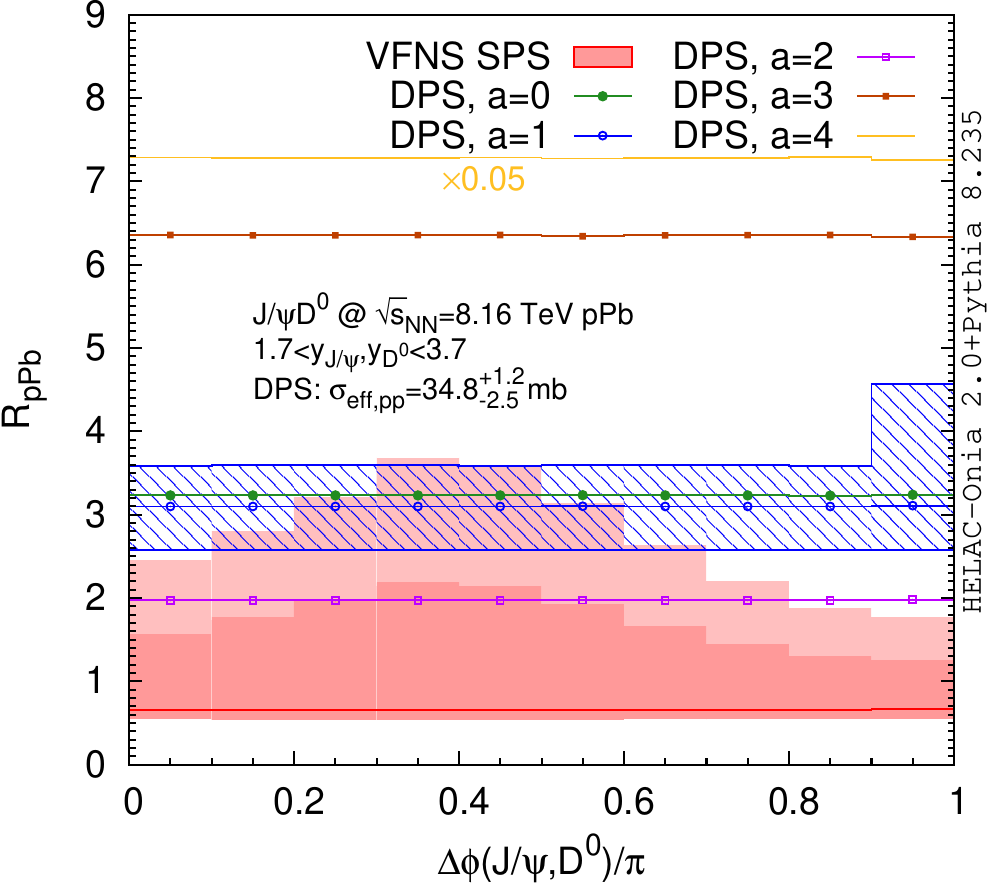}\label{fig:RpPbD0dphia}}
  \subfloat[Backward]{\includegraphics[width=0.45\textwidth,valign=c,draft=false]{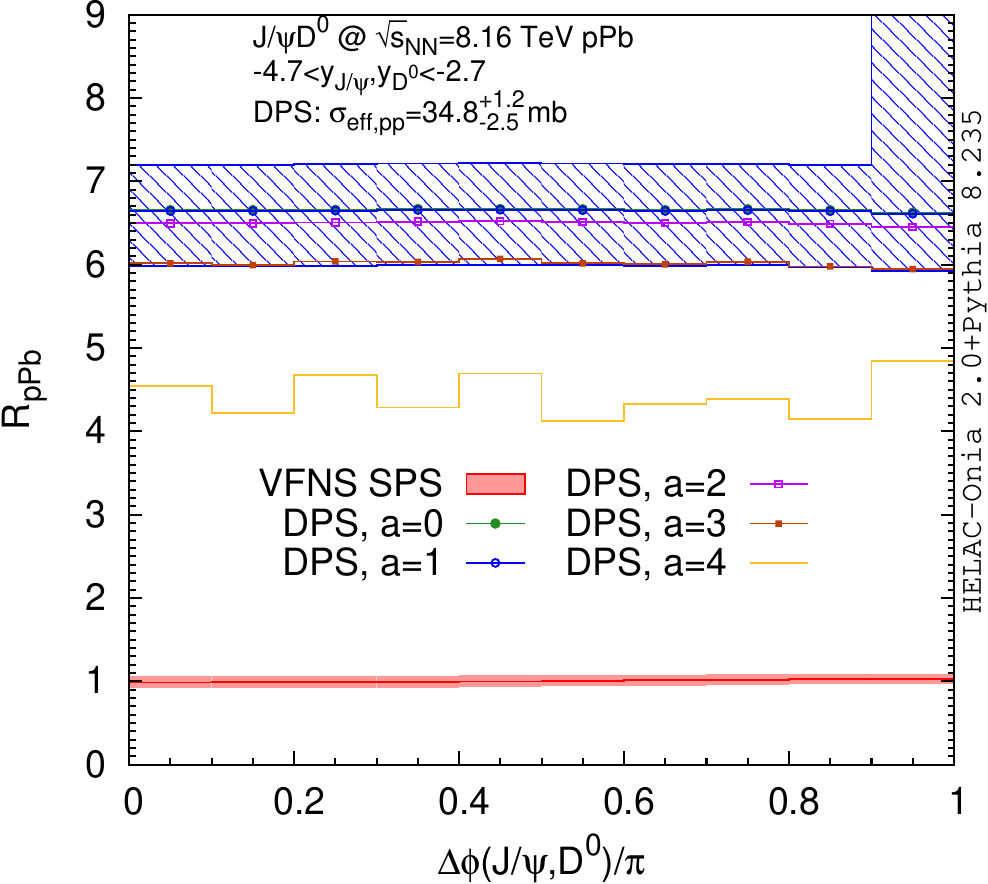}\label{fig:RpPbD0dphib}}
  \caption{Same as Fig.~\ref{fig:RpPbD0dy} but for the azimuthal angle difference between $J/\psi$ and  $D^0$.
  \label{fig:RpPbD0dphi}}
\end{figure}

The $R_{p{\rm Pb}}$ versus $\Delta y(J/\psi,D^0)$ and $\Delta \phi(J/\psi,D^0)$ predicted in Figs.~\ref{fig:RpPbD0dy} and \ref{fig:RpPbD0dphi} are rather simple. $R_{p{\rm Pb}}$ is in general flat for DPS and close to flat for SPS. From Fig.~\ref{fig:RpPbD0dyb}, if the $J/\psi$ and $D^0$ mesons are more asymmetrically produced in rapidity, the discrimination power for the different curves in $a$ of the DPS events becomes larger.

Since our predictions are plagued with the very large theoretical uncertainties, in particular in SPS, we do not show the combined $R_{p{\rm Pb}}$ after encompassing both SPS and DPS contributions. Before closing this section, we would like to comment on how to obtain $R_{p{\rm Pb}}$ based on our results. There are at least two ways. If the fraction $f^{p{\rm Pb}}_{{\rm DPS}}$ of DPS events in $p{\rm Pb}$ collisions can be better known (e.g., in some control regions), we can obtain the combined $R_{p{\rm Pb}}$ via
\begin{eqnarray}
R_{p{\rm Pb}}^{{\rm SPS+DPS}}&=&\frac{1}{\frac{f^{p{\rm Pb}}_{{\rm DPS}}}{R_{p{\rm Pb}}^{{\rm DPS}}}+\frac{1-f^{p{\rm Pb}}_{{\rm DPS}}}{R_{p{\rm Pb}}^{{\rm SPS}}}}.
\end{eqnarray}
On the other hand, if one is able to determine the DPS fraction $f^{pp}_{{\rm DPS}}$ in a corresponding $pp$ measurement, ideally under the same condition, we are able to derive $R_{p{\rm Pb}}$ via
\begin{eqnarray}
R_{p{\rm Pb}}^{{\rm SPS+DPS}}&=&f^{pp}_{{\rm DPS}}R_{p{\rm Pb}}^{{\rm DPS}}+\left(1-f^{pp}_{{\rm DPS}}\right)R_{p{\rm Pb}}^{{\rm SPS}}.
\end{eqnarray}

\section{Conclusions}

In this paper, we have carried out a thorough study on the process of a $J/\psi$ meson production in association with an open (anti-)charm hadron at the LHC. As our first step, we tried to understand the existing LHCb measurement in $pp$ collisions at $\sqrt{s}=7$ TeV, where there were apparent tensions between the DPS dominance picture and the experimental data. We proposed a new DPS mechanism and revised the SPS Monte Carlo simulation based on \ho\ 2.0+\Pythia8 in order to resolve the discrepancies. The novel DPS mechanism DPS$_2$ has peculiar features with respect to the conventional DPS procedure DPS$_1$. The former is strongly correlated among the particles in the final state, while the latter is expected to correlate only in a weak way between the two distinct scatterings. In our specific process, DPS$_2$, however, turns out to be small. Nevertheless, such a new DPS$_2$ process is ubiquitous for processes involving more-than-one pair of same-flavor heavy quarks with at least a quarkonium, like $J/\psi+c\bar{c}$ and $J/\psi+J/\psi$, etc. The DPS$_2$ contributions in the processes other than $J/\psi+c\bar{c}$ deserve being investigated in the future. On the other hand, in order to scrutinize the LHCb data, we have performed a proper matching between 3FS and 4FS calculations in the SPS simulation. Our calculations demonstrate that the 4FS part, which resums the large logarithms of initial gluon-to-charm splitting, is indispensable to account for the robust SPS predictions and also the LHCb data. The matched VFNS  calculation significantly enhances the SPS cross sections and alters the DPS-dominance conclusion, which was based on the 3FS-alone simulation. To the best our knowledge, it is also a first VFNS calculation for a quarkonium process in the literature. After encompassing all components, we (re)determine the effective cross section $\sigma_{{\rm eff},pp}$ entering into the DPS formula from the LHCb data with a likelihood-based approach. Without big surprise, the new determinations yield $\sigma_{{\rm eff},pp}\simeq 30$ mb with six different proton PDFs, which are a factor $2$ larger than the previous fit based on the DPS-alone hypothesis. These improvements allow us to fill the gaps between theory and experiment, albeit with large theoretical uncertainties. This clearly calls for an improved simulation in the future that contains higher-order quantum corrections in order to ensure whether the discrepancies have really been gone.

Furthermore, we have also presented our predictions for the same process but in $p{\rm Pb}$ minimum-bias collisions at $\sqrt{s_{NN}}=8.16$ TeV. The corresponding experimental measurement is carrying on by the LHCb Collaboration. We identify a few kinematical regions that would be very useful in improving the purity of the DPS events. Namely, they are the large $M(J/\psi+D^0)$ and high $P_T(J/\psi+D^0)$ regions. We also exploited the potential to extract the spatial-dependent nPDF information from this process. The nuclear modification factors $R_{p{\rm Pb}}$ are very different between DPS and SPS, and are also significantly depend on the (unknown) impact-parameter dependence of nPDF. Therefore, with enough statistics, the LHC can definitely tell us something on how initial partons are impacted when they are located at different positions in a nucleus. We look forward to see the emergence of more measurements at the LHC in the near future.

\vfill




\acknowledgments

I would like to thank Yanxi Zhang for useful discussions. I am also grateful to Hannu Paukkunen and Ilkka Helenius for pointing out that the $\sigma_{{\rm eff},pp}$ values determined here would further improve the agreement in the double-D case in Ref.~\cite{Helenius:2019uge}.
The work is supported by
the ILP Labex (ANR-11-IDEX-0004-02, ANR-10-LABX-63).




\clearpage

\appendix

\section{Likelihood approach to determine $\sigma_{{\rm eff},pp}$\label{app:likelihood}}

We employ the likelihood-based approach to estimate $\sigma_{{\rm eff},pp}$ from the LHCb $7$ TeV $pp$ data. The marginalized joint Bayesian posterior follows the similar form in the CMS analysis of $\alpha_s$ and top-quark mass determinations from the top-quark pair cross section~\cite{Chatrchyan:2013haa}. For a single cross section $\sigma_{J/\psi+C}$ ($C: D^0/\bar{D}^0, D^\pm,D_s^\pm,\Lambda_c^\pm$), the construction of the Bayesian prior is a convolution of a probability function for the theoretically predicted cross section $\rho_{\rm th}\left(\sigma_{J/\psi+C}\left|\sigma_{{\rm eff},pp}\right.\right)$ and another probability function for the experimental measured cross section $\rho_{\rm exp}\left(\sigma_{J/\psi+C}\left|\sigma_{{\rm eff},pp}\right.\right)$. The first one accounts for all the theoretical uncertainties, while the second one takes into account the experimental uncertainties. In the theory part $\rho_{\rm th}\left(\sigma_{J/\psi+C}\left|\sigma_{{\rm eff},pp}\right.\right)$, we use a standard Gaussian distribution of width $\delta_{\rm PDF}$ to describe the PDF parameterization uncertainty. On the other hand, because there is no particular probability distribution is known which is adequate for the confidence interval obtained from the renormalization/factorization scale variation, the corresponding scale uncertainty on $\sigma_{J/\psi+C}$ prediction is using a flat prior. Same argument applies to the LDME uncertainty in the rescaling factor representing of color-octet and excited state feed-down contributions. In order to be more conservative, we combine the scale uncertainty and the LDME uncertainty linearly. The resulting probability function for the theoretical prediction is given by
\begin{eqnarray}
&&\rho_{\rm th}\left(\sigma_{J/\psi+C}\left|\sigma_{{\rm eff},pp}\right.\right)=\nonumber\\
&&\frac{1}{2\left(\sigma_{J/\psi+C}^{({\rm pred.~max})}-\sigma_{J/\psi+C}^{({\rm pred.~min})}\right)}\left[{\rm erf}{\left(\frac{\sigma_{J/\psi+C}^{({\rm pred.~max})}-\sigma_{J/\psi+C}}{\sqrt{2}\delta_{\rm PDF}}\right)}-{\rm erf}{\left(\frac{\sigma_{J/\psi+C}^{({\rm pred.~min})}-\sigma_{J/\psi+C}}{\sqrt{2}\delta_{\rm PDF}}\right)}\right],
\end{eqnarray}
where $\sigma_{J/\psi+C}^{({\rm pred.~max/min})}$ denotes the maximal/minimal predicted cross section considering the scale uncertainty and the LDME uncertainty. The $\rho_{\rm exp}\left(\sigma_{J/\psi+C}\left|\sigma_{{\rm eff},pp}\right.\right)$ is independent of the value of $\sigma_{{\rm eff},pp}$ and can be approximated as a Gaussian function
\begin{eqnarray}
\rho_{\rm exp}\left(\sigma_{J/\psi+C}\left|\sigma_{{\rm eff},pp}\right.\right)&=&\frac{1}{\sqrt{2\pi}\delta_{\rm exp}}\exp{\left[-\frac{\left(\sigma_{J/\psi+C}-\sigma_{J/\psi+C}^{({\rm exp})}\right)^2}{2\delta_{\rm exp}^2}\right]},
\end{eqnarray}
with $\sigma_{J/\psi+C}^{({\rm exp})}$ and $\delta_{\rm exp}$ being the LHCb measured cross section and its $68\%$ CL error. The Bayesian confidence interval of $\sigma_{{\rm eff},pp}$ is computed through marginalization of the joint posterior by integration over $\sigma_{J/\psi+C}$,
\begin{eqnarray}
L\left(\sigma_{{\rm eff},pp}\right)=\int{\rho_{\rm th}\left(\sigma_{J/\psi+C}\left|\sigma_{{\rm eff},pp}\right.\right)\rho_{\rm exp}\left(\sigma_{J/\psi+C}\left|\sigma_{{\rm eff},pp}\right.\right)d\sigma_{J/\psi+C}}.
\end{eqnarray}
In particular, the central value of $\sigma_{{\rm eff},pp}$ is determined by maximizing $L\left(\sigma_{{\rm eff},pp}\right)$, which we denote as $\sigma_{{\rm eff},pp}=\delta_0$, and its $68\%$ CL error is determined via
\begin{eqnarray}
\int_{\delta_0-\delta_-}^{\delta_0+\delta_+}{L\left(\sigma_{{\rm eff},pp}\right)d\sigma_{{\rm eff},pp}}&=&0.6827
\end{eqnarray}
and
\begin{eqnarray}
L\left(\delta_0+\delta_+\right)&=&L\left(\delta_0-\delta_-\right).
\end{eqnarray}
The final $68\%$ confidence interval is $\sigma_{{\rm eff},pp}=\delta_0~^{+\delta_+}_{-\delta_-}$. In general, $\delta_+$ and $\delta_-$ should be not identical, and $\sigma_{{\rm eff},pp}$ poses an asymmetric error.

In the presence of several (differential) cross section measurements, we have to opt for an approach to combine the individual determinations of $\sigma_{{\rm eff},pp}$. We adopt a likelihood-based approach as advocated in Refs.~\cite{Cowan:2010js,Klijnsma:2017eqp}, in which a global likelihood function is constructed from the probability distribution functions of individual extractions. For simplicity, we do not breakdown the errors of individual determinations into different sources and take all of them uncorrelated. Such a simple and conservative treatment allows us to avoid introducing additional nuisance parameters. Let us assume that we want to combine $n$ different determinations with the $i$th one as $\delta_{0}^{(i)}~^{+\delta^{(i)}_+}_{-\delta^{(i)}_-}$. The probability distribution function of the $i$th determination is given by
\begin{eqnarray}
\xi_{i}\left(\sigma_{{\rm eff},pp}\right)&=&\frac{1}{\sqrt{2\pi}\delta_i}\exp{\left[-\frac{\left(\sigma_{{\rm eff},pp}-\delta_0^{(i)}\right)^2}{2\delta_i^2}\right]},
\end{eqnarray}
where
\begin{eqnarray}
\delta_i&=&\delta_{-}^{(i)}\theta\left(\delta_{0}^{(i)}-\sigma_{{\rm eff},pp}\right)+\delta_{+}^{(i)}\theta\left(\sigma_{{\rm eff},pp}-\delta_{0}^{(i)}\right).
\end{eqnarray}
The global likelihood function is then
\begin{eqnarray}
L_{n}\left(\sigma_{{\rm eff},pp}\right)&=&\prod_{i=1}^{n}{\xi_{i}\left(\sigma_{{\rm eff},pp}\right)},\label{eq:globalLn}
\end{eqnarray}
and the test statistic is
\begin{eqnarray}
q_n\left(\sigma_{{\rm eff},pp}\right)&=&-2\log{\frac{L_{n}\left(\sigma_{{\rm eff},pp}\right)}{L_{n}\left(\bar{\sigma}_{{\rm eff},pp}\right)}},
\end{eqnarray}
where $L_{n}\left(\sigma_{{\rm eff},pp}\right)$ is maximized when $\sigma_{{\rm eff},pp}=\bar{\sigma}_{{\rm eff},pp}$. The test statistic $q_n$ is always positive or zero. The 1$\sigma$ (i.e., $68\%$) confidence interval is extracted from $q_n=1$.

\section{More on differential theory-data comparison\label{app:diffcomp}}

In this Appendix, we collect the additional theory-data comparison plots for the differential distributions of $J/\psi+D^+$ (Fig.~\ref{fig:ppDp}), $J/\psi+D_s^+$ (Fig.~\ref{fig:ppDs}) and $J/\psi+\Lambda_c^+$ (Fig.~\ref{fig:ppLc}). All of LHCb measurements are in reasonable agreement with our Monte Carlo simulations, though persisting large theoretical errors.

\begin{figure}
  \centering
  \subfloat[Transverse momentum of $J/\psi$]{
    \includegraphics[width=0.45\textwidth,valign=c,draft=false]{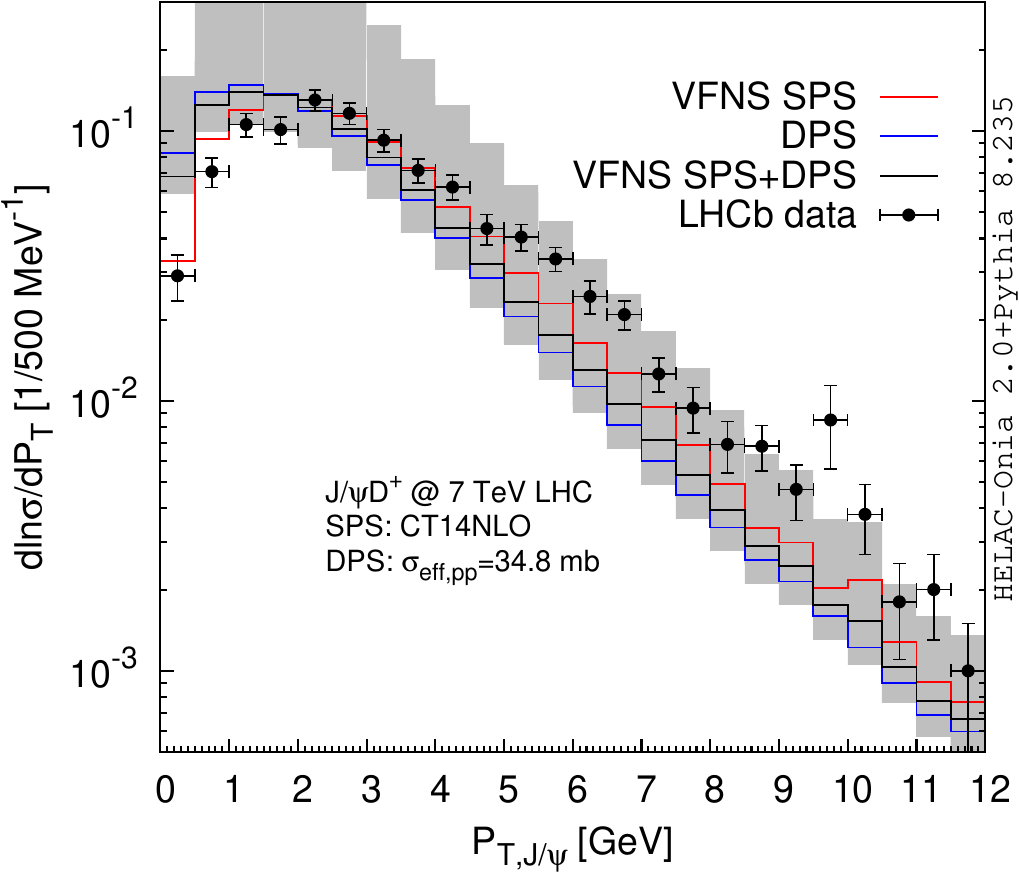}\label{fig:ppDpa}}
  \subfloat[Invariant mass of $J/\psi$ and $D^+$]{\includegraphics[width=0.45\textwidth,valign=c,draft=false]{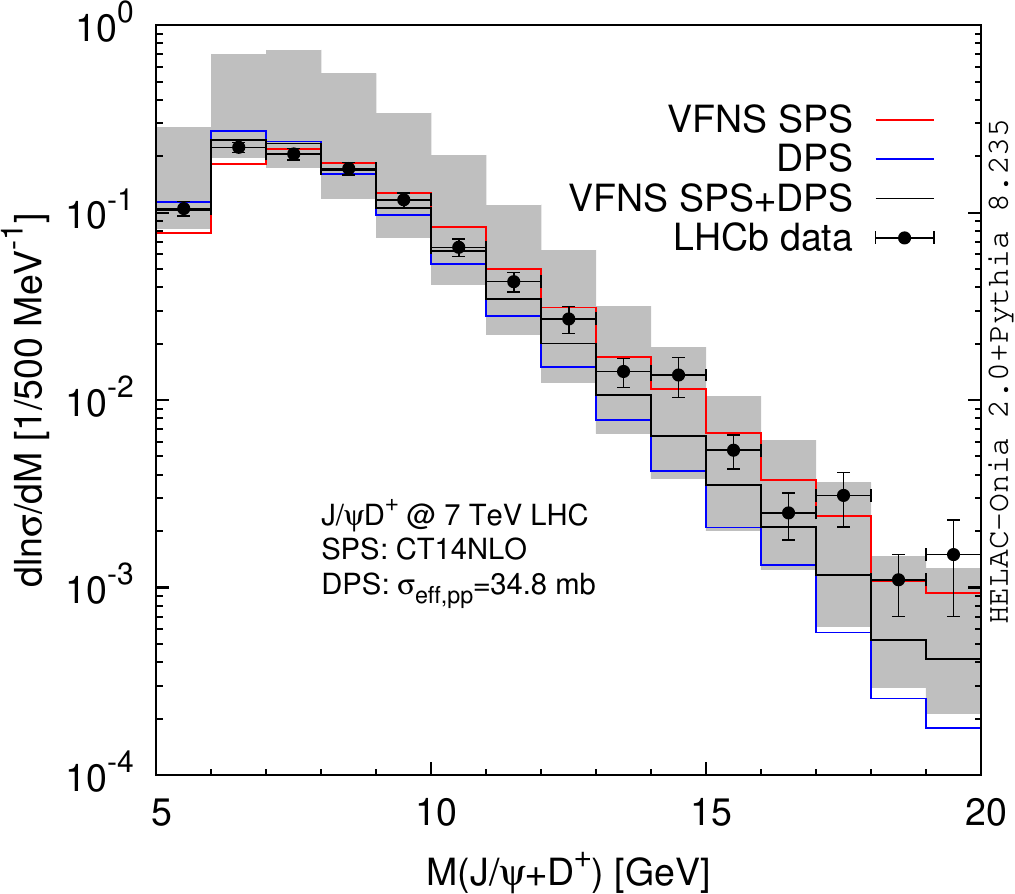}\label{fig:ppDpb}}\\
  \subfloat[Invariant mass of $J/\psi$ and $D^+$]{\includegraphics[width=0.45\textwidth,valign=c,draft=false]{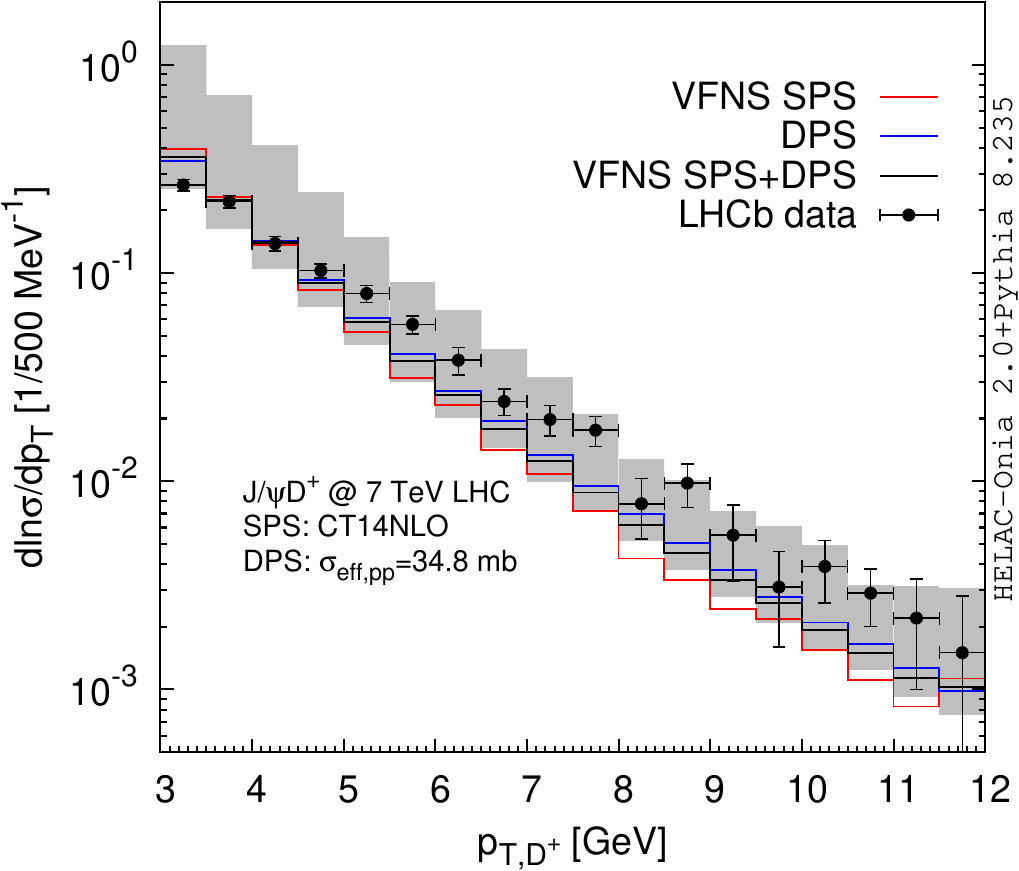}\label{fig:ppDpc}}
  \subfloat[Rapidity gap between $J/\psi$ and $D^+$]{\includegraphics[width=0.45\textwidth,valign=c,draft=false]{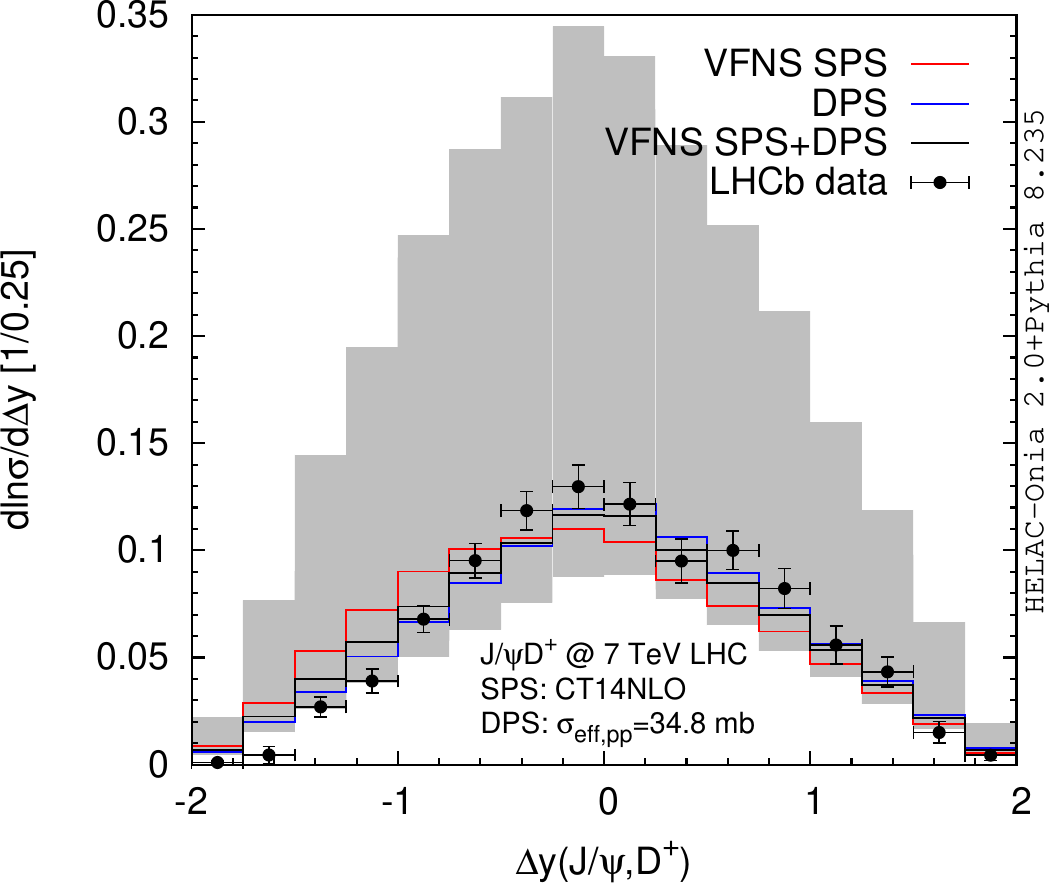}\label{fig:ppDpd}}\\
   \subfloat[Azimuthal angle difference between $J/\psi$ and $D^+$]{\includegraphics[width=0.45\textwidth,valign=c,draft=false]{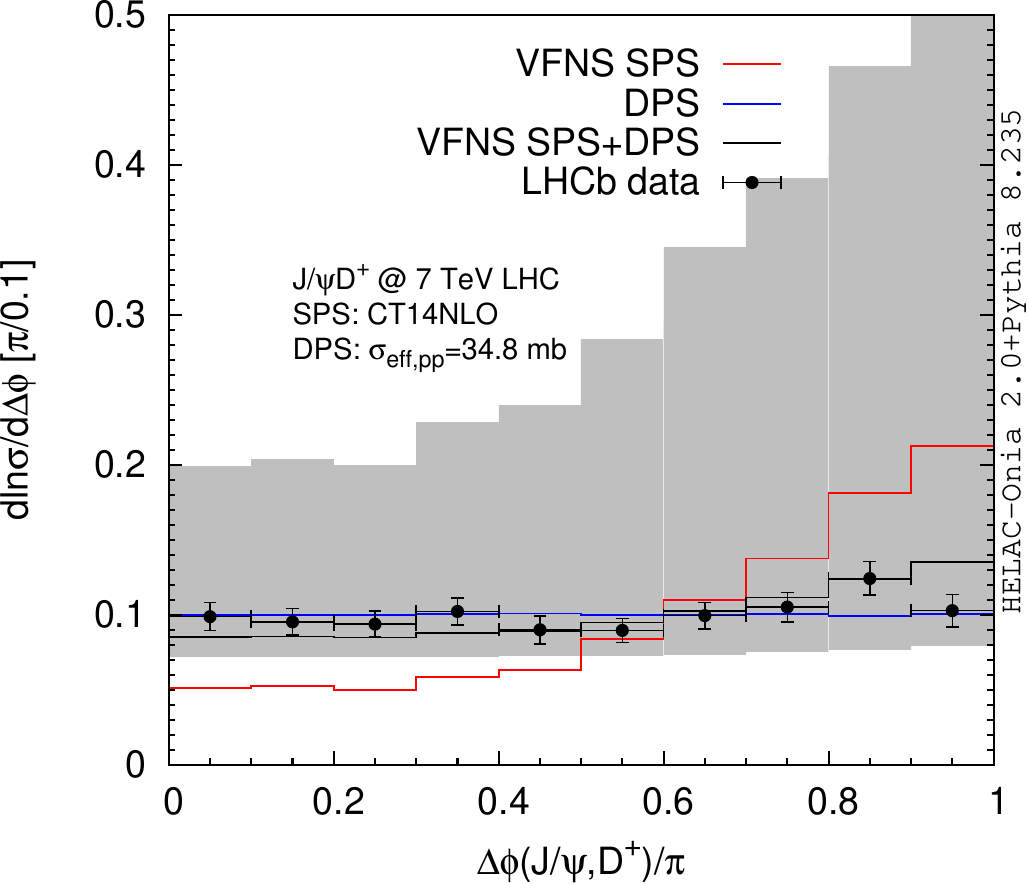}\label{fig:ppDpe}}
 \caption{Same as Fig.~\ref{fig:ppD0} but for $J/\psi+D^+$.\label{fig:ppDp}}
\end{figure}

\begin{figure}
  \centering
  \subfloat[Transverse momentum of $J/\psi$]{
    \includegraphics[width=0.45\textwidth,valign=c,draft=false]{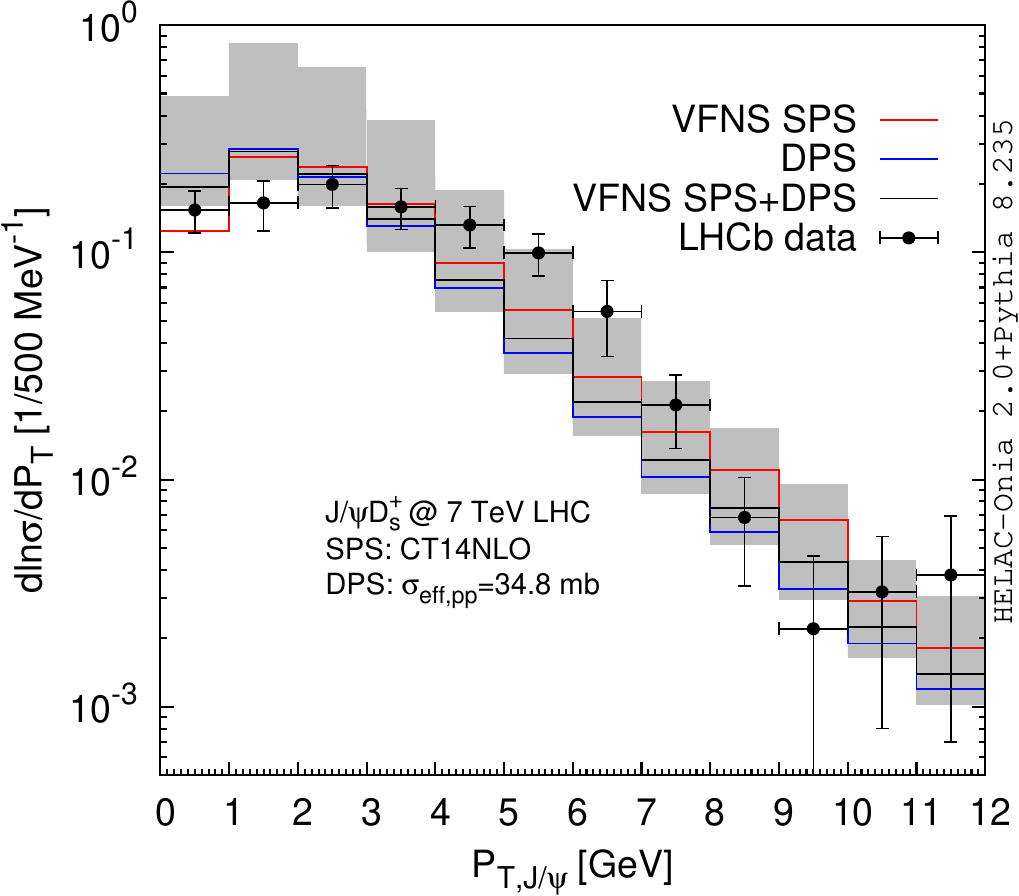}\label{fig:ppDsa}}
  \subfloat[Invariant mass of $J/\psi$ and $D_s^+$]{\includegraphics[width=0.45\textwidth,valign=c,draft=false]{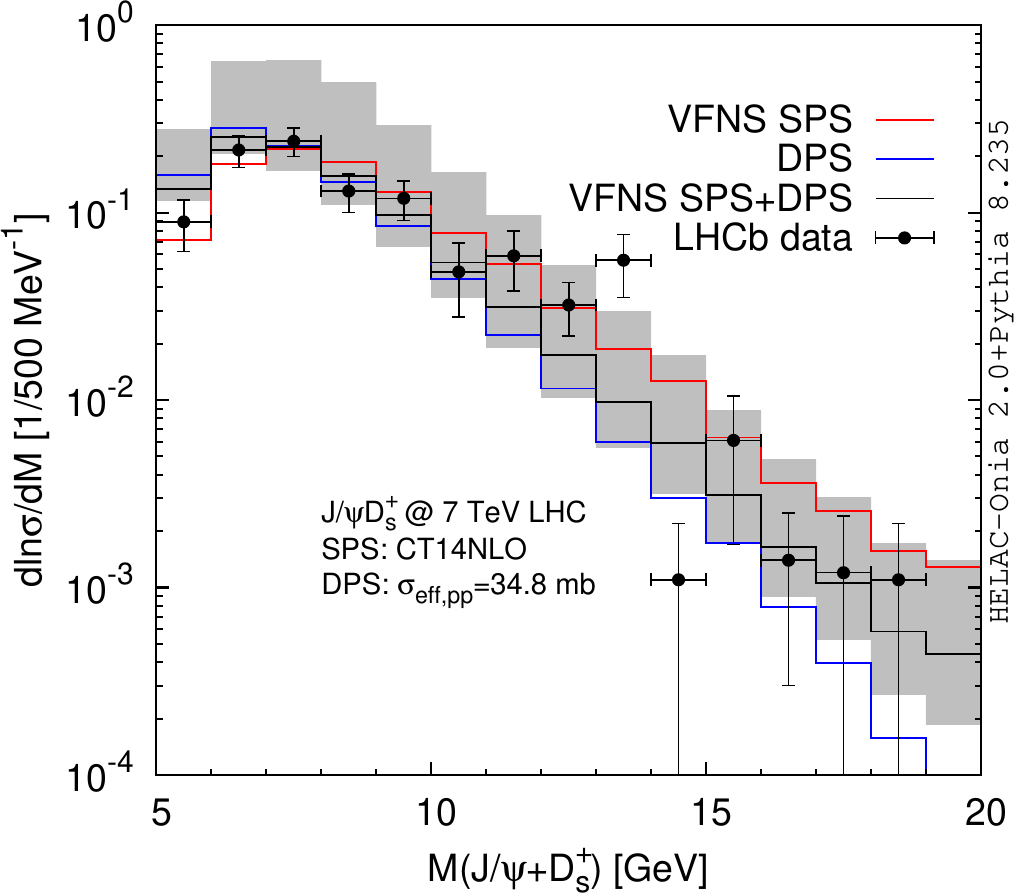}\label{fig:ppDsb}}\\
  \subfloat[Invariant mass of $J/\psi$ and $D_s^+$]{\includegraphics[width=0.45\textwidth,valign=c,draft=false]{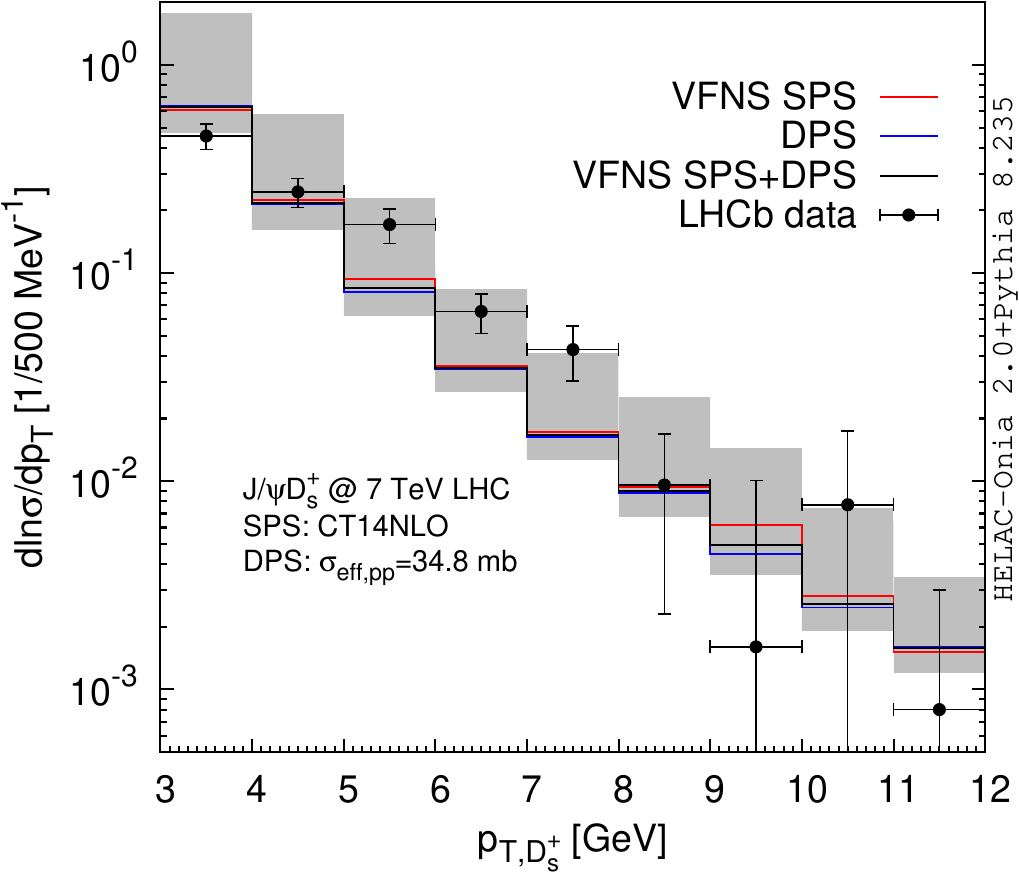}\label{fig:ppDsc}}
  \subfloat[Rapidity gap between $J/\psi$ and $D_s^+$]{\includegraphics[width=0.45\textwidth,valign=c,draft=false]{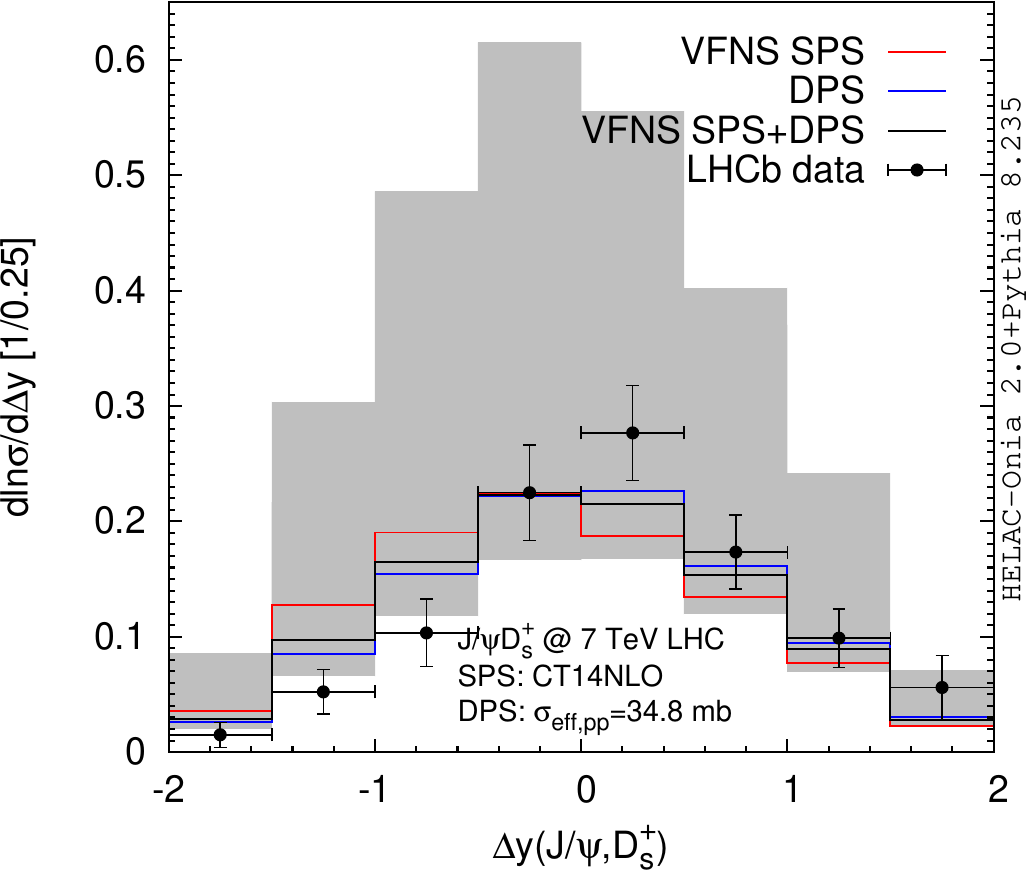}\label{fig:ppDsd}}\\
   \subfloat[Azimuthal angle difference between $J/\psi$ and $D_s^+$]{\includegraphics[width=0.45\textwidth,valign=c,draft=false]{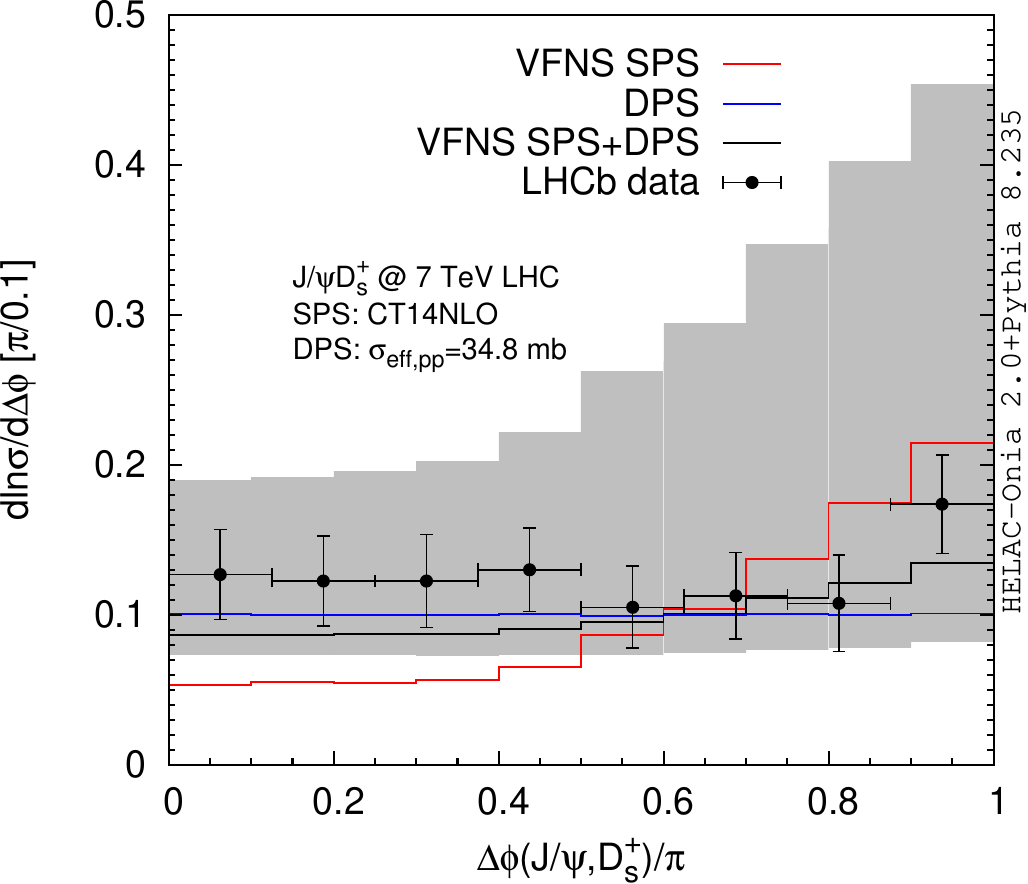}\label{fig:ppDse}}
  \caption{Same as Fig.~\ref{fig:ppD0} but for $J/\psi+D_s^+$.\label{fig:ppDs}}
\end{figure}

\begin{figure}
  \centering
  \subfloat[Transverse momentum of $J/\psi$]{
    \includegraphics[width=0.45\textwidth,valign=c,draft=false]{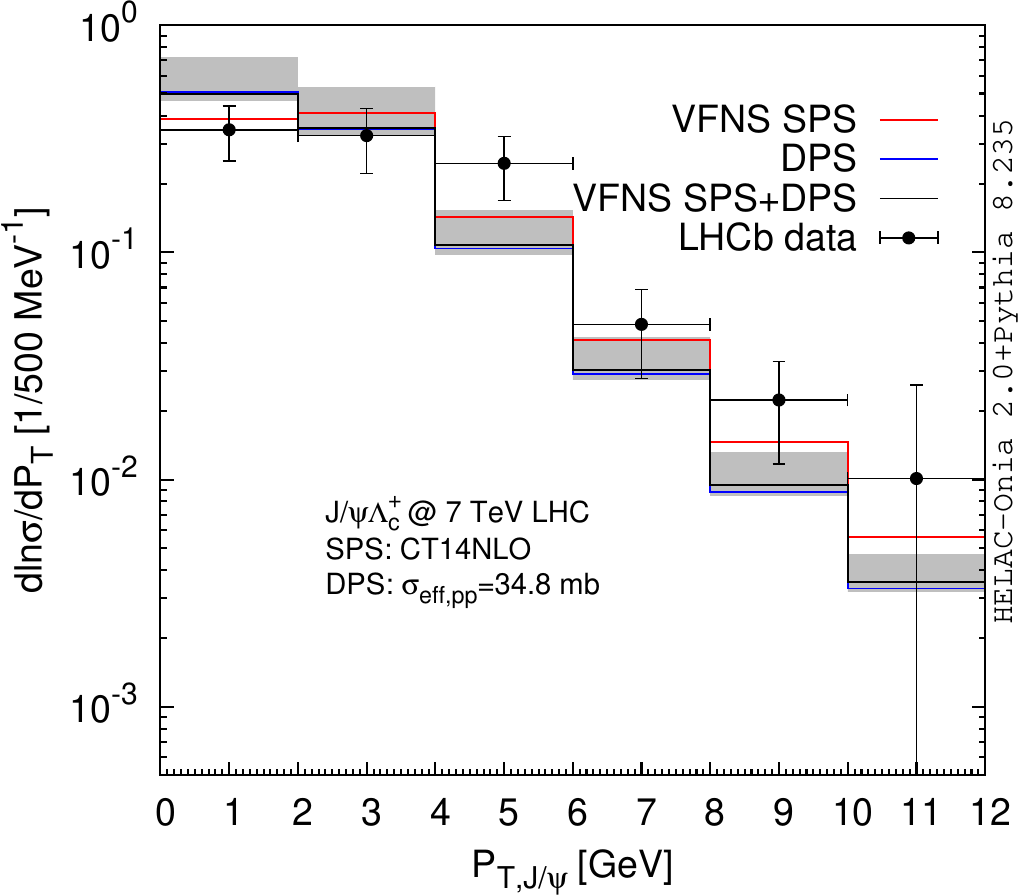}\label{fig:ppLca}}
  \subfloat[Invariant mass of $J/\psi$ and $\Lambda_c^+$]{\includegraphics[width=0.45\textwidth,valign=c,draft=false]{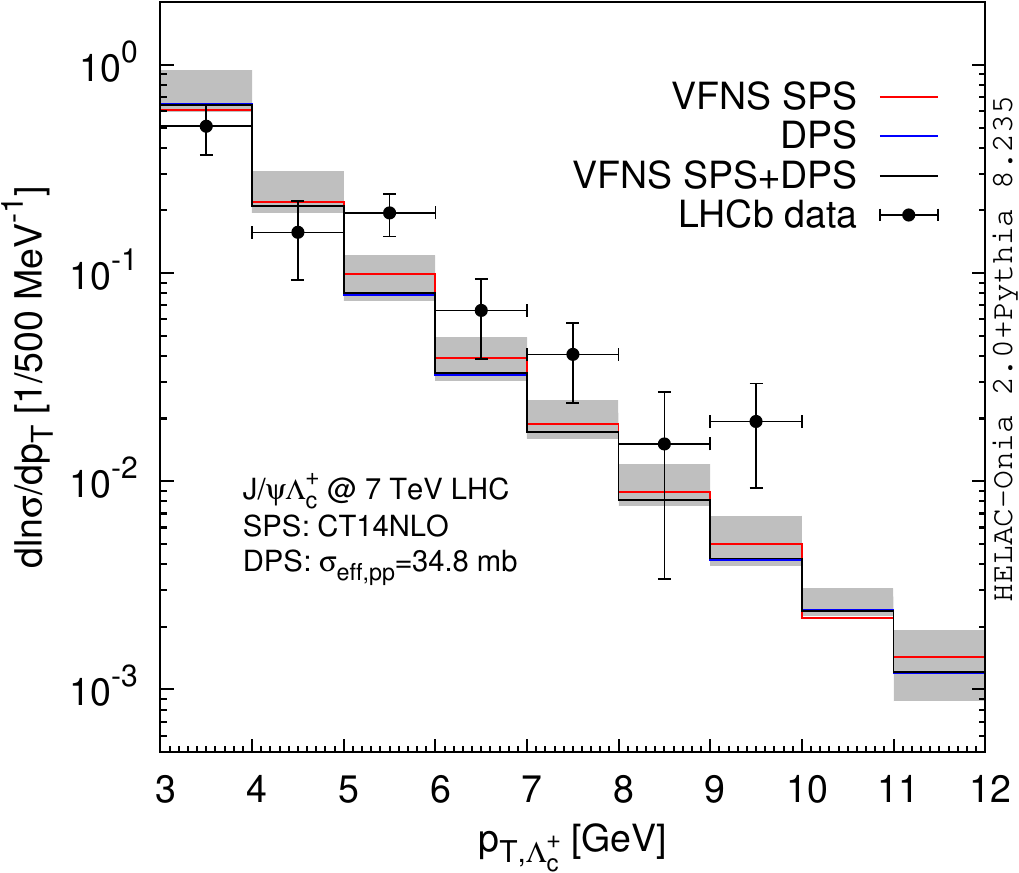}\label{fig:ppLcb}}
  \caption{The differential shape comparison between the LHCb $J/\psi+\Lambda_c^+$ data and the corresponding theoretical calculations. There are only (a) $P_{T,J/\psi}$ and (b) $p_{T,\Lambda_c^+}$ distributions. The differential cross sections have been divided by the corresponding integrated cross sections.
  \label{fig:ppLc}}
\end{figure}

\bibliography{LHCb_JpsiC}

\end{document}